\documentclass[iop,numberedappendix,apj]{emulateapj}
\usepackage{color}
\usepackage{amssymb}
\usepackage{hyperref}

\usepackage{natbib}
\usepackage{graphicx}
\usepackage{epsfig}
\usepackage{afterpage}

\usepackage[tbtags]{amsmath}
\usepackage{hyperref,xcolor}
\hypersetup{colorlinks,linkcolor={blue!50!black},citecolor={blue!50!black},urlcolor={blue!80!black}}
\newcommand {\apgt} {\ {\raise-.5ex\hbox{$\buildrel>\over\sim$}}\ }
\newcommand {\aplt} {\ {\raise-.5ex\hbox{$\buildrel<\over\sim$}}\ }

\newcommand{\asec}{$^{\prime \prime}$}
\newcommand{\asecs}{$^{\prime \prime}\ $}
\newcommand{\amin}{$^{\prime}$}
\newcommand{\amins}{$^{\prime}\ $}

\newcommand{\etal}{{\it et al.}}

\defcitealias{nagai2007}{N07}
\defcitealias{sayers2013}{S13}
\defcitealias{arnaud2010}{A10}
\defcitealias{planck2013a}{P13}
\defcitealias{cavagnolo2009}{C09}
\defcitealias{baldi2014}{B14}
\defcitealias{bulbul2010}{B10}
\defcitealias{vikhlinin2006b}{V06}

\newcommand{\quotes}[1]{``#1''}

\slugcomment{}
\shortauthors{Romero \etal}
\shorttitle{Joint SZE Map Fitting with MUSTANG and Bolocam}
\begin{document}

\title{Galaxy Cluster Pressure Profiles as Determined by Sunyaev Zel'dovich Effect 
  Observations with MUSTANG and Bolocam II: Joint Analysis of Fourteen Clusters}
\author{
%  Order TBD ,
  Charles E. Romero\altaffilmark{1,2},
  Brian S. Mason\altaffilmark{3},
  Jack Sayers\altaffilmark{4},
  Tony Mroczkowski\altaffilmark{5,6},
  Craig Sarazin\altaffilmark{7},
  Megan Donahue\altaffilmark{8},
  Alessandro Baldi\altaffilmark{8},
  Tracy E. Clarke\altaffilmark{6},
  Alexander H.\ Young\altaffilmark{9},
  Jonathan Sievers\altaffilmark{10},
  Simon R. Dicker\altaffilmark{11},
  Erik D.\ Reese\altaffilmark{12},
  Nicole Czakon \altaffilmark{4,13},
  Mark Devlin\altaffilmark{11},
  Phillip M.\ Korngut\altaffilmark{4},
  Sunil Golwala\altaffilmark{4}
} 
\date{\today}

%%%%%%%%%%%%%%%%%%%%%%%%%%%%%%%%%%%%%%%%%%%%%%%%%%%%%%%%%%%%%%%%%%%%%%%%%%%%%%%
\altaffiltext{1}{Institut de Radioastronomie Millim\'{e}trique
300 rue de la Piscine, Domaine Universitaire
38406 Saint Martin d'H\`{e}res, France} 
\altaffiltext{2}{Author contact: \email{romero@iram.fr}}
\altaffiltext{3}{National Radio Astronomy Observatory, 520 Edgemont Rd.,
Charlottesville, VA 22903, USA}
\altaffiltext{4}{Department of Physics, Math, and Astronomy,
  California Institute of Technology, Pasadena, CA 91125, USA}
\altaffiltext{5}{ESO - European Organization for Astronomical Research in the Southern hemisphere, 
Karl-Schwarzschild-Str. 2, D-85748 Garching b. M\"unchen, Germany} 
\altaffiltext{6}{U.S.\ Naval Research Laboratory,
  4555 Overlook Ave SW, Washington, DC 20375, USA}
\altaffiltext{7}{Department of Astronomy, University of Virginia,
  P.O. Box 400325, Charlottesville, VA 22904, USA}
\altaffiltext{8}{Physics and Astronomy Department, Michigan State University,
  567 Wilson Rd., East Lansing, MI 48824, USA}
\altaffiltext{9}{MIT Lincoln Laboratories}
\altaffiltext{10}{Astrophysics \& Cosmology Research Unit, University of KwaZulu-Natal,
  Private Bag X54001, Durban 4000, South Africa}
\altaffiltext{11}{Department of Physics and Astronomy, University of
  Pennsylvania, 209 South 33rd Street, Philadelphia, PA, 19104, USA}
\altaffiltext{12}{Department of Physics, Astronomy, and Engineering, 
  Moorpark College, 7075 Campus Rd., Moorpark, CA 93021, USA} 
\altaffiltext{13}{Academia Sinica, 128 Academia Road, Nankang, Taipei 115, Taiwan}
%\begin{document}

%%%%%%%%%%%%%%%%%%%%%%%%%%%%%%%%%%%%%%%%%%%%%%%%%%%%%%%%%%%%%%%%%%%%%%%%%%%%%%%

\begin{abstract}
We present pressure profiles of galaxy clusters determined from high resolution 
Sunyaev-Zel'dovich (SZ) effect observations of fourteen clusters, which span the
redshift range $ 0.25 < z < 0.89$. 
%We compare our pressure profiles to those from the X-ray data presented by the ACCEPT 
%collaboration, both under the assumption of spherical symmetry. 
The procedure simultaneously fits spherical cluster models to MUSTANG and Bolocam data. In this analysis, 
we adopt the generalized NFW parameterization of pressure profiles to produce our models.
Our constraints on ensemble-average pressure profile parameters, in this study $\gamma$, $C_{500}$, and $P_0$,
are consistent with those in previous studies, but for individual clusters we find discrepancies 
with the X-ray derived pressure profiles from the ACCEPT2 database. We investigate potential sources of these 
discrepancies, especially cluster geometry, electron temperature of the intracluster medium, and substructure.
We find that the ensemble mean profile for all clusters in our sample is described by the parameters: 
$[\gamma,C_{500},P_0] = [0.3_{-0.1}^{+0.1}, 1.3_{-0.1}^{+0.1}, 8.6_{-2.4}^{+2.4}]$, 
for cool core clusters: $[\gamma,C_{500},P_0] = [0.6_{-0.1}^{+0.1}, 0.9_{-0.1}^{+0.1}, 3.6_{-1.5}^{+1.5}]$,
and for disturbed clusters: 
$[\gamma,C_{500},P_0] = [0.0_{-0.0}^{+0.1}, 1.5_{-0.2}^{+0.1},13.8_{-1.6}^{+1.6}]$.
Four of the fourteen clusters have clear substructure in our SZ observations, while an additional
two clusters exhibit potential substructure.
\end{abstract}


\maketitle

%%%%%%%%%%%%%%%%%%%%%%%%%%%%%%%%%%%%%%%%%%%%%%%%%%%%%%%%%%%%%%%%%%%%%%%%%%%%%%%
\section{Introduction}
\label{sec:intro}
%%%%%%%%%%%%%%%%%%%%%%%%%%%%%%%%%%%%%%%%%%%%%%%%%%%%%%%%%%%%%%%%%%%%%%%%%%%%%%%

%\textcolor{red}{Trimming my own notes now.}

Galaxy clusters are the largest gravitationally bound objects in the universe and thus serve as excellent cosmological probes 
and astrophysical laboratories. Within a galaxy cluster, the gas in the intracluster medium (ICM) constitutes 90\% of the
baryonic mass \citep{vikhlinin2006b} and is directly observable in the X-ray due to bremsstrahlung emission. 
At millimeter and sub-millimeter wavelengths, the ICM is observable via the Sunyaev-Zel'dovich (SZ) effect 
\citep{sunyaev1972}: the inverse Compton scattering of cosmic microwave background (CMB) photons off of
the hot ICM electrons. The thermal SZ is observed as an intensity decrement relative to the CMB at wavelengths longer 
than $\sim$1.4 mm (frequencies less than $\sim$220 GHz). The amplitude of the thermal SZ is proportional to the integrated
line-of-sight electron pressure, and is often parameterized as Compton $y$: $y = (\sigma_T / m_e c^2) \int P_e dl$, where
$\sigma_T$ is the Thomson cross section, $m_e$ is the electron mass, $c$ is the speed of light, and $P_e$ is the electron
pressure.
%At longer radio wavelengths, if relativistic electrons are present, parts of the ICM may emit synchrotron emission.

%\textcolor{red}{[I need to revise this paragraph.]}
Cosmological constraints derived from galaxy cluster samples are generally limited by the accuracy of mass calibration of 
galaxy clusters \citep[e.g.][]{hasselfield2013, reichardt2013}, which is often calculated via a scaling relation with 
respect to some integrated observable quantity. Scatter in the scaling relations will then depend on the regularity of 
clusters and the adopted integration radius of the clusters. Determining pressure profiles of galaxy clusters provides an 
assessment of the relative impact and frequency of various astrophysical processes in the ICM and can refine the choice of 
integration radius of galaxy clusters to reduce the scatter in scaling relations.

In the core of a galaxy cluster, some observed astrophysical processes include shocks and cold fronts 
\citep[e.g.][]{markevitch2007}, sloshing \citep[e.g.][]{fabian2006}, and X-ray cavities \citep{mcnamara2007}. 
It is also theorized that helium sedimentation should occur, most noticeably in low redshift, dynamically-relaxed 
clusters \citep{abramopoulos1981, gilfanov1984} 
and recently the expected helium enhancement via sedimentation has been numerically simulated \citep{peng2009}. 
This would result in an offset between X-ray and SZ derived pressure profiles if not accounted for correctly.

At large radii ($R \gtrsim R_{500}$),\footnote{$R_{500}$ is the radius at which the enclosed average mass density is 
500 times the critical density, $\rho_c(z)$, of the universe} equilibration timescales are longer, accretion is ongoing, 
and hydrostatic equilibrium (HSE) can be a poor approximation. Several numerical simulations show that the fractional contribution
from non-thermal pressure increases with radius \citep{shaw2010,battaglia2012,nelson2014}. 
For all three studies, non thermal pressure fractions between 15\% and 30\% are found at ($R \sim R_{500}$)
for redshifts $0 < z < 1$. Additionally, clumping is expected to increase with radius \citep{kravtsov2012}, and is expected to
increase the scatter of pressure profiles at large radii \citep{nagai2011} as well as biasing X-ray derived gas density high,
and thus X-ray derived thermal pressure low \citep{battaglia2015}.

%However, 
By contrast, the intermediate region, between the core and outer regions of the galaxy cluster, 
is often the best region to apply self-similar scaling relations derived from HSE \citep[e.g.][]{kravtsov2012}. 
Moreover, both simulations and observations find low
cluster-to-cluster scatter in pressure profiles within this intermediate radial range \citep[e.g.][]{borgani2004,
nagai2007,arnaud2010,bonamente2012,planck2013a,sayers2013}.

In recent years, the SZ community has often adopted the pressure profile
presented in \citet{arnaud2010} (hereafter, A10), who derive their pressure profiles from X-ray data from the 
REXCESS sample of 31 nearby ($z < 0.2$) clusters out to $R_{500}$ and numerical simulations for larger radii. The
adoption of the A10 pressure profile allows for the extraction of an integrated observable quantity which,
via scaling relations, can then be used to determine the mass of the clusters. In this paper, we use high resolution
SZ data to test the validity of this pressure profile in our sample of 14 clusters at intermediate redshifts.

There are many existing facilities capable of making SZ observations, but most have
angular resolutions of one arcminute or larger. The MUSTANG camera \citep{dicker2008}
on the 100 meter Robert C. Byrd Green Bank Telescope \citep[GBT, ][]{jewell2004} with its angular resolution of 9\asec 
(full-width, half-maximum FWHM) is one of only a few SZ effect instruments with sub-arcminute resolution.
However, MUSTANG's instantaneous field of view (FOV) of 42\asecs means that it is not sensitive to scales over $\sim1$\amin. 
To probe a wider range of scales we complement our MUSTANG data with SZ data from Bolocam \citep{glenn1998}. 
Bolocam is a 144-element bolometer
array on the Caltech Submillimeter Observatory (CSO) with a beam FWHM of 58\asecs at 140 GHz and circular FOV with 8\amins 
diameter, which is well matched to the angular size of $R_{500}$ ($\sim 4$\amin) of the clusters in our sample. 

This paper is organized as follows. In Section~\ref{sec:obs} we describe the MUSTANG and Bolocam observations and reduction. 
In Section~\ref{sec:jointfitting} we review the method used to jointly fit pressure profiles to MUSTANG and Bolocam data. We
present results from the joint fits in Section~\ref{sec:pp_constraints} and compare our results to X-ray derived pressures 
in Section~\ref{sec:xray_comp}. 
Throughout this paper we assume a $\Lambda$CDM cosmology with $\Omega_m = 0.3$, $\Omega_{\lambda} = 0.7$, and $H_0 = 70$ 
km s$^{-1}$ Mpc$^{-1}$. For the remainder of the paper we denote the electron pressure as $P$, electron density as $n_e$, 
and electron temperature as $T$. The errors we report are $1\sigma$ (68.5\% confidence) unless otherwise noted.
%consistent with the 9-year \emph{Wilkinson Microwave Anisotropy Probe} (WMAP) results reported in \cite{hinshaw2013}.

%%%%%%%%%%%%%%%%%%%%%%%%%%%%%%%%%%%%%%%%%%%%%%%%%%%%%%%%%%%%%%%%%%%%%%%%%%%%%%%
\section{Observations and Data Reduction}
\label{sec:obs}
%%%%%%%%%%%%%%%%%%%%%%%%%%%%%%%%%%%%%%%%%%%%%%%%%%%%%%%%%%%%%%%%%%%%%%%%%%%%%%%

\subsection{Sample}

Our cluster sample is based primarily on the Cluster Lensing And Supernova survey with Hubble (CLASH) sample
%, which is a 524-orbit multi-cycle treasury program 
\citep{postman2012}.
%One of its main goals is to ``measure the profiles and substructures of dark matter 
%in galaxy clusters with unprecedented precision and resolution'' \citep{postman2012}. 
The CLASH sample has 25 massive galaxy clusters, 20 of which are selected from X-ray data 
(from \emph{Chandra X-ray Observatory}, hereafter \emph{Chandra}), and 5 based on exceptional lensing strength. 
These clusters have the following properties: 
$0.187 < z < 0.890$, $5.5 < k_B T$ (keV)$ < 15.5$, and $6.7 \times 10^{44} < L_{bol}$ 
(erg s$^{-1}$) $<90.8$. Thus, these clusters are large enough that we should expect to detect 
them with MUSTANG with a reasonable 
amount of time on the sky (on average, $<$25 hours per cluster).

%While these clusters are not a complete sample, many already have SZ effect observations from the Sunyaev-Zel'dovich 
%Array (SZA), AMiBA, or Bolocam, making them well studied, and deserving of high resolution SZ effect measurements. 
%The wealth of observations on these clusters will allow us to constrain pressure and 
%mass profiles of clusters as well as the impact of substructure. Additionally, we will be able to assess discrepancies
%between X-ray derived properties, shown in Table~\ref{tbl:cluster_properties} 
%and compare to SZ derived properties. 

\begin{figure*}[!h]
  \centering
  \begin{tabular}{cccc}
   \epsfig{file=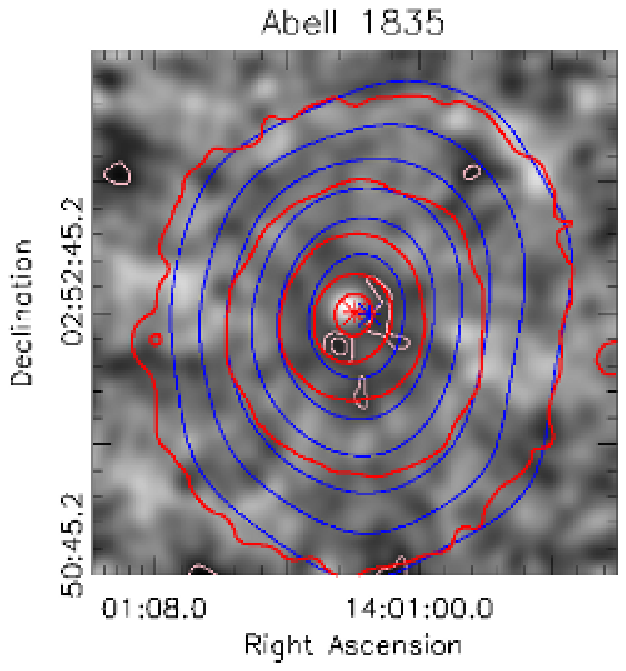,width=0.25\linewidth,clip=}   &
   \epsfig{file=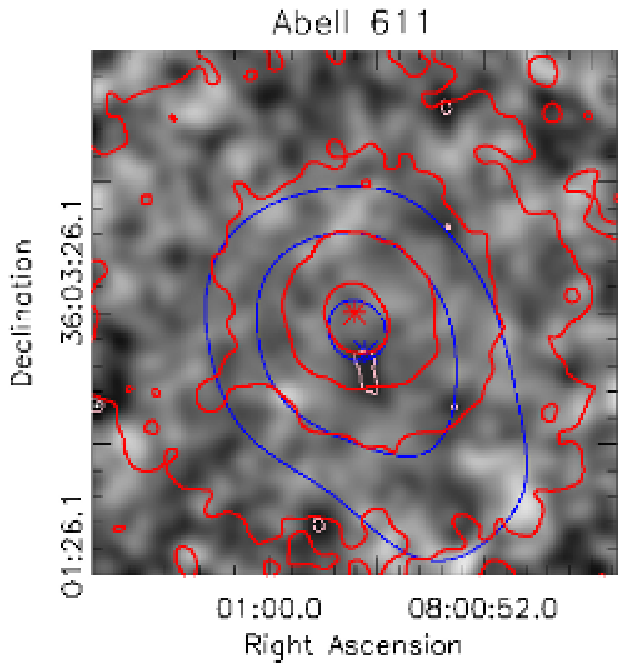,width=0.25\linewidth,clip=}    &
   \epsfig{file=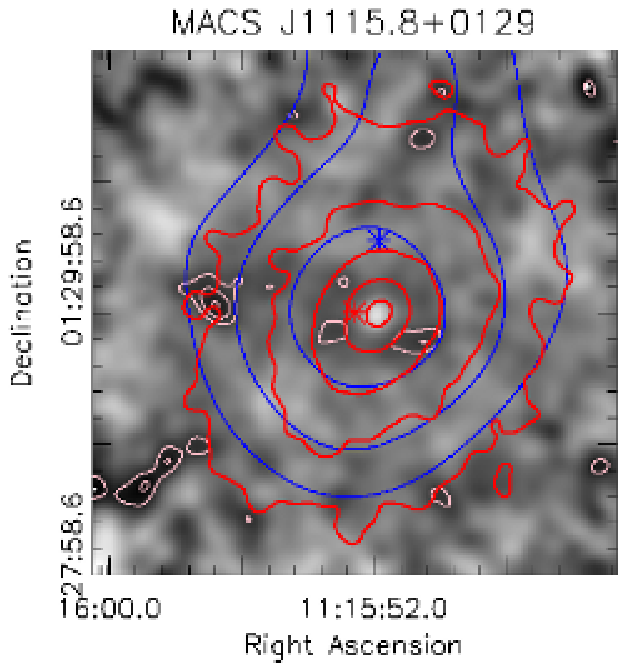,width=0.25\linewidth,clip=}    &
   \epsfig{file=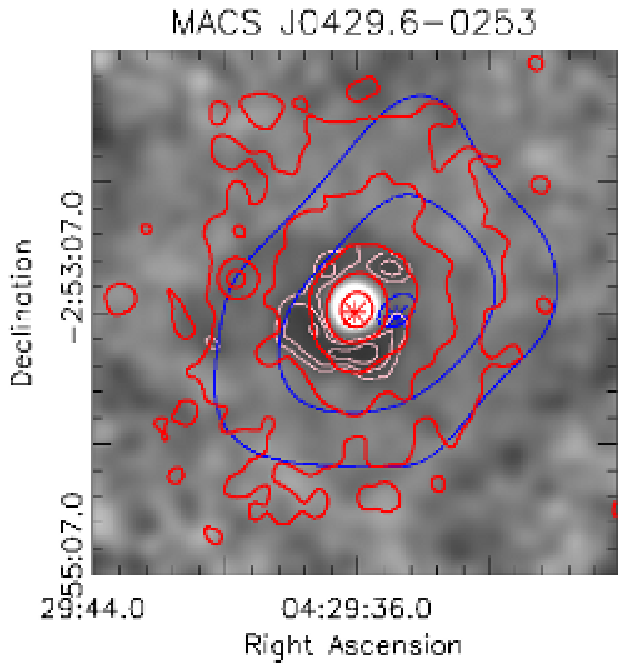,width=0.25\linewidth,clip=}    \\
   \epsfig{file=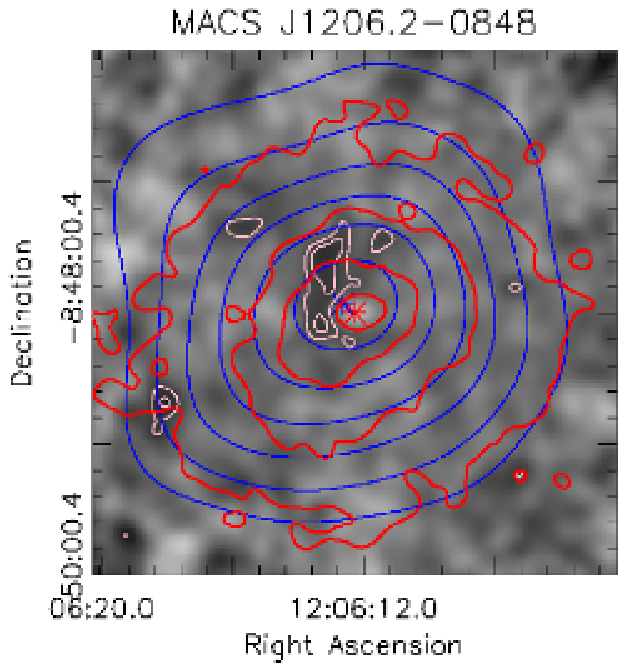,width=0.25\linewidth,clip=}    &
   \epsfig{file=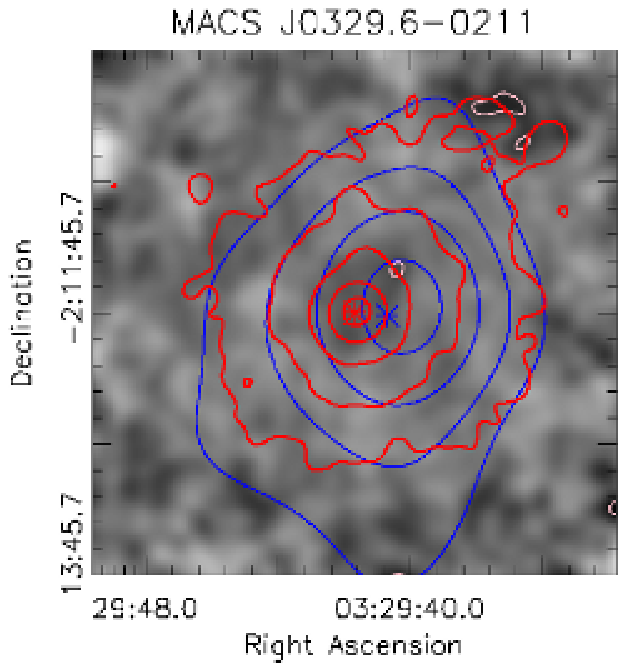,width=0.25\linewidth,clip=}    &
   \epsfig{file=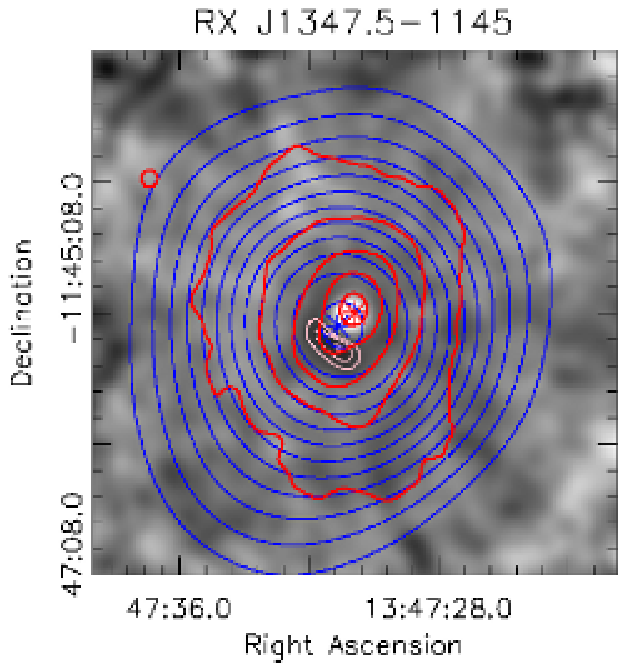,width=0.25\linewidth,clip=}    &
   \epsfig{file=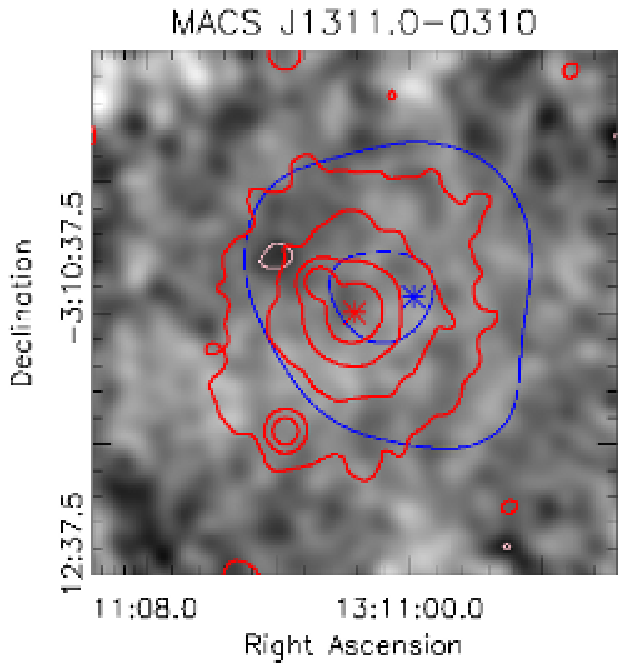,width=0.25\linewidth,clip=}    \\
   \epsfig{file=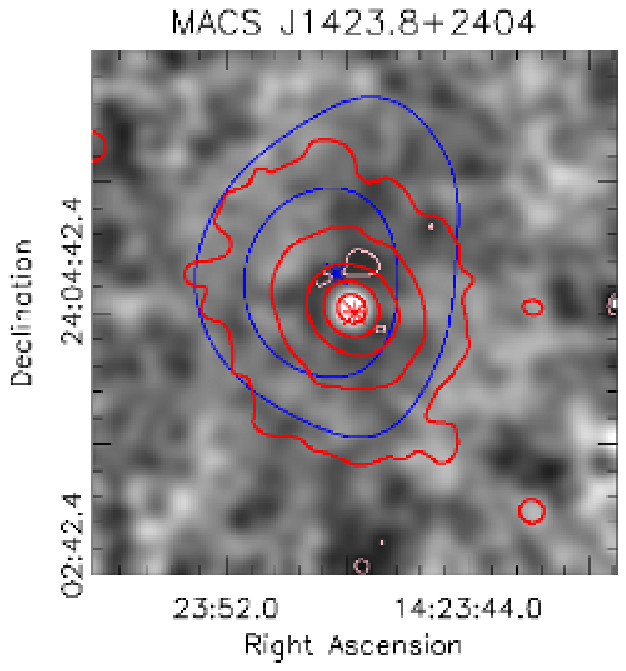,width=0.25\linewidth,clip=}    &
   \epsfig{file=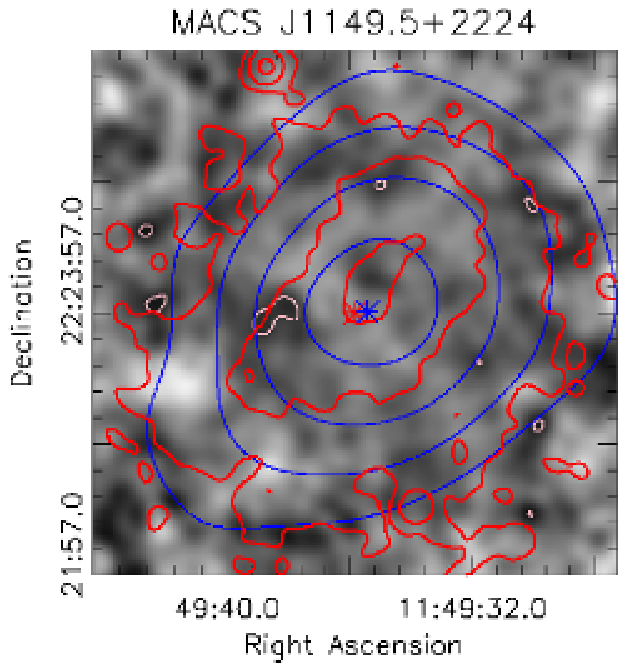,width=0.25\linewidth,clip=}    &
   \epsfig{file=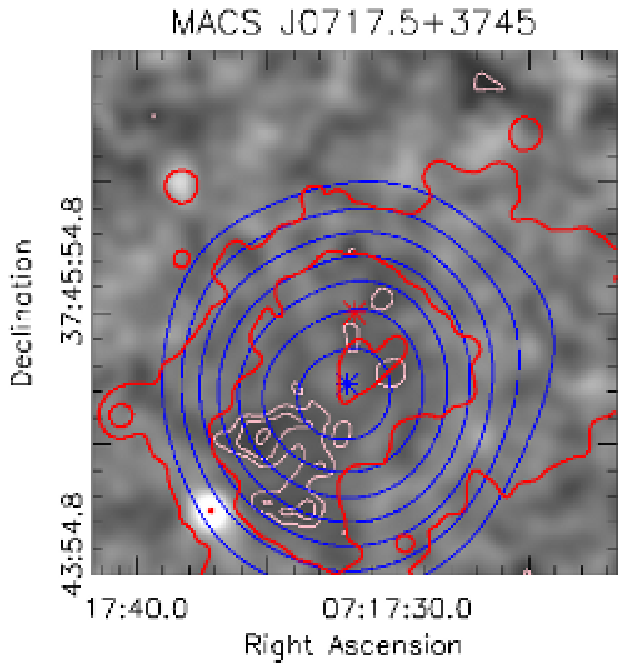,width=0.25\linewidth,clip=}    &
   \epsfig{file=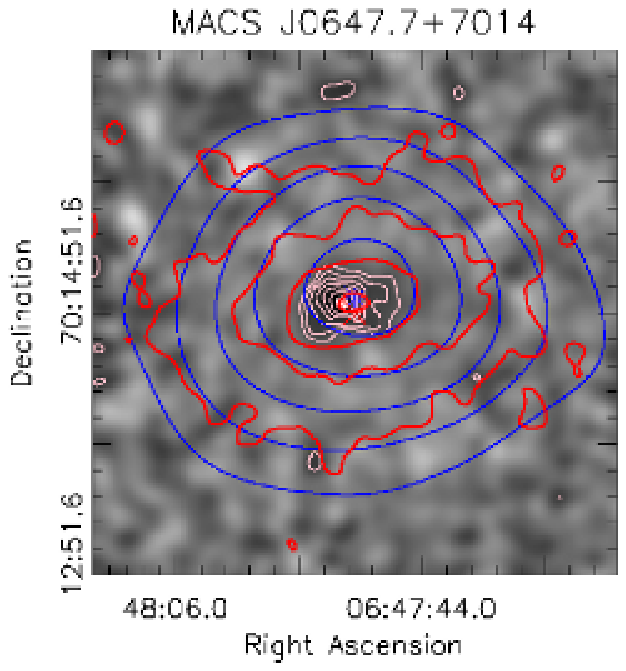,width=0.25\linewidth,clip=}    \\
   \epsfig{file=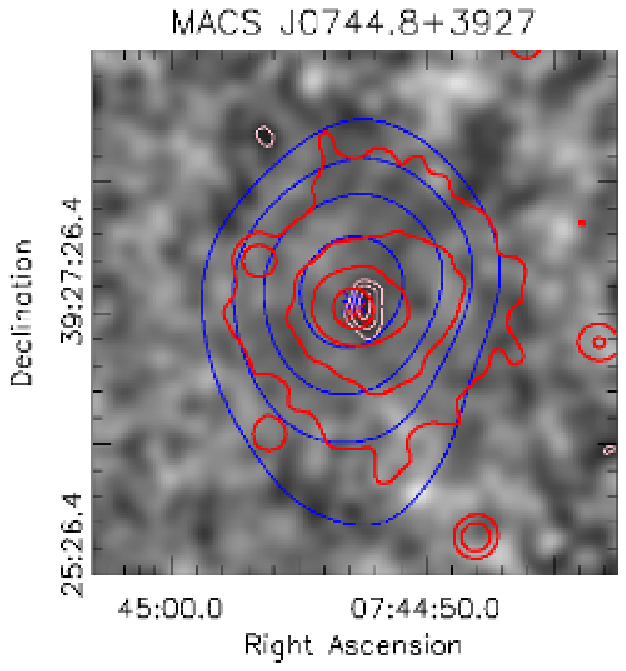,width=0.25\linewidth,clip=}    &
   \epsfig{file=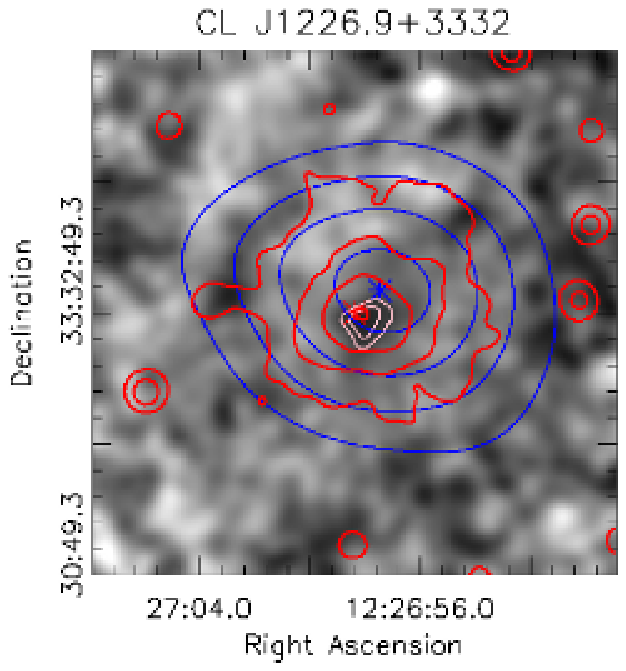,width=0.25\linewidth,clip=}    &
     &
  \end{tabular}
  \caption{The grey scale shows MUSTANG maps of the clusters in our sample, in Jy/beam. The color scaling spans 
    the range $\pm5\times$Noise$_{M}$, where Noise$_{M}$ (for MUSTANG) is given in Table~\ref{tbl:cluster_obs}. 
    Pale contours are MUSTANG contours; blue contours are Bolocam. Both start at $3\sigma$ decrement 
    (i.e. negative), with $1\sigma$ intervals for MUSTANG and $2\sigma$ intervals for Bolocam.
    Red contours are X-ray surface brightness contours at arbitrary levels. The red asterisk is the 
    ACCEPT centroid; the blue asterisk is the Bolocam centroid.}
% Currently Figure 1. (Nov. 2016)
  \label{fig:mustang_maps_sample}
\end{figure*}

\begin{deluxetable*}{llclllllllll}
\tabletypesize{\scriptsize}
\tablecolumns{10}
\tablewidth{0pt}
% Currently Table 1. (Nov. 2016)
\tablecaption{Cluster properties \label{tbl:cluster_properties}}
\tablehead{ 
  \colhead{Cluster} & \colhead{$z$} & \colhead{$M_{500}$} & \colhead{$P_{500}$} & \colhead{$R_{500}$} & \colhead{$R_{500}$} &
  \colhead{$T_x^a$} & \colhead{$T_x^b$} & \colhead{$T_{mg}$} & \colhead{Dynamical} & \colhead{$\Delta r_{X,SZ}$} \\
  \colhead{}  & \colhead{} & \colhead{($10^{14} M_{\odot}$)} & \colhead{(keV/cm$^{3}$)} & \colhead{(kpc)} & 
  \colhead{(\amin)} & \colhead{(keV)} & \colhead{(keV)} & \colhead{(keV)} & \colhead{state} & \colhead{(\asec)}
}
\startdata
    \textbf{Abell 1835}  & 0.253 & 12   & 0.00594   & 1490   & 6.30 & 9.0  & 10.0 & 7.49 & CC      & 6.8   \\
    \textbf{Abell 611}   & 0.288 & 7.4  & 0.00445   & 1240   & 4.75 & 6.8  & --   & 6.71 & --      & 18.7  \\
    \textbf{MACS1115}    & 0.355 & 8.6  & 0.00545   & 1280   & 4.28 & 9.2  & 9.14 & 7.04 & CC      & 34.8  \\
    \textbf{MACS0429}    & 0.399 & 5.8  & 0.00448   & 1100   & 3.41 & 8.3  & 8.55 & 5.56 & CC      & 18.7  \\
    \textbf{MACS1206}    & 0.439 & 19   & 0.01059   & 1610   & 4.73 & 10.7 & 11.4 & 10.0 & --      & 6.9   \\
    \textbf{MACS0329}    & 0.450 & 7.9  & 0.00596   & 1190   & 3.44 & 6.3  & 5.85 & 5.64 & CC \& D & 14.8  \\
    \textbf{RXJ1347}     & 0.451 & 22   & 0.01171   & 1670   & 4.83 & 10.8 & 13.6 & 9.86 & CC      & 9.6   \\
    \textbf{MACS1311}    & 0.494 & 3.9  & 0.00399   & 930    & 2.56 & 6.0  & 6.36 & 5.18 & CC      & 27.7  \\
    \textbf{MACS1423}    & 0.543 & 6.6  & 0.00612   & 1090   & 2.85 & 6.9  & 6.81 & 5.50 & CC      & 19.8  \\
    \textbf{MACS1149}    & 0.544 & 19   & 0.01228   & 1530   & 4.01 & 8.5  & 8.76 & 7.70 & D       & 6.0   \\
    \textbf{MACS0717}    & 0.546 & 25   & 0.01490   & 1690   & 4.40 & 11.8 & 10.6 & 9.06 & D       & 32.4  \\
    \textbf{MACS0647}    & 0.591 & 11   & 0.00923   & 1260   & 3.17 & 11.5 & 12.6 & 8.06 & --      & 6.9   \\
    \textbf{MACS0744}    & 0.698 & 13   & 0.01199   & 1260   & 2.96 & 8.1  & 8.90 & 6.85 & D       & 4.9   \\
    \textbf{CLJ1226}     & 0.888 & 7.8  & 0.01184   & 1000   & 2.15 & 12.0 & 11.7 & 11.3 & --      & 15.3  
%    \hline
%    Abell 383            & 0.187 & 4.7  & 0.00285   & 1110   & 5.4  & 5.47 & --   & CC      & --    \\
%    Abell 209            & 0.206 & 13   & 0.00564   & 1530   & 8.2  & 8.69 & --   & --      & --    \\
%    Abell 1423           & 0.213 & 8.7  & 0.00445   & 1350   & 5.8  & 6.61 & --   & --      & --    \\
%    Abell 2261           & 0.224 & 14   & 0.00632   & 1590   & 6.1  & 8.09 & --   & CC      & --    \\
%    RXJ2129              & 0.234 & 7.7  & 0.00423   & 1280   & 6.3  & 7.78 & --   & CC      & --    \\
%    MS 2137              & 0.313 & 4.7  & 0.00342   & 1060   & 4.7  & --   & --   & CC      & --    \\
%    RXC J2248            & 0.348 & 22   & 0.01014   & 1760   & 10.9 & 11.5 & --   & --      & --    \\
%    MACS1931             & 0.352 & 9.9  & 0.00595   & 1340   & 7.5  & 7.92 & --   & CC      & --    \\
%    MACS1532             & 0.362 & 9.5  & 0.00589   & 1310   & 6.8  & 6.47 & --   & CC      & --    \\
%    MACS1720             & 0.387 & 6.3  & 0.00465   & 1140   & 7.9  & 6.50 & --   & CC      & --    \\
%    MACS0416             & 0.397 & 9.1  & 0.00625   & 1270   & 8.2  & 8.14 & --   & --      & --    \\
%    MACS2129             & 0.570 & 11   & 0.00903   & 1250   & 8.6  & 8.11 & --   & D       & --    
\enddata
\tablecomments{$z$, $M_{500}$, $R_{500}$, and $T_X^a$ are taken from \citet{mantz2010}:  
  $T_X^a$ is calculated from a 
  single spectrum over $0.15 R_{500} < r < R_{500}$ for each cluster. $T_X^b$ is from \citet{morandi2015},
  and is calculated over $0.15 R_{500} < r < 0.75 R_{500}$.  $T_{mg}$ is a fitted gas mass weighted temperature,
  (Section~\ref{sec:temp_profiles}) determined by fitting the ACCEPT2 \citep{baldi2014} temperature profiles to 
  the gas mass weighted profile found in \citet{vikhlinin2006b}. 
  The dynamical states: cool core (CC) and disturbed (D) are taken from (and defined in) \citet{sayers2013}. 
  %The bolded clusters are the 14 clusters in our sample.
  $\Delta r_{X,SZ}$ denotes the offset between the ACCEPT and Bolocam centroids.}
\end{deluxetable*}

Of the 25 clusters in the CLASH sample, four are too far south to be observed with MUSTANG from Green Bank, WV.
Of the remaining 21, we were able to observe fourteen given the available good weather and their limited visibility
during the observational campaign from 2009 to 2014.
Abell 209 was observed, but was relatively noisy and showed no trace of any detection. Our final sample
includes thirteen CLASH clusters. We also include Abell 1835, a cluster of similar mass and redshift as the CLASH
clusters, which was observed under the program GBT/09A-052. These clusters (see Table~\ref{tbl:cluster_properties}
and Figure~\ref{fig:mustang_maps_sample}) were also observed with Bolocam, and have been analyzed in 
\citet{sayers2012, sayers2013, czakon2015}. The
\emph{Archive of Chandra Cluster Entropy Profile Tables} \citep[ACCEPT][]{cavagnolo2009}) and Bolocam centroids
are indicated in Figure~\ref{fig:mustang_maps_sample} with red and blue asterisks respectively, and their separations
($\Delta r_{X,SZ}$) are also listed in Table~\ref{tbl:cluster_properties}. The total integration times of
MUSTANG and Bolocam observations, along with detection significances, of our sample are listed in Table~\ref{tbl:cluster_obs}. Bolocam and MUSTANG significances, A10$_B$ and A10$_M$ respectively, are taken
  as the significance of the fitted spherical A10 \citep{arnaud2010} profile
  (see Section~\ref{sec:bulk_ICM}) based on the amplitude of the fit ($P_0/\sigma_{P_0}$) to the respective
  dataset (separately). Aside from fixing the pressure profile shape, the fits are performed as described
  in Section~\ref{sec:jointfitting}, with relevant (point source and/or residual) components fit simultaneously.
  This calculation of cluster significance is better than a peak surface brightness measure as it incorporates signal,
  even if weak, within the entire fitted region. As this metric is intended to measure the strength of an overall cluster
  detection, negative values are permitted. Null detections with MUSTANG set upper limits on the slope of the inner
  pressure profile, which are stronger than those from Bolocam data.

\begin{deluxetable*}{llllllllllll}
\tabletypesize{\footnotesize}
\tablecolumns{10}
\tablewidth{0pt} 
% Currently Table 2. (Nov. 2016)
\tablecaption{Bolocam and MUSTANG observational properties. \label{tbl:cluster_obs}}
\tablehead{ 
    \colhead{Cluster} & \colhead{$z$} & \colhead{R.A.} & \colhead{Decl.} & 
              \colhead{$t_{obs,B}$} & \colhead{Noise$_{B}$} & \colhead{A10$_{B}$} & 
              \colhead{$\tilde{\chi}_{B}^2$} & 
              \colhead{$t_{obs,M}$} & \colhead{Noise$_{M}$} & \colhead{A10$_{M}$} &
              \colhead{$\tilde{\chi}_{M}^2$} \\
            & \colhead{} & \colhead{(J2000)} & \colhead{(J2000)} &  
              \colhead{(hours)} & \colhead{$\mu K\dag$} &
              \colhead{($P_0/\sigma_{P_0}$)} & \colhead{} & 
              \colhead{(hours)} & \colhead{$\mu$Jy/bm}  &
              \colhead{($P_0/\sigma_{P_0}$)} & \colhead{} 
}
\startdata
    \textbf{Abell 1835}  & 0.253 & 14:01:01.9 & +02:52:40 & 14.0 & 16.2 & 28.9 & 1.05 & 8.6  & 53.4 & 10.0 & 0.99 \\
    \textbf{Abell 611}   & 0.288 & 08:00:56.8 & +36:03:26 & 18.7 & 25.0 & 13.9 & 0.97 & 12.0 & 46.2 & 1.73 & 1.03 \\
    \textbf{MACS1115}    & 0.355 & 11:15:51.9 & +01:29:55 & 15.7 & 22.8 & 16.3 & 1.08 & 10.0 & 56.4 & 8.66 & 1.04 \\
    \textbf{MACS0429}    & 0.399 & 04:29:36.0 & -02:53:06 & 17.0 & 24.1 & 13.2 & 1.05 & 11.6 & 47.2 & -0.02& 1.03 \\
    \textbf{MACS1206}    & 0.439 & 12:06:12.3 & -08:48:06 & 11.3 & 24.9 & 28.7 & 0.97 & 13.3 & 42.5 & 8.89 & 1.02 \\
    \textbf{MACS0329}    & 0.450 & 03:29:41.5 & -02:11:46 & 10.3 & 22.5 & 17.4 & 1.09 & 13.1 & 39.9 & 8.63 & 0.98 \\
    \textbf{RXJ1347}     & 0.451 & 13:47:30.8 & -11:45:09 & 15.5 & 19.7 & 45.3 & 1.04 & 1.9  & 276. & 8.90 & 0.98 \\
    \textbf{MACS1311}    & 0.494 & 13:11:01.7 & -03:10:40 & 14.2 & 22.5 & 11.3 & 1.06 & 10.6 & 64.5 & 0.71 & 1.00 \\
    \textbf{MACS1423}    & 0.543 & 14:23:47.9 & +24:04:43 & 21.7 & 22.3 & 11.8 & 0.88 & 11.2 & 35.7 & 6.15 & 1.00 \\
    \textbf{MACS1149}    & 0.544 & 11:49:35.4 & +22:24:04 & 17.7 & 24.0 & 22.0 & 0.99 & 13.9 & 32.7 & -1.47& 1.01 \\
    \textbf{MACS0717}    & 0.546 & 07:17:32.1 & +37:45:21 & 12.5 & 29.4 & 31.3 & 1.09 & 14.6 & 27.1 & 3.05 & 1.05 \\
    \textbf{MACS0647}    & 0.591 & 06:47:49.7 & +70:14:56 & 11.7 & 22.0 & 24.1 & 1.03 & 16.4 & 20.3 & 11.3 & 1.01 \\
    \textbf{MACS0744}    & 0.698 & 07:44:52.3 & +39:27:27 & 16.3 & 20.6 & 17.8 & 1.19 & 7.6  & 48.5 & 7.67 & 1.01 \\
    \textbf{CLJ1226}     & 0.888 & 12:26:57.9 & +33:32:49 & 11.8 & 22.9 & 13.7 & 1.20 & 4.9  & 85.6 & 9.43 & 1.00  
\enddata
\tablecomments{Subscripts $_{B}$ and $_{M}$ denote Bolocam and MUSTANG properties respectively. Noise$_{B}$
  and $t_{obs,B}$ are those reported in \citet{sayers2013}. \dag$\mu K$ is more precisely $\mu K_{CMB}$-amin.
  Noise$_{M}$ is calculated on MUSTANG maps with $10$\asecs smoothing, in the central arcminute.
  $t_{obs}$ are the integration times (on source) for the 
  given instruments. A10$_B$ and A10$_M$ are the Bolocam and MUSTANG significances, respectively.
    The quality of the fits is respectable, as indicated by the $\tilde{\chi}^2$ values being close to 1.}
  %For these fits, the degrees of freedom in Bolocam maps are always 1599, and in the MUSTANG maps
  %  the degrees of freedom are between 11275 and 11282.}
\end{deluxetable*}

%The clusters were observed with
%MUSTANG over the projects AGBT08A\_056, AGBT09A\_052, AGBT09C\_059, AGBT10A\_056, AGBT10C\_017, AGBT10C\_026, AGBT10C\_042, 
%AGBT10C\_031, AGBT11A\_009, and AGBT11B\_001.

%%%%%%%%%%%%%%%%%%%%%%%%%%%%%%%%%%%%%%%%%%%%%%%%%%%%%%%%%%%%%%%%%%%%%%%%%%%%%%%
\subsection{MUSTANG Observations and Reduction}
\label{sec:musobs}
%%%%%%%%%%%%%%%%%%%%%%%%%%%%%%%%%%%%%%%%%%%%%%%%%%%%%%%%%%%%%%%%%%%%%%%%%%%%%%%

MUSTANG is a 64 pixel array of Transition Edge Sensor (TES) bolometers arranged in an $8 \times 8$ array
located at the Gregorian focus on the 100 m GBT. Operating at 90 GHz (81--99~GHz),
MUSTANG has an angular resolution of 9\asecs and pixel spacing of 0.63$f \lambda$ resulting in a FOV
of 42\asec. More detailed information about the instrument can be found in \citet{dicker2008}.

Our observations and data reduction are described in detail in \citet{romero2015a}, and we briefly review them
here. Absolute flux calibrations are based on the planets Mars, Uranus, and Saturn; or the star Betelgeuse 
($\alpha_{Ori}$). At least one of these flux calibrators was observed at least once per night, and we find our 
calibration is accurate to a 10\% RMS uncertainty. We also observe bright point sources every half hour
to track our pointing and beam shape. To observe the target galaxy clusters, we employ Lissajous daisy scans 
with a $3\arcmin$ radius and in many of the clusters we broadened our coverage with a hexagonal pattern of 
daisy centers (with $1\arcmin$ offsets). 
For most clusters, the coverage (weight) drops to 50\% of its peak value at a radius of $1.3\arcmin$.
%'

Processing of MUSTANG data is performed using a custom IDL pipeline. Raw data is recorded as time ordered data (TOD)
from each of the 64 detectors. An outline of the data processing for each scan on a galaxy cluster is as follows:
  
  (1) We define a pixel mask from the nearest preceding CAL scan; unresponsive detectors are masked out.
  The CAL scan provides us with unique gains to be applied to each of the responsive detectors.

  (2) A common mode template, polynomial, and sinusoid are fit to the data and then subtracted. The common mode is
  calculated as the arithmetic mean of the TOD across detectors.

  (3) After the common mode and polynomial subtraction each scan is subjected to spike (glitch), skewness, and Allan 
  variance tests and are flagged according to the following criteria. Glitches are flagged as $4\sigma$ excursions based
  on the median absolute deviation; The skewness threshold for flagging is 0.4. Flags based on Allan variance require
  the variance over a two second interval to be greater than 9 times the variance between each integration. Typical scan
  integration times were 150 seconds.

  (4) Individual detector weights are calculated as $1/ \sigma_i^2$, where $\sigma_i$ is the RMS of the non-flagged
  TOD for that detector. 

  (5) Maps are produced by gridding the TOD in 1\asec pixels in Right Ascension (R.A.) and Declination (Dec). A weight 
  map is produced in addition to the signal map.

\begin{figure}
  \begin{center}
  \includegraphics[width=0.5\textwidth]{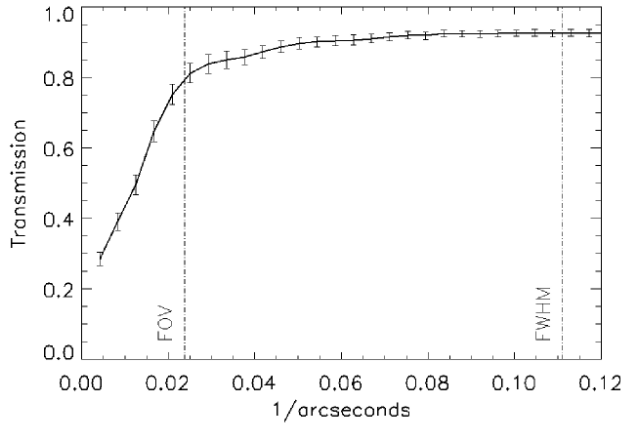}
  \end{center}
  \caption{Effective average transfer function of our MUSTANG data reduction over our sample. 
    The variations between cluster are less than 3\%. For each cluster, attenuation is
    calculated based on simulated observations of 25 fake skies. The plotted one-dimensional
    transfer function is the weighted average of the transfer functions
    of individual clusters. The error bars show the scatter among cluster transfer functions. 
    The transfer functions (transmission) of individual clusters are calculated as the square
    root of the ratio of the one dimensional power spectra of the observed fake sky
    and input fake sky. We have labelled the relevant angular wavenumbers for the FOV and FWHM.}
% Currently Figure 2. (Nov. 2016)
  \label{fig:xfer_all}
\end{figure}

%The effective transfer function over our sample has been quantified and is shown in Figure~\ref{fig:xfer_all}.
The effect of the MUSTANG data processing results in the transfer function shown in 
  Figure~\ref{fig:xfer_all}. Specifically, it is the average across our sample. This transfer function is
  very stable as little scatter is seen across our sample.
  
\subsection{Bolocam Observations and Reduction}
\label{sec:bolocamredox}

Bolocam is a 144-element camera that was a facility instrument on the Caltech Submillimeter Observatory (CSO) from
2003 until 2012. Its field of view is 8\amins in diameter, and at 140 GHz it has a resolution of 58\asecs FWHM
(\citet{glenn1998,haig2004}). The clusters were observed with a Lissajous pattern that results in a tapered
coverage dropping to 50\% of the peak value at a radius of roughly 5\amin, and to 0 at a radius of 10\amin.
The Bolocam maps used in this analysis are $14\arcmin \times 14\arcmin$. The Bolocam data 
\footnote{Bolocam data is publicly available at 
\href{http://irsa.ipac.caltech.edu/data/Planck/release\_2/ancillary-data/bolocam/}
{http://irsa.ipac.caltech.edu/data/Planck/release\_2/ancillary-data/} 
\href{http://irsa.ipac.caltech.edu/data/Planck/release\_2/ancillary-data/bolocam/}{bolocam/}.} 
are the same as those used in \citet{czakon2015} and \citet{sayers2013}; the details of the reduction are 
given therein, along with \citet{sayers2011}. 
%Bolocam observed Abell 1835 for 14.0 hours resulting in a noise of 16.2 $\mu K_{CMB}$-arcminute, and observed 
%MACS 0647 for 11.4 hours resulting in a noise of 22.0 $\mu K_{CMB}$-arcminute.
%In addition to the data maps, for each cluster 1000 noise maps are also provided, which
%included relevant sources of instrumental, atmospheric, and astronomical noise. 
The reduction and calibration is similar to that used for MUSTANG, and Bolocam achieves a 
5\% calibration accuracy and 5\asecs pointing accuracy.

%%%%%%%%%%%%%%%%%%%%%%%%%%%%%%%%%%%%%%%%%%%%%%%%%%%%%%%%%%%%%%%%%%%%%%%%%%%%%%%
\section{Joint Map Fitting Technique}
\label{sec:jointfitting}
%%%%%%%%%%%%%%%%%%%%%%%%%%%%%%%%%%%%%%%%%%%%%%%%%%%%%%%%%%%%%%%%%%%%%%%%%%%%%%%

\subsection{Overview}
\label{sec:jf_overview}

The joint map fitting technique used in this paper is described in detail in \citet{romero2015a}. We review
it briefly here. The general approach follows that of a least squares fitting procedure, which assumes that
we can make a model map as a linear combination of model components. 

This linear combination can be written as:
\begin{equation}
  \vec{d}_{mod} = \mathbf{A} \vec{a}_{mod},
  \label{eqn:model_array}
\end{equation}
where $d_{mod}$ is the total model, each column in $\mathbf{A}$ is a filtered model component (Section~\ref{sec:components}), 
and $\vec{a}_{mod}$ is 
an array of amplitudes of the components. There are up to four types of components for which we fit: 
a bulk component, point source(s), residual component(s), and a mean level. From these, we produce a
sky model for the bulk component and point source to be filtered. The residual component is calculated 
directly as a filtered component.

We wish to fit $\vec{d}_{mod}$ to our data, $\vec{d}$, and allow for a calibration offset between Bolocam and
MUSTANG data. To accomplish this, we define our data vector as:
\begin{equation}
  \vec{d} = [ \vec{d}_{B}, k \vec{d}_{M}, k ] ,
  \label{eqn:data_arr}
\end{equation}
where $\vec{d}_{B}$ is the Bolocam data, taken as the provided map (14\amins sides), $\vec{d}_{M}$ is the MUSTANG data,
taken as the inner (radial) arcminute of MUSTANG maps. $k$ is the calibration offset of
MUSTANG relative to Bolocam, to which we apply an 11.2\% Gaussian prior derived from the MUSTANG and Bolocam
calibration uncertainties. 

We use the $\chi^2$ statistic as our goodness of fit:
\begin{equation}
  \chi^2 = (\overrightarrow{d} - \overrightarrow{d}_{mod})^T \mathbf{N}^{-1} (\overrightarrow{d} - \overrightarrow{d}_{mod}),
  \label{eqn:chi_sq}
\end{equation}
where $\mathbf{N}$ is the covariance matrix; however, because we wish to fit for $k$ in addition to the 
amplitude of model components, we no longer have completely linearly independent variables, and thus we 
employ MPFIT \citep{markwardt2009} to solve for these variables. Confidence intervals are derived from 
$\chi^2$ values over the parameter space searched (Section~\ref{sec:param_space}), and adjusted based
on monte carlo simulations \citep{romero2015a}.

Our approach is to fix the shape and position of point sources and residuals (if any), fitting only their amplitudes.
We explore the shape of the bulk ICM component parametrically, where each point in the parametric space may be forward 
modeled (Section~\ref{sec:param_space}).
%changing only the shape of the
%bulk ICM component, which must be forward modeled for each instance(Section~\ref{sec:param_space}). 
At each point in the
parameter space we do a linear least squares fit followed by a nonlinear minimization over $k$, the Bolocam pointing, 
and $\vec{a}_{mod}$.

%%% Suggestion to remove this section. Do so?
%\subsection{Simulated Observations}
%\label{sec:jf_filtering}

%Simulated observations are performed by converting an input sky into input TOD and processing the simulated TOD
%as the true TOD are processed (Section~\ref{sec:musobs}). The time requirements for simulated observations are  
%reduced by storing the necessary telescope and detector information from the true TOD in an IDL structure. 
%Due to the time requirements to cover the
%necessary parameter space (Section~\ref{sec:param_space}), TOD were produced with a fraction of the scans (termed
%short TOD) on a given cluster, where care was taken to ensure the same coverage (relative weight distribution) 
%in the map as the full observation. 
%The filtering is observed to be the same between full TOD and short TOD (Figure~\ref{fig:long_vs_short_qv}). 

%\begin{figure}[!h]
%  \centering
%  \includegraphics[width=0.5\textwidth]{Long_vs_short_qv_bestfit_m0647_22_Jun_2015_v2.eps}
%  \caption{MUSTANG simulated observations of a fitted model to MACS 0647. 
%    The scatter in the short TOD is $\lesssim 3$\% 
%    of the peak Compton $y$. In absolute terms, this translates to roughly $2\times 10^{-5}$ in Compton $y$. 
%    Typical pixel noise in maps is 7 to 8 times greater.}
%  \label{fig:long_vs_short_qv}
%\end{figure}

\subsection{Components}
\label{sec:components}

In order to produce component maps, it is necessary to account for the response of both instruments and
imaging pipeline filter functions. For Bolocam, we use the transfer function provided. For MUSTANG, we
perform simulated observations, processing the sky models in the same manner that real data is processed.

\subsubsection{Bulk ICM}
\label{sec:bulk_ICM}

 As in \citet{romero2015a}, the bulk component is taken to be a 
spherically symmetric 3D electron pressure profile as parameterized by a generalized Navarro, Frenk,
and White profile \citep[hereafter, gNFW][]{navarro1997,nagai2007}:
\begin{equation}
  \Tilde{P} = \frac{P_0}{(C_{500} X)^{\gamma} [1 + (C_{500} X)^{\alpha}]^{(\beta - \gamma)/\alpha}}
  \label{eqn:gnfw}
\end{equation}
where $X = R / R_{500}$, and $C_{500}$ is the concentration parameter; one can also write ($C_{500} X$) as
($R / R_s$), where $R_s = R_{500}/C_{500}$. $\Tilde{P}$ is the electron pressure in units of the characteristic
pressure $P_{500}$. This pressure profile is integrated along the line of sight to produce 
a Compton $y$ profile, given as 
\begin{equation}
  y(r) = \frac{P_{500} \sigma_{T}}{m_e c^2} \int_{-\infty}^{\infty} \Tilde{P}(r,l) dl
  \label{eqn:compton_y}
\end{equation}
where $R^2 = r^2 + l^2$, $r$ is the projected radius, and $l$ is the distance from the center of the cluster
along the line of sight. Once integrated, $y(r)$ is gridded as $y(\theta)$ and is realized as two maps with
the same astrometry as the MUSTANG and Bolocam data maps (pixels of 1\asecs and 20\asecs on a side, respectively). 
%From here, we produce two model maps: one for Bolocam and one for MUSTANG. 
In each case, we convolve the Compton $y$ map by the appropriate beam shape. For Bolocam we use a Gaussian with FWHM
$= 58$\asec, and for MUSTANG we use the double Gaussian, representing the GBT main beam and stable error beam
\citep{romero2015a}. Subsequently, we account for the filtering effects of data processing for each instrument,
as described in \citet{romero2015a}.

\subsubsection{Point Sources}
\label{sec:ptsrcs}

Point sources are treated in the same manner as in \citet{romero2015a}. All compact sources in our sample
are well modelled as a point source. We clearly detect point sources in Abell 1835, MACS 1115, MACS 0429, 
MACS 1206, RXJ1347, MACS 1423, and MACS 0717 in the MUSTANG maps. While no point source is evident from our
raw MUSTANG map, a point source is identified by NIKA \citep{adam2015} in CLJ1226, which is posited to be a 
submillimeter galaxy (SMG) behind the cluster. That point source is distinct from the point source seen in 
\citet{korngut2011}, which is not evident in our map. 
%and has a lower fitted significance ($3.0\sigma$) to the MUSTANG data than the one identified by NIKA ($3.3\sigma$). 
The fitted point source in MACS 0717 is due to a foreground elliptical galaxy and was fit in \citet{mroczkowski2012};
it is not within the central arcminute, our nominal MUSTANG region considered (Section~\ref{sec:jf_overview}).
Therefore, we extend the fitted region of the MUSTANG map to include the point source in MACS 0717 
(see Figure~\ref{fig:resid_maps}).
All of the remaining point sources (six) are coincident (within 3\asecs of reported coordinates) with the BCGs 
of their respective clusters \citep[][]{crawford1999,donahue2015}. 
%%%%%%% Suggestion to rewrite this text:
Moreover, of these six BCGs, four of them exhibit \quotes{unambiguous UV excess} \citep{donahue2015}. 
The remaining two are Abell 1835 and MACS 1206. The UV excess in MACS 1206 may be due to lensed
background systems \citep{donahue2015}. Abell 1835 is not in the CLASH sample and thus was not
included in \citet{donahue2015}. However, it was observed by \citet{odea2010} and found to have a 
far UV flux corresponding to a star formation rate of 11.7 $M_{\odot}$ per year, which fits within the SFR range 
(5 - 80 $M_{\odot}$ yr$^{-1}$) of the UV excess BCGs found in \citep{donahue2015}.
%%%%%%%%%%%%%%%%%%%%%%%%%%%%%%%%%%%%%%
For the Bolocam images, the point sources in Abell 1835, MACS 0429, RXJ1347, and MACS 1423
have been subtracted based on an extrapolation of a power law fit to the 1.4 GHz NVSS \citep{condon1998}
and 30 GHz SZA \citep{bonamente2012} measurements
%the extrapolation and account for potential breaks in the power law are detailed in \citet{sayers2013c}.
as detailed in \citet{sayers2013c}; they found that the Bolocam measurements were consistent with a
30\% scatter in the extrapolated flux densities from the fits to the lower frequency data.
This additional uncertainty is applied to all extrapolated flux densities and accounts for potential breaks
in the spectral index. The flux densities for these point sources are shown in Table~\ref{tbl:sample_ptsrc};
the MUSTANG flux densities provide support for the extrapolated flux densities at 140 GHz.

\begin{deluxetable}{c c c c c}
\tabletypesize{\footnotesize}
\tablecolumns{5}
\tablewidth{0pt} 
% Currently Table 3. (Nov. 2016)
\tablecaption{Point source flux densities \label{tbl:sample_ptsrc}}
\tablehead{ 
    \colhead{Cluster} & \colhead{R.A. (J2000)} & \colhead{Dec (J2000)} & 
    \colhead{$S_{90}$ (mJy)}  & \colhead{$S_{140}$ (mJy)}
}
\startdata
Abell 1835  & 14:01:02.07  &  +2:52:47.52  & $1.37 \pm 0.08$ & $0.7 \pm 0.2$ \\
MACS 1115   & 11:15:51.82  &  +1:29:56.82  & $1.04 \pm 0.11$ & --            \\  
MACS 0429   & 04:29:35.97  &  -2:53:04.74  & $7.67 \pm 0.84$ & $6.0 \pm 1.8$ \\
MACS 1206   & 12:06:12.11  &  -8:48:00.85  & $0.75 \pm 0.08$ & --            \\  
RXJ1347     & 13:47:30.61  & -11:45:09.48  & $7.40 \pm 0.58$ & $4.0 \pm 1.2$ \\  
MACS 1423   & 14:23:47.71  & +24:04:43.66  & $1.36 \pm 0.13$ & $0.7 \pm 0.2$ \\  
MACS 0717   & 07:17:37.03  & +37:44:24.00  & $2.08 \pm 0.25$ & --            \\   
CLJ1226     & 12:27:00.01  & +33:32:42.00  & $0.36 \pm 0.11$ & --   
\enddata
  \tablecomments{$S_{90}$ is the best fit flux density to MUSTANG, and $S_{140}$ is the extrapolated flux density in
  the Bolocam maps (at 140 GHz). The location of the point source is reported from the fitted centroid to the
  MUSTANG data. The conversion from mJy to the equivalent uK$_{CMB}$ is given as: 
  $S_{140} (\text{mJy/bm}) \sim S_{140} / 20 (\mu\text{K}_{CMB})$.}
\end{deluxetable}

\subsubsection{Residual Components}
\label{sec:residuals}

Residual components are selected primarily based on peak decrements exceeding $4 \sigma$  within the central arcminute 
of smoothed MUSTANG \emph{first-pass} residual maps, which are not well fitted by a bulk model. For clarity,
any subsequent residual maps (after fitting any residual component described here), are simply referred to as
residual maps.
We fit residual components for MACS 1206, RXJ 1347, MACS 0717 and MACS 0744. The residual component for MACS 0717
is coincident with subcluster C as identified in \citet{ma2009}, and has a centroid outside the central arcminute, but
the features in the MUSTANG map extend into the central arcminute.
Although we do not fit for residual components in Abell 611 and MACS 1115, we report properties of potential
residual components for these two clusters. We do not fit the residual component for Abell 611 because the peak
significance is not $4\sigma$. For MACS 1115, the centroid of the residual component is just outside the central 
arcminute and does not affect our fit.

To model the shape of residual component, we fit a two dimensional Gaussian to the selected pixels 
(those below $-3\sigma$). This Gaussian is then fit to the unsmoothed MUSTANG data map 
(in units of Compton $y$) with only its amplitude is allowed to vary to obtain the results presented
in Table~\ref{tbl:resid_comps}.
%To create the residual component, we first select the feature of interest based on the MUSTANG signal-to-noise (SNR) map
%of the cluster \citep[see][]{romero2015a}. All pixels below $-3 \sigma$ pertaining to the feature are selected, and the 
%shape is determined by fitting a two dimensional Gaussian. This Gaussian is then fit to the unsmoothed MUSTANG data map 
%(in units of Compton $y$), where only its amplitude is allowed to vary.

\begin{deluxetable*}{c | c c c c c c c}
\tabletypesize{\footnotesize}
\tablecolumns{8}
\tablewidth{0pt} 
% Currently Table 4. (Nov. 2016)
\tablecaption{Parameters of Residual Components from MUSTANG \label{tbl:resid_comps}}
\tablehead{
Cluster & RA      & Dec     & Modelled Peak $y$    & FWHM$_B$ & FWHM$_A$ & $\theta$ & Fitted Peak $y$ \\
        & (J2000) & (J2000) & ($10^{-5}$) & (\asec) &  (\asec) & (deg.) & ($10^{-5}$)     
}
\startdata
Abell 611 &  8:00:56.20 & 36:03:00.08 &  8.4  &  20.7 &  35.3 &    70 & --        \\ 
MACS 1115 & 11:15:56.66 &  1:30:02.82 &  14   &  17.8 &  28.8 &    48 & --        \\ 
MACS 1206 & 12:06:12.91 & -8:47:33.48 &  7.6  &  23.5 &  23.5 &   155 & $3.6 \pm 0.7$   \\ 
RXJ1347   & 13:47:31.06 &-11:45:18.38 &  42   &  12.2 &  30.1 &    48 & $52  \pm 9$     \\ 
MACS 0717 &  7:17:34.01 & 37:44:49.73 &  4.4  &  58.9 &  58.9 &    -- & $4.6 \pm 1.1$   \\
MACS 0744 &  7:44:52.22 & 39:27:28.71 &  11   &  17.0 &  23.5 &    91 & $9.0 \pm 2.8$    
\enddata
% Uncorrected thetas: 160, 138, -115, -52, 1
\tablecomments{Residual components modeled with a two dimensional Gaussian. $\theta$ is 
  measured CCW (going east) from due north. The modelled peak $y$ is the peak when fit to the first-pass
  residual map, and the fitted peak $y$ is the re-normalized peak when fit, with the other components,
  to the data map. FWHM$_A$ and FWHM$_B$ correspond to the widths of the major and minor axes,
  respectively.}
\end{deluxetable*}

\subsubsection{Mean Level}
\label{sec:mean_level}

%%% JSayers: Bolocam too? Yes...if done, both are done. But I removed the mean level fit in the simple sense.
%%% I don´t think I fit out a Bolocam mean level, because it was so low.
Similar to \citet{czakon2015}, we wish to account for a mean level (signal offset) in the MUSTANG maps.
We do not wish to fit for a mean level simultaneously as a bulk component given the degeneracies. Therefore,
to determine the mean level independent of the other components, we create a MUSTANG noise map
% from time-flipped TOD 
and calculate the mean within the inner arcminute for each cluster. This mean is then subtracted before 
the other components are fit. 

\subsection{Parameter Space}
\label{sec:param_space}

As in \citet{romero2015a}, we fix MUSTANG's centroid, but allow Bolocam's pointing to vary by $\pm 10$\asecs 
in RA and Dec with a prior on Bolocam's radial pointing accuracy with an RMS uncertainty of $5$\asec. Our 
approach to find the absolute calibration offset between Bolocam and MUSTANG is the same as in
\citet{romero2015a} (see also Section~\ref{sec:jf_overview}). 

In \citet{romero2015a}, we performed a grid search over $\gamma$ and $C_{500}$, marginalizing over $P_0$,
where $\alpha$ and $\beta$ are fixed to values determined from \citetalias{arnaud2010}. To determine the
impact of our choice of fixed $\alpha$ and $\beta$, we explored how the profile shapes change when different,
fixed, values of $\alpha$ and $\beta$ are adopted. In all cases, we find the pressure profile shapes are in very 
good agreement with one another and that the differences in $\chi^2$ values are statistically consistent. Thus, 
our fits are not sensitive to the exact choice of $\alpha$ and $\beta$.

We adopt $R_{500}$ from \citet{mantz2010} and we search over $0 < \gamma < 1.3$ in steps of $\delta \gamma = 0.1$, 
and over $0.1 < C_{500} < 3.3$ in steps of $\delta C_{500} = 0.1$. 
This choice of parameter space searched is determined by computation requirements (largely in
filtering maps) and covering a sufficient range of values. Our choice to limit $\gamma \ge 0$ is 
motivated by its implications to hydrostatic equilibrium under thermal pressure support. We revisit this 
choice in Section~\ref{sec:pp_error}. To create models in finer steps than $\delta \gamma$ 
and $\delta C_{500}$, we interpolate filtered model maps from nearest neighbors from the grid of original 
filtered models.

All of the gNFW parameters ($P_0$, $C_{500}$, $\alpha$, $\beta$, and $\gamma$) have some degeneracy 
with each other. $C_{500}$ relates the scaling radius, $R_s$, which is directly constrained by the SZ data, 
to $R_{500}$ as $R_s = C_{500} R_{500}$. Because we take $R_{500}$, $\alpha$, and $\beta$ from \citetalias{arnaud2010},
which used X-ray data and numerical simulations to derive their values of $\alpha$, and $\beta$,
the values constrained by our SZ data in this analysis are not entirely independent of X-ray data. However, given
the insensitivity to $\alpha$ and $\beta$ found in \citet{romero2015a}, and the independent nature of $R_s$, 
the profile shapes themselves should be considered approximately independent from X-ray data, if not
the constrained shape parameter values as well.
%It is worth noting that only $\gamma$ and $P_0$ are truly independent of X-ray data. Our choice of $R_{500}$ does
%not imply a dependence of our profile shape on X-ray data, but merely a dependence on the value of $C_{500}$ on
%X-ray data. The restriction of $\alpha$ and $\beta$ to \citetalias{arnaud2010} values will impose a dependence on
%X-ray data, but as noted, the shape of our pressure profiles do not appear to be sensitive to the choice of 
%$\alpha$ and $\beta$.

\subsubsection{Centroid Choice}
\label{sec:centroids}

The default centroids used when gridding our bulk ICM component are the ACCEPT centroids. Given the offsets
between ACCEPT and Bolocam centroids (Table~\ref{tbl:cluster_properties}), we perform a second set of
fits where we grid the bulk ICM component using the Bolocam centroids and we do not find significant changes in
the fitted gNFW parameters (Section~\ref{sec:pp_constraints}). The ACCEPT centroids are taken to be the
X-ray peaks unless the centroiding algorithm produced a centroid more than 70 kpc from the X-ray peak, in which
case they adopt that centroid \citep{cavagnolo2008a}. 
%We do not find significant changes in
%the fitted gNFW parameters (Section~\ref{sec:pp_constraints}), 

\subsection{Robustness of the Joint Fitting Technique}

%\textbf{
%\textcolor{red}{[The referee would like to see ``some residual maps, a discussion on the goodness of fit, and any
%major sources of error. How much do my results depend on the mean level subtraction?]}}

%\subsubsection{Goodness of fits}

Our goodness of fits are tabulated as reduced $\chi^2$ in Table~\ref{tbl:pressure_profile_results}. 
%Given our degrees of freedom in each fit, deviations from $\pm 0.01$ in $\tilde{\chi}^2$ are not trivial. 
%$\tilde{\chi}^2 < 1$ appear to be due to a slight overestimation in the noise, in the MUSTANG or Bolocam maps. 
  The residual MUSTANG and Bolocam maps indicate that a spherically symmetric gNFW pressure profile provides an adequate
  description of the data. In several residual MUSTANG maps, especially for those clusters with $\tilde{\chi}^2 > 1.02$, 
some $3\sigma$ features remain (within the fitted region). MUSTANG residuals in MACS 1115, and MACS 0717, which have 
significances beyond $4\sigma$, are  well away from the X-ray cluster centroid and nearly outside of the fitted region.
Thus, these residuals will not impact the fitted cluster profiles, but can still elevate the overall $\tilde{\chi}^2$.

Another potential source of noise worth considering is the primary CMB anisotropies. Bolocam accounts for
  CMB anisotropies in their noise model by adding astronomical sky realizations based on \citet{keisler2011,reichardt2012}
SPT measurements \citep{sayers2013}. In the MUSTANG data, we are not concerned with the primary CMB
anisotropies as these are negligible beyond $\ell \gtrsim 6000$ ($\theta \lesssim 2$\amin) \citep{george2015}.
Given the MUSTANG transfer function, we estimate that the expected CMB contamination will fall below
4 $\mu$Jy/beam, which is well below our noise level, and therefore negligible.

Our pressure profile fits do not change significantly between the chosen Bolocam or X-ray centroids. As seen
  in Figure~\ref{fig:resid_maps}, MACS 0647 is the only cluster to show significant residuals near the centroid,
  indicative of a centroid offset. However, given MUSTANG's sensitivity to substructure, and potential degeneracy,
  it is possible that this apparent centroid offset could be due to substructure that is not well separated from
  the cluster core. We note that the reduced $\chi^2$ (Table~\ref{tbl:pressure_profile_results}) indicates that
  MACS 0647 is still well fit.

% but none reach $4\sigma$ within the fitted region, which was the threshold for modelling residual components 
%(Section~\ref{sec:residuals}). The $\tilde{\chi}^2$

\subsubsection{Impact of MUSTANG mean level on the Pressure Profile}

The mean levels in the MUSTANG maps are typically $\lesssim 15 \mu$Jy/beam in amplitude. The subtraction of a mean level 
within the MUSTANG maps results in a minimal change in the fitted pressure profile shapes, but in every case, it reduces
$\chi^2$ (as compared to subtracting no mean level). In the case of MACS 1206, 
when subtracting a mean level, the parameters $\gamma$ and $C_{500}$ change by $\sim 0.1$, creating a steeper inner 
pressure profile, where the pressure profile is elevated by $\sim 50$\% in the innermost 10\asecs and elevated by 
$\sim 5$\% at 240\asec. However, in all other clusters, the changes in the parameters $\gamma$ and $C_{500}$ are less 
than $0.05$, and corresponding pressure profile changes, between 5\asecs and 240\asec, are generally less than 5\%.

\subsubsection{Impact of Potential Substructure on the Pressure Profile}

\begin{figure}
  \begin{center}
  \begin{tabular}{cc}
    \epsfig{file=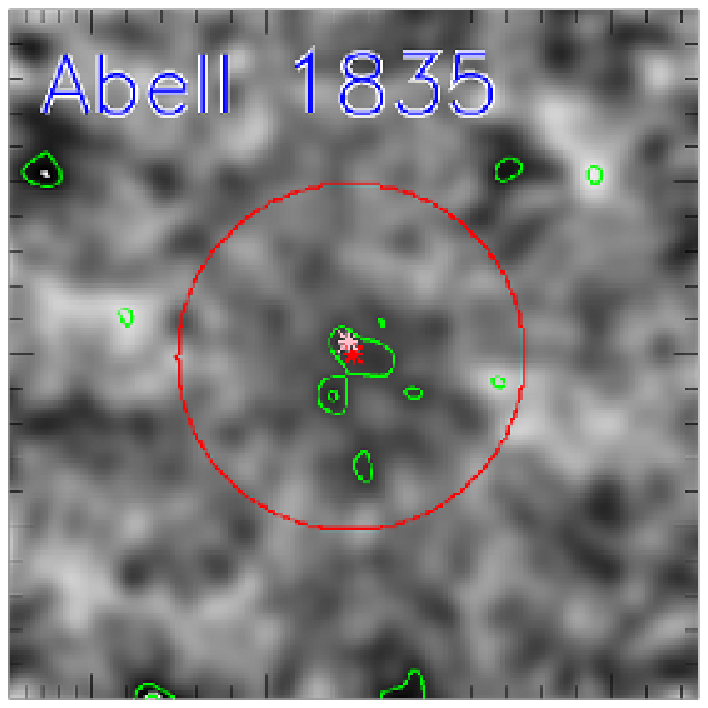,width=0.50\linewidth,clip=}   &
    \epsfig{file=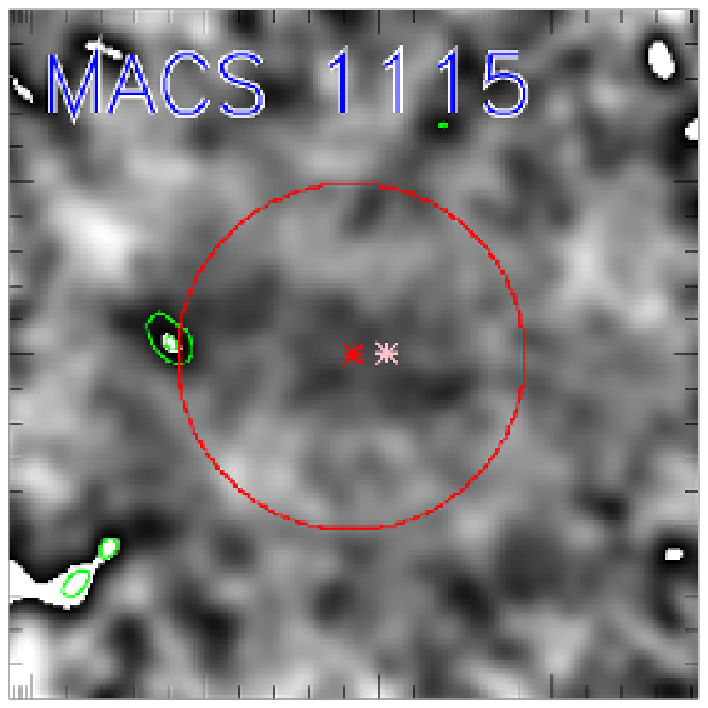,width=0.50\linewidth,clip=}   \\
    \epsfig{file=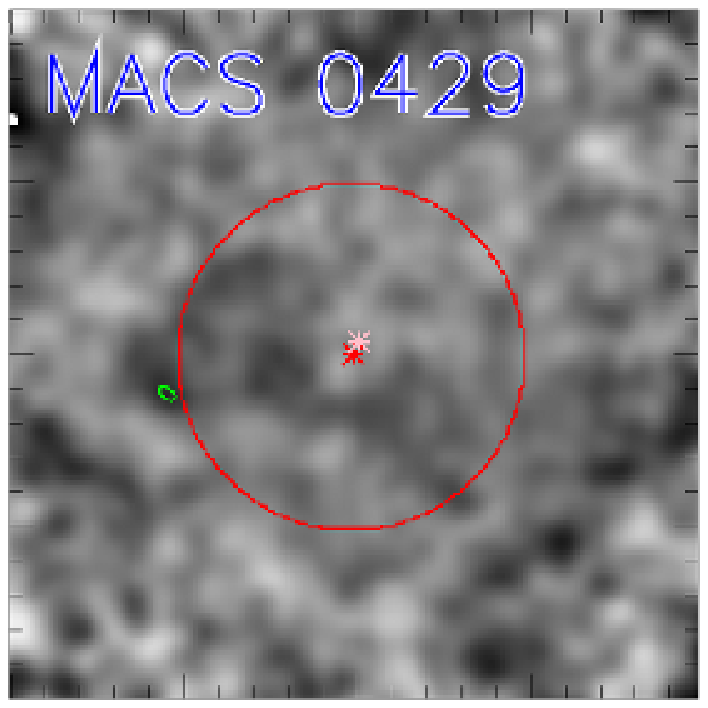,width=0.50\linewidth,clip=}   &
    \epsfig{file=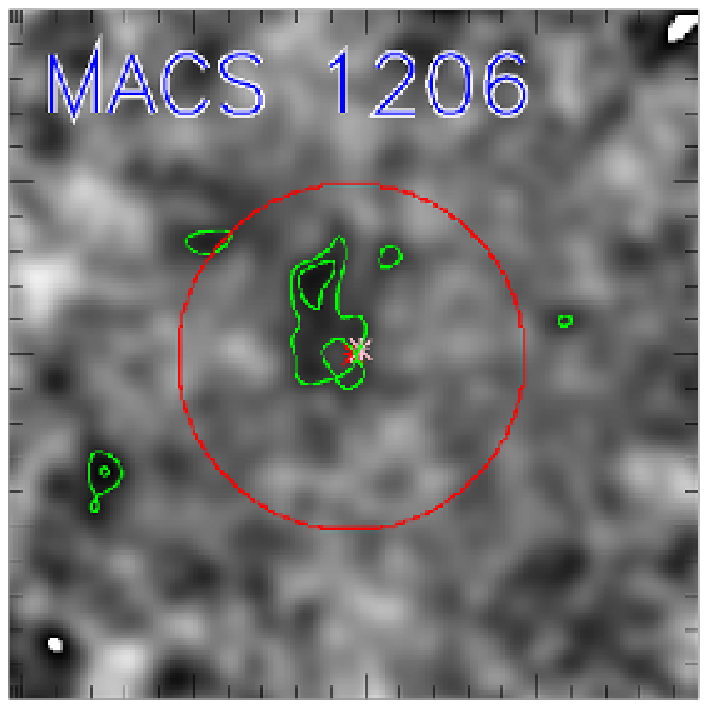,width=0.50\linewidth,clip=}   \\
    \epsfig{file=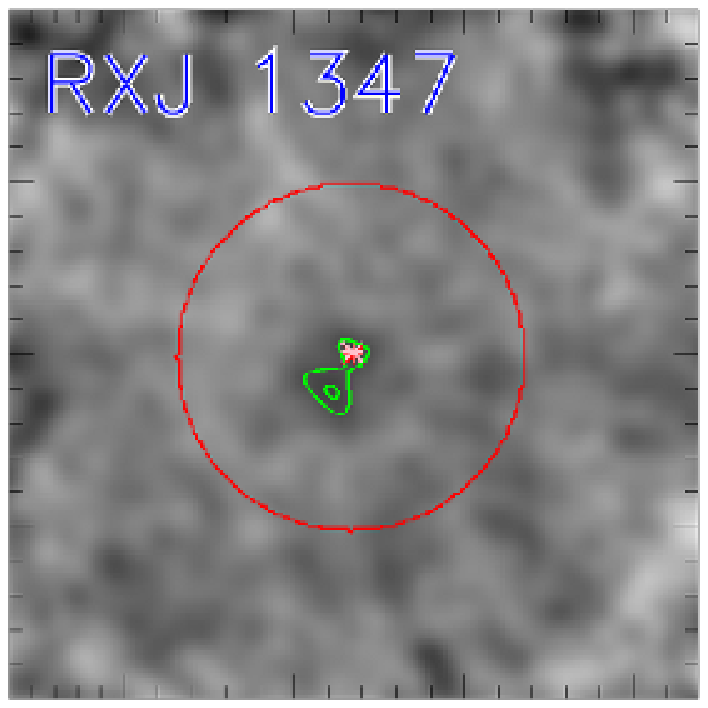,width=0.50\linewidth,clip=} & 
    \epsfig{file=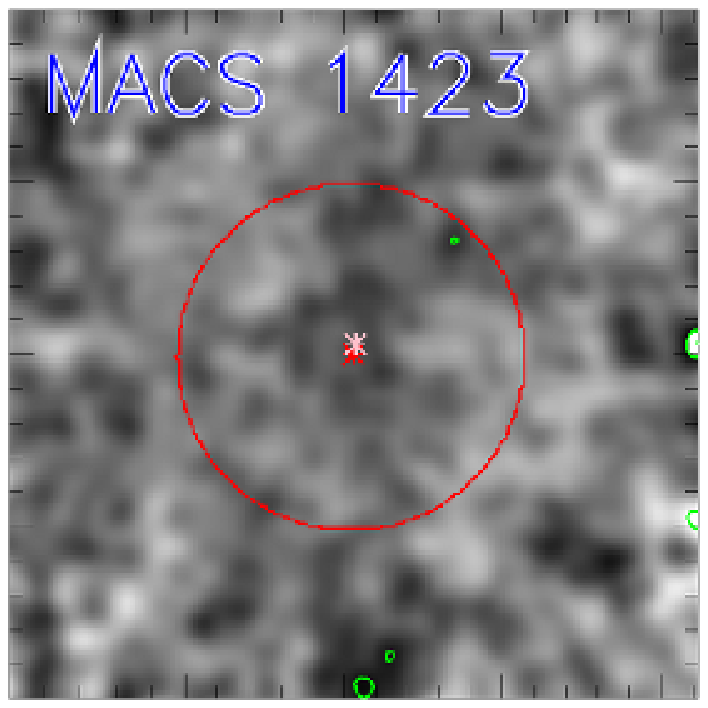,width=0.50\linewidth,clip=}   \\
    \epsfig{file=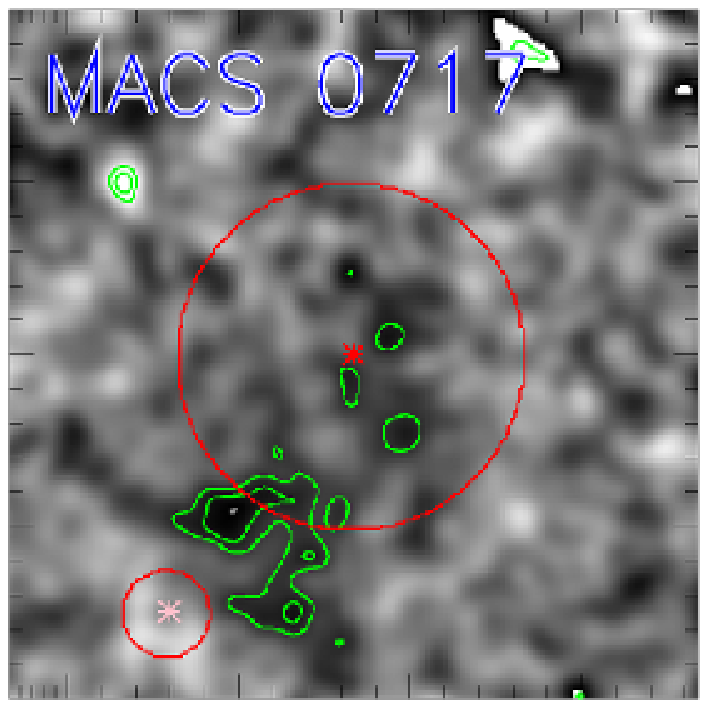,width=0.50\linewidth,clip=}   &
    \epsfig{file=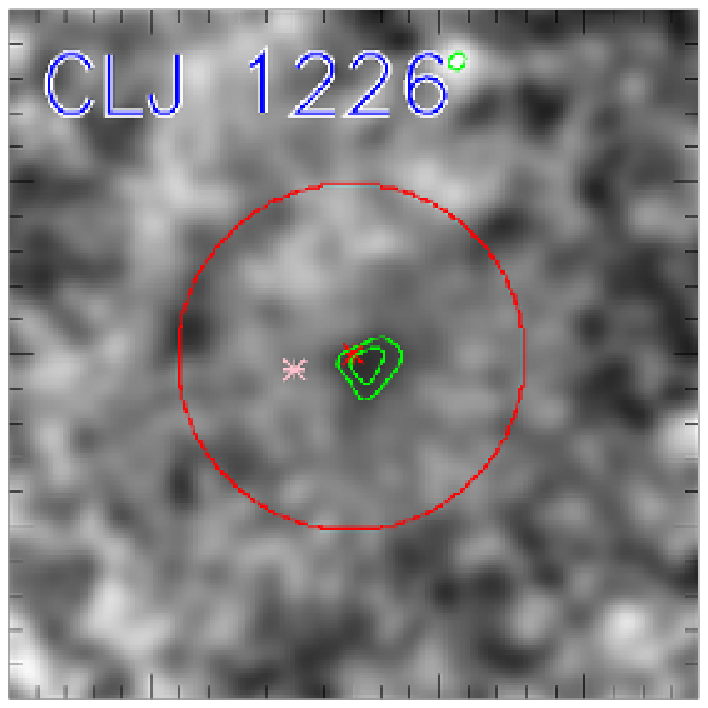,width=0.50\linewidth,clip=} 
 \end{tabular}
  \end{center}
  \caption{Point source subtracted MUSTANG flux maps for clusters with fitted point sources (Table~\ref{tbl:sample_ptsrc})
    in MUSTANG data. The color scaling spans the range $\pm5\times$Noise$_{M}$, where Noise$_{M}$ (for MUSTANG) is given in 
    Table~\ref{tbl:cluster_obs}. As Noise$_{M}$ was calculated in the inner arcminute, the increase in noise
    with radius is evident with this scaling. The contours are calculated from a signal-to-noise map (i.e. noise-corrected) 
    and start at $\pm3\sigma$, with $1\sigma$ intervals. The red asterisk is the 
    ACCEPT centroid; the pink asterisk is the point source centroid. 
    All relevant components were fit, but we have subtracted only the point source model here.}
% Currently Figure 3. (Nov. 2016)
  \label{fig:ptsrc_subs}
\end{figure}

The impact of residual components and point sources is heterogeneous given the varying relevancy of these components.
The residual maps and $\chi^2$ suggest that the point source components are appropriate models for the sources
we see in our clusters. MUSTANG maps, removing the fitted point sources, are shown in Figure~\ref{fig:ptsrc_subs}.
In Section~\ref{sec:pp_error}, we revisit the impact of point sources on the pressure profile. 
The residual components generally appear to be sufficient, despite the simplicity of a 2D Gaussian. In the case of MACS 0717,
the structure is not well modelled by a 2D Gaussian, but its modelling is minimally impactful as it is sufficiently far from
the center. In contrast, if a residual component is not fit in RXJ 1347 and MACS 0744, we find a maximum of 20\% and 100\%
increase in the pressure profile, occurring towards the center (at 4.5\asecs radius: MUSTANG's half width at half maximum, HWHM).
This increase drops to 3\% and 60\% at 240\asecs (half of Bolocam's FOV). These two clusters exhibit this strong dependence
on the treatment of substructure due to the substructures' proximity to the core, where azimuthal averaging does not dilute
the signal. In the other clusters, MACS 1206 and MACS 0717, the omission of a residual component results in a difference in
fitted pressure profiles by less the 10\%. Residual MUSTANG maps, i.e. maps with all components
(including residual components) subtracted, are shown in Figure~\ref{fig:resid_maps}.

%\subsubsection{Point Sources}
%\textbf{Some initial discussion.}

%\subsubsection{Residual Components}
%\subsubsection{Overall goodness of fits}
%\textcolor{red}{[Respond to concerns of CMB anisoptropies here.]}
% Calculated using the estimation in \citet{adam2016a}, acknowledging the transfer function cited in
% \citet{adam2014b}, and accounting for similar power spectra between 95 and 150 GHz in \citet{george2015}.

%%%%%%%%%%%%%%%%%%%%%%%%%%%%%%%%%%%%%%%%%%%%%%%%%%%%%%%%%%%%%%%%%%%%%%%%%%%%%%%
\section{SZ Pressure Profile Constraints}
\label{sec:pp_constraints}
%%%%%%%%%%%%%%%%%%%%%%%%%%%%%%%%%%%%%%%%%%%%%%%%%%%%%%%%%%%%%%%%%%%%%%%%%%%%%%%

\begin{figure}[!h]
  \centering
  \begin{tabular}{cc}
     \epsfig{file=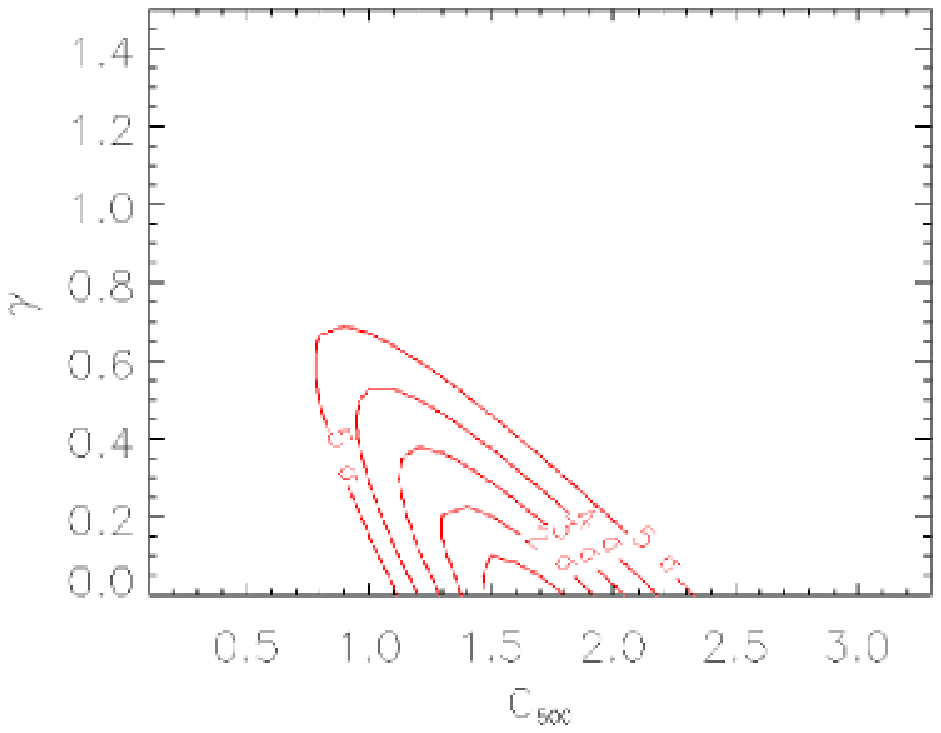,width=0.50\linewidth,clip=} &
     \epsfig{file=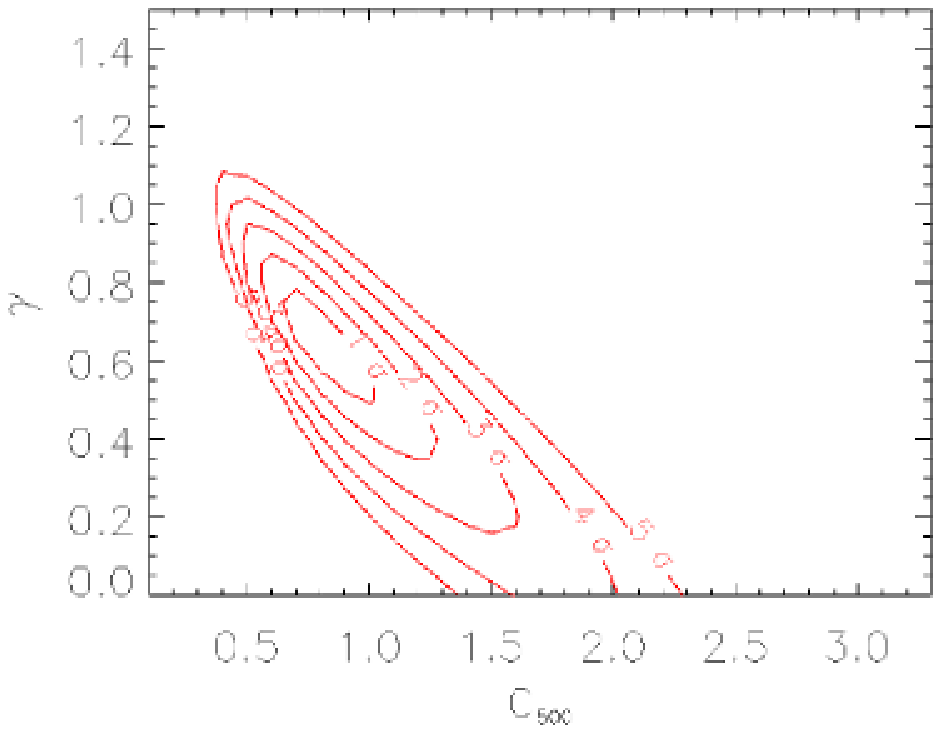,width=0.50\linewidth,clip=} \\
     \epsfig{file=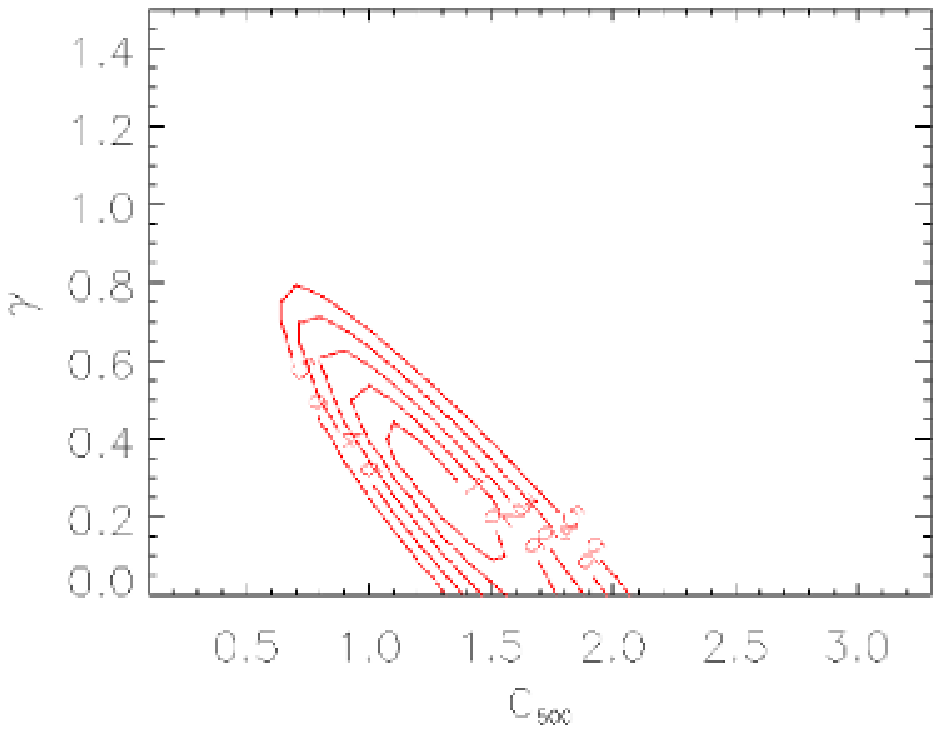,width=0.50\linewidth,clip=} &
   \end{tabular}
  \caption{Confidence intervals over all disturbed clusters (upper left panel), 
    cool-core clusters (upper right panel), and the entire sample (lower left panel).
    Cool core clusters include: Abell 1835, MACS 1115, MACS 0429, MACS 0329, RXJ 1347, 
    MACS 1311 and MACS 1423. Disturbed clusters include: MACS 0329, MACS 1149, MACS 0717, and
    MACS 0744.}
           %Well behaved clusters include: Abell 1835, MACS 1115, MACS 1206, RXJ 1347,
%           MACS 0647, MACS 0744, and CLJ 1226. Well behaved clusters are identified above.
% Currently Figure 4. (Nov. 2016)
  \label{fig:ensemble_cis}
\end{figure}

We have constrained the gNFW parameters $P_0$, $C_{500}$, and $\gamma$ for fourteen individual clusters and present 
these constraints in Table~\ref{tbl:pressure_profile_results}. Given that we find minimal differences between the 
fitted parameters using either the ACCEPT or Bolocam centroids, we report the results using the ACCEPT centroids.
%We find that six of the fourteen clusters are best fit by $\gamma = 0$. 
We find that six of our sample of fourteen have a best fit $\gamma = 0$, where we do not allow $\gamma <0$. 
We find that our range of $C_{500}$ is sufficient, and that it is generally found to be $0.5 < C_{500} < 2.0$. 
Across our entire sample, we find the best fitted gNFW parameters to be
$[\gamma,C_{500},P_0] = [0.3_{-0.1}^{+0.1}, 1.3_{-0.1}^{+0.1}, 8.6_{-2.4}^{+2.4}]$. Cool core clusters show a steeper
inner pressure profile, and are fitted with $[\gamma,C_{500},P_0] = [0.6_{-0.1}^{+0.1}, 0.9_{-0.1}^{+0.1}, 3.6_{-1.5}^{+1.5}]$,
and disturbed clusters show a flatter inner pressure profile with fitted parameters:
$[\gamma,C_{500},P_0] = [0.0_{-0.0}^{+0.1}, 1.5_{-0.2}^{+0.1},13.8_{-1.6}^{+1.6}]$. These constraints are visualized in
Figure~\ref{fig:ensemble_cis}.
%, where $\gamma$ and $C_{500}$ contours are marginalized over $P_0$.}

\begin{figure}
  \begin{center}
  \includegraphics[width=0.5\textwidth]{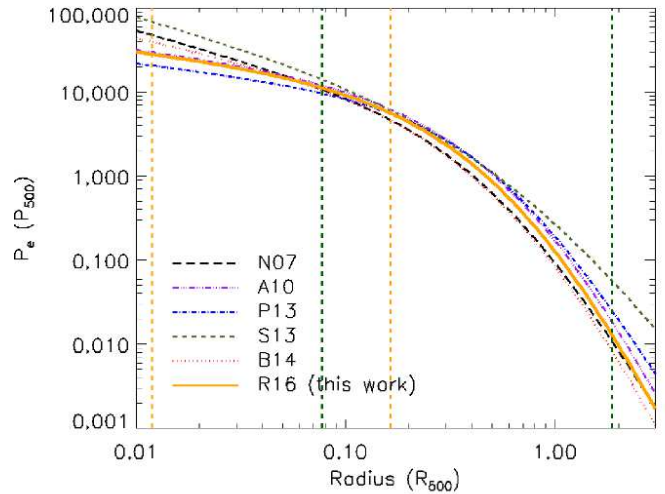}
  \end{center}
  \caption{Pressure profiles from this (R16) and other works. We observe that for our
    fourteen clusters, the ACCEPT2 data \citepalias{baldi2014} falls below R16, whereas
    \citetalias{arnaud2010,planck2013a}, and \citetalias{sayers2013} show higher pressure 
    at large radii. The pressure profile \citetalias{nagai2007} also agrees
    well with our work, but shows a steeper inner profile. The dark green dashed lines indicate
    the extent of Bolocam's nominal coverage (from HWHM to the radial FOV), and the orange
    dashed lines indicate the extent of MUSTANG's nominal coverage.}
% Currently Figure 5. (Nov. 2016)
  \label{fig:pp_sets}
\end{figure}

%For each cluster, we compare our SZ-derived pressure profiles with X-ray derived pressure profiles from ACCEPT2 
%\citep{baldi2014}. Specifically, 
We are further interested in comparing our pressure profile constraints, individually, and as a sample, to previous
constraints. To compare to the pressure profiles from ACCEPT2 \citep{baldi2014},
we fit gNFW profiles to the deprojected pressure profiles of our cluster sample (Section~\ref{sec:ellgeo}). 
We adopt B14 to refer to the ensemble pressure profiles fit to ACCEPT2 data for our sample of 14 clusters. 
Individually, we find discrepancies in pressure profiles, but as an ensemble there is relatively good agreement. 
Moreover, the average pressure profile for the 14 clusters has parameter values which are very similar 
to those found using X-ray data in \citet{arnaud2010}. This can also be seen in Figure~\ref{fig:pp_sets}, 
where the A10 and B14 pressure profiles are generally consistent with the profile from this work (R16),
where deviations are $<30$\% over $0.03 R_{500} < r < R_{500}$ for A10 and $<50$\% for B14. While all 14 clusters
in this work are in \citet{sayers2013} (hereafter S13), we note that they find a consistently higher average
pressure profile. Furthermore, the average pressure profile found by \citet{planck2013a} (hereafter P13) is 
higher than our average profile at large radii. In Figure~\ref{fig:pp_sets} we also include
a comparison to the pressure profile determined from simulations in \citet{nagai2007}, denoted as N07.
The pressure profiles of N07, A10, and B14 broadly cover the same spatial scales as our work ($0.03 R_{500} < r < R_{500}$),
while P13 and S13 generally loose sensitivity below $0.1 R_{500}$ and are senstive to scales beyond $R_{500}$.
%\textcolor{red}{[I think I want to expand on this - perhaps postulating why we
%find these trends.]}

%%%%%%%%%%%%%%%%%%%%%%%%%%%%%%%%%%%%%%%%%%%%%%%%%%%%%%%%%%%%%%%%%%%%%%%%%%%%%%%%%%%%%%%%%%%%%%%%%%%%%%%%%%%
%%%                                                SOME FIGURES                                         %%%
%%%%%%%%%%%%%%%%%%%%%%%%%%%%%%%%%%%%%%%%%%%%%%%%%%%%%%%%%%%%%%%%%%%%%%%%%%%%%%%%%%%%%%%%%%%%%%%%%%%%%%%%%%%

%%% Remove references to well behaved clusters.

% Now I want to compare to A10, P13, S13? OK...this is done.
%%% What else to say?

While our average pressure profiles are in excellent agreement with the previously derived pressure profiles
in the region $0.1 R_{500} < r < R_{500}$, we see deviations at small and large radii. In Figures~\ref{fig:pp_sets}
we indicate the nominal coverage of each instrument as the minimum HWHM, expressed in $R_{500}$, and maximum radial FOV,
expressed in $R_{500}$. As demonstrated in \citet{romero2015a}, the greatest constraints from individual instruments tends
to be at the center (geometric mean) of these two values. It is not too surprising
that our fits agree with A10 at large radii, as we have fixed $\alpha$ and $\beta$ to the A10 values.
Despite our fourteen clusters being included in the BOXSZ sample \citep{sayers2013}, we see that S13 shows
higher pressure at all radii. 
S13, fixing the slope of $\gamma$, present a higher pressure at small radii than found here, where the MUSTANG
data provide stronger constraints on the pressure gradients in the cluster core and suggest they are often weaker
than previously thought.

%We note that S13 presents a higher pressure at small radii than found in this work, where the addition of MUSTANG 
%data contributes significantly to constraining the pressure at small radii (towards smaller pressures). 
%At larger radii, our restriction of $\alpha$ and $\beta$ again explain our reduced pressure relative
%to S13.

%Figure~\ref{fig:ppr_ensembles} shows the ratios of the pressure profiles derived from this work to those from 
%other works, when clusters are characterized by dynamical type. We calculate these average ratios by weighting 
%the ratios of individual clusters, where the specific fitted gNFW pressure profile for
%each cluster is taken for ACCEPT2, but for the other sets (A10, P13, and S13), we assume the gNFW profile 
%found for all clusters in their sample. In fitting profiles to ACCEPT2, we impose the same restriction on 
%$\alpha$ and $\beta$ (fixing them to A10 values). In this manner, this imposition no longer biases our results
%at large radii, and we consistently see higher pressures at large radii in the SZ as compared to X-ray (ACCEPT2) 
%data. 

\begin{figure}
  \begin{center}
  \begin{tabular}{cc}
    \epsfig{file=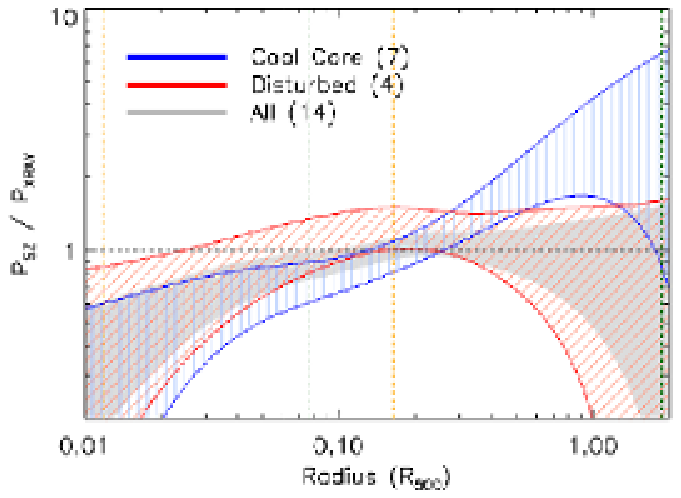,width=0.50\linewidth,clip=}   &
    \epsfig{file=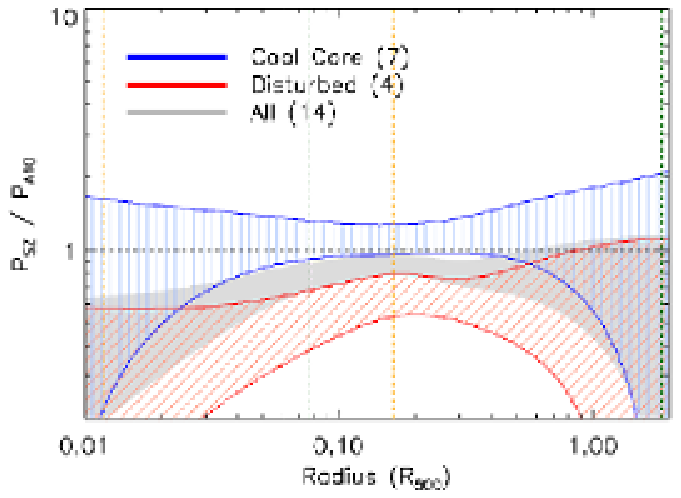,width=0.50\linewidth,clip=}  \\
    \epsfig{file=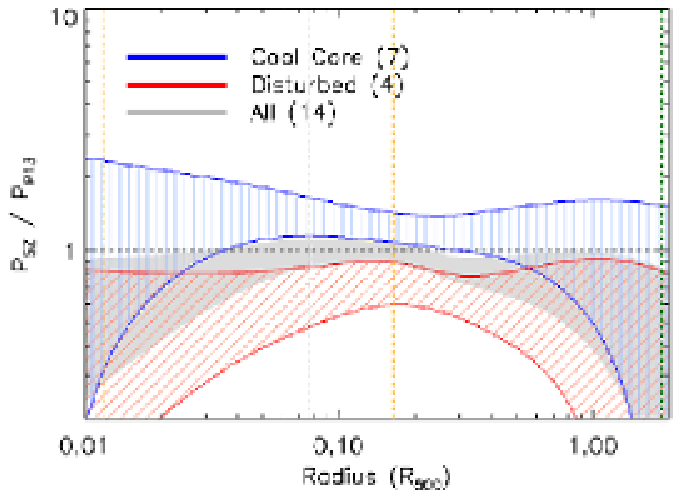,width=0.50\linewidth,clip=}   &
    \epsfig{file=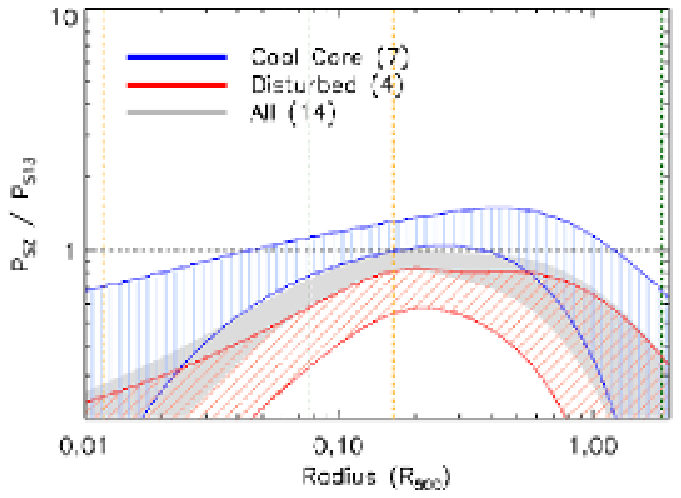,width=0.50\linewidth,clip=}  
  \end{tabular}
  \end{center}
  \caption{Pressure ratios as compared to different sets, plotted as the central 68\% confidence intervals. 
    The ensemble pressure ratios relative to ACCEPT2 ($P_{X}$) are calculated per cluster and weighted by
    the error in the ratio (per radial bin). For the other pressure ratios, the ratio is again calculated per
    individual cluster, but the comparison pressure profile is the gNFW profile for the entire dataset,
    respectively (i.e. A10, P13, or S13). These ratios are weighted in the same manner.
    As in Figure~\ref{fig:pp_sets}, the dark green dashed lines indicate the extent of Bolocam's nominal
    coverage, and the orange dashed lines indicate the extent of MUSTANG's nominal coverage.}
% Currently Figure 6. (Nov. 2016)
  \label{fig:ppr_ensembles}
\end{figure}

We further consider the ratio of individual cluster pressure profiles from our work ($P_{SZ}$) to the pressure profiles from 
other works. For comparisons with A10, P13, and S13, we take $P_{A10}$, $P_{P13}$, and $P_{S13}$ to be the gNFW
profile which each respective work had fit to their entire sample. For any of these sets (A10, P13, or S13), 
the ratio $P_{SZ} / P_{set}$ is calculated for each cluster, where only $P_{SZ}$ changes for each cluster. To
compare $P_{SZ}$ to ACCEPT2 ($P_{X}$), we fit a gNFW profile to ACCEPT2 data (Section~\ref{sec:ellgeo}) for
each cluster, and thus compare unique $P_{SZ}$ and $P_{X}$ pressure profiles for each cluster. These ratios are
shown in Figure~\ref{fig:ppr_ensembles}, where the shaded regions are influenced both by statistical errors and
scatter.

%\textcolor{red}{I also started thinking about \emph{XMM-Newton} vs. \emph{Chandra} data. However,
%\citet{donahue2014} would indicate that \emph{XMM-Newton} data tends to be lower in $n_e$ and $T_e$, which is
%\textbf{opposite} what we see in C09 vs. A10.}

%%%%%%%%%%%%%%%%%%%%%%%%%%%%%%%%%%%%%%%%%%%%%%%%%%%%%%%%%%%%%%%%%%%%%%%%%%%%%%%
\subsection{Fits with $\gamma = 0$}
%\subsection{Potential Sources of Error}
\label{sec:pp_error}
%%%%%%%%%%%%%%%%%%%%%%%%%%%%%%%%%%%%%%%%%%%%%%%%%%%%%%%%%%%%%%%%%%%%%%%%%%%%%%%

%%% Look at Jack´s comments. 
In 6 of our 14 clusters we find best fit pressure profiles with $\gamma = 0$, the limit we impose as a prior. 
There is no clear segregation based on dynamical state or presence of central point source.
Here, we consider two effects which could spuriously bias the cluster central pressures: our choice of centroid
and the mis-subtraction of a central point source.

%Several of our fits results in which $\gamma = 0$, the limit we imposed as a prior.

As it stands, finding slopes in the cores of galaxy clusters that are fit with $\gamma = 0$ is not unprecedented; 
\citetalias{arnaud2010} find six of their 31 analyzed clusters in the REXCESS sample have $\gamma=0$, where all gNFW parameters 
except $\beta$ were fit for individual clusters. They find a similar range in $C_{500}$ as we do, They fit for $\alpha$, 
which is fit by the range $0.3 < \alpha < 2.5$ over their sample. While \citetalias{arnaud2010} is a local ($z < 0.2$) sample,
\citet{mantz2016} find $\gamma=-0.01$, using \emph{Chandra} data, for their sample of 40 galaxy clusters of $0.07 < z < 1.10$.
Moreover, in an analysis of X-ray (\emph{Chandra}) data from observations of 80 clusters, 
\citet{mcdonald2014} find $\gamma=0$ for their low redshift ($0.3 < z < 0.6$), non-cool-core clusters, and similarly shallow 
inner pressure profile slope ($\gamma=0.05$) for the high redshift ($0.6 < z < 1.2$) non-cool-core clusters.
Although these previous studies indicate that $\gamma = 0$ is relatively common, we explore whether systematics related to
either our data or analysis methods may produce these results in our fits.
%Given our analysis and care to not be biased towards $\gamma=0$ and the precedence for other studies to find $\gamma=0$ in
%either individual clusters or cluster ensembles, we conclude that our study is not systematically biased towards $\gamma = 0$.

Shallow slopes in the cores of clusters could be suggestive of a centroid offset either between MUSTANG and Bolocam or
between SZ and X-ray data. Given the MUSTANG and Bolocam pointing accuracies (2\asec and 5\asec, respectively),
it is unlikely that the centroid offsets between MUSTANG and Bolocam are driving the fits to shallow slopes. 
The difference between SZ (Bolocam) centroids and ACCEPT centroids (Table~\ref{tbl:cluster_properties}) 
are large relative to pointing accuracies and thus potentially more important. However, when we adopt Bolocam's centroid 
%(\textcolor{red}{[Is this worth another figure? I'm OK without another figure for this.]}) 
we find negligible change to the SZ pressure profile as compared to adopting the ACCEPT centroid. 

Individually, the Bolocam and MUSTANG data sets yield consistent fits with each other, where changes in best fit 
parameters generally occur along the shallow gradient in confidence intervals (i.e. along the degeneracy
between  $C_{500}$, and $\gamma$).

Additionally, we consider the impact of the assumed flux densities of point sources in the Bolocam maps. 
%There are four clusters where Bolocam assumes a point source flux density: Abell 1835, MACS 0429, RXJ 1347, and MACS 1423. 
There are four clusters (Abell 1835, MACS 0429, RXJ 1347, and MACS 1423) where it was necessary to extrapolate a 140 GHz flux
density from lower frequency measurements in order to analyze the Bolocam data \citep{sayers2013}. 
For points sources other than those in MACS 0429 and RXJ 1347, their uncertainty is less than the noise in the Bolocam maps.
Moreover, in all but MACS 0429, the point source uncertainty is considerably less than the peak Bolocam decrement 
%(the clusters are very well detected Table~\ref{tbl:cluster_obs}) 
and mis-estimations of the point source flux densities in these clusters 
will not significantly change our results. Therefore, we are left with only MACS 0429 where
we believe that the treatment of the point source may affect our results non-trivially.
%The conversion for $S_{140}$ values from mJy to the equivalent $\mu K_{CMB-amin}$ 
%is $\sim20$, which puts the uncertainties of these point sources at 6, 52, 35, and 6 $\mu K_{CMB-amin}$ respectively. 
%From Table~\ref{tbl:cluster_obs}, we see that the noise in the Bolocam maps of these clusters are 16.2, 24.1, 19.7, 
%and 22.3 $\mu K_{CMB-amin}$ respectively. Thus, for MACS 0429 and RXJ 1347, which have point source flux density 
%uncertainties larger than the measurement noise, the assumed point sources could have a noticeable impact on the cluster
%fits. However, we must also consider the strength of cluster detections (Table~\ref{tbl:cluster_obs}), 
%relative to the point source significances (Table~\ref{tbl:sample_ptsrc}). With the point sources being equally significant
%between RXJ 1347 and MACS 0429, but the cluster decrement in RXJ 1347 being detected much more significantly %($>3\times$)
%than in MACS 0429, the errors in the assumed point source flux density will impact RXJ 1347 less. 
%Therefore, we are left with only MACS 0429 where
%we believe that the treatment of the point source may affect our results non-trivially.

If we utilize the same low-frequency point source flux densities in \citet{sayers2013c} and add in the MUSTANG data,
%which was used to estimate the flux density of point sources at 140 GHz, and add in the MUSTANG data, 
we can recalculate the expected flux densities of point sources at 140 GHz, still assuming one power law. 
We find that the current Bolocam estimates, with reported uncertainties,
are within $1\sigma$ of this recalculated value, except for MACS 1423, whose current value falls $1.3\sigma$
below the recalculated expectation. This does not address the potential for a break in the power
law, which appears to be the case for RXJ 1347.

In RXJ 1347, flux densities of $S_{86} = 4.16 \pm 0.03 \pm 0.25$ mJy and $S_{98} = 3.96 \pm 0.03 \pm 0.24$ mJy
have been reported from ALMA \citep{kitayama2016} and $S_{86} = 4.9 \pm 0.1$ mJy from
CARMA \citep{plagge2013}. The flux densities reported in \citet{kitayama2016} come from a baseline
cutoff to separate the point source from signal beyond roughly 5\asec.
%, and they show (their Figure 3) that
%the total flux density in their map rises with increasing maximum scale of Gaussians used in Multi-Scale CLEAN
%out to a maximum near 45\asec. This is likely due to an extended nature of the source
%(being an AGN, which should have accompanying jets),
%and a known minihalo \citep{ferrari2011} in the cluster center.
Additionally, we note that
\citet{adam2014} used an extrapolated flux of $S_{140} = 4.4 \pm 0.3$, deduced from the power law shown in
\citet{pointecouteau2001}, which was calculated from data between 1.4 GHz and 300 GHz.

Only two (MACS 0429 and MACS 1423) of these four clusters are fit by notably low $\gamma$ values.
In addition, for the remaining four clusters in which MUSTANG detects a point source, the Bolocam maps assume no 
point source contamination. Of these remaining clusters, only MACS 0717 is fit by a notably low $\gamma$, and that is best 
attributed to the dynamics of the cluster (Section~\ref{sec:results_m0717}). 

We also consider that our treatment of point sources in the MUSTANG maps may leave residual emission from point sources,
either due to our fitting procedure or the assumption that our assumed point source has a non-trivial extent. Our point
source treatment was designed and extensively tested \citep[e.g.][]{romero2015a} to accurately remove point sources. In
the case of Abell 1835, \citet{romero2015a} find good agreement between the MUSTANG and ALMA \citep{mcnamara2014} 
point source flux density. 
\begin{deluxetable*}{l|llllllllllll}
\tabletypesize{\footnotesize}
\tablecolumns{13}
%\tablewidth{\columnwidth} 
\tablewidth{0pt} 
% Currently Table 5. (Nov. 2016)
\tablecaption{Summary of Fitted Pressure Profiles \label{tbl:pressure_profile_results}}
\tablehead{
\colhead{Cluster} & \colhead{$R_{500}^a$} & \colhead{$Y_{cyl}(R_{500})$} & \colhead{$Y_{sph}(R_{500})$} & \colhead{$10^3 P_{500}^a$} & 
        \colhead{$P_0$} & \colhead{$C_{500}$} & \colhead{$\alpha$} & \colhead{$\beta$} & \colhead{$\gamma$} & 
                  \colhead{$k$} & \colhead{$\tilde{\chi}^2$} &  \colhead{d.o.f.}        \\ 
      & \colhead{(Mpc)} & \colhead{($10^{-5}$ Mpc$^2$)} & \colhead{($10^{-5}$ Mpc$^2$)} & \colhead{keV cm$^{-3}$} & \colhead{}  & 
      \colhead{} & \colhead{}  & \colhead{}  & \colhead{}  &   \colhead{}    
}
\startdata
Abell 1835 & 1.49 & $26.75_{-6.15}^{+6.05}$ & $21.81_{-4.49}^{+4.12}$ &  5.94 & $2.15 \pm 0.07$ & $0.77_{-0.17}^{+0.23}$ & 
1.05 & 5.49 & $0.78_{-0.13}^{+0.12}$ & 1.08 & 0.99 & 12880   \\ 
Abell 611  & 1.24 & $9.67_{-2.57}^{+4.85}$ & $8.73_{-2.21}^{+3.68}$ &  4.45 & $35.43 \pm 2.46$ & $2.00_{-0.30}^{+0.40}$ & 
1.05 & 5.49 & $0.00^{+0.15}$ & 0.96 & 1.02 & 12882   \\ 
MACS 1115  & 1.28 & $30.28_{-6.30}^{+7.32}$ & $20.10_{-3.52}^{+3.84}$ &  5.45 & $0.67 \pm 0.04$ & $0.35_{-0.10}^{+0.15}$ & 
1.05 & 5.49 & $0.87_{-0.27}^{+0.18}$ & 1.11 & 1.04 & 12875   \\ 
MACS 0429  & 1.10 & $30.41_{-6.88}^{+7.72}$ & $19.57_{-3.74}^{+4.00}$ &  4.48 & $11.01 \pm 0.77$ & $0.59_{-0.09}^{+0.11}$ & 
1.05 & 5.49 & $0.00^{+0.15}$ & 1.00 & 1.03 & 12875   \\ 
MACS 1206  & 1.61 & $61.52_{-12.63}^{+12.49}$ & $48.16_{-8.27}^{+8.19}$ & 10.59 & $2.39 \pm 0.10$ & $0.74_{-0.14}^{+0.16}$ & 
1.05 & 5.49 & $0.51_{-0.16}^{+0.14}$ & 1.09 & 1.01 & 12874   \\ 
MACS 0329  & 1.19 & $13.38_{-2.99}^{+3.83}$ & $11.86_{-2.37}^{+2.93}$ &  5.93 & $9.30 \pm 0.50$ & $1.18_{-0.28}^{+0.72}$ & 
1.05 & 5.49 & $0.41_{-0.41}^{+0.19}$ & 1.03 & 0.99 & 12876   \\ 
RXJ1347    & 1.67 & $42.47_{-6.81}^{+8.29}$ & $37.80_{-5.11}^{+5.78}$ & 11.71 & $3.24 \pm 0.08$ & $1.18_{-0.48}^{+1.02}$ & 
1.05 & 5.49 & $0.80_{-0.70}^{+0.30}$ & 1.15 & 0.99 & 12880   \\ 
MACS 1311  & 0.93 & $17.18_{-3.49}^{+3.80}$ & $10.08_{-1.73}^{+1.79}$ &  3.99 & $2.75 \pm 0.22$ & $0.35_{-0.05}^{+0.15}$ & 
1.05 & 5.49 & $0.41_{-0.41}^{+0.34}$ & 0.98 & 1.00 & 12881   \\ 
MACS 1423  & 1.09 & $10.35_{-2.73}^{+4.00}$ & $8.89_{-2.07}^{+2.53}$ &  6.12 & $22.39 \pm 1.71$ & $1.58_{-0.48}^{+0.22}$ & 
1.05 & 5.49 & $0.00^{+0.35}$ & 1.04 & 0.98 & 12876   \\ 
MACS 1149  & 1.53 & $56.87_{-9.00}^{+8.04}$ & $41.62_{-5.67}^{+4.99}$ & 12.28 & $5.50 \pm 0.25$ & $0.83_{-0.03}^{+0.07}$ & 
1.05 & 5.49 & $0.00^{+0.05}$ & 0.87 & 1.00 & 12876   \\ 
MACS 0717  & 1.69 & $54.16_{-8.72}^{+9.56}$ & $48.06_{-6.95}^{+7.71}$ & 14.90 & $21.88 \pm 0.68$ & $2.00_{-0.20}^{+0.20}$ & 
1.05 & 5.49 & $0.00^{+0.05}$ & 0.49 & 1.03 & 13583   \\ 
%MACS 0717  & 1.69 & $64.50_{-10.18}^{+12.42}$ & $55.72_{-8.00}^{+9.28}$ & 14.90 & $21.28 \pm 0.68$ & $1.97_{-0.37}^{+0.53}$ & 
%1.05 & 5.49 & $0.00^{+0.25}$ & 0.48 & 1.04 & 12876   \\ 
MACS 0647  & 1.26 & $34.06_{-7.76}^{+10.21}$ & $26.33_{-4.72}^{+5.37}$ &  9.23 & $2.78 \pm 0.11$  & $0.70_{-0.20}^{+0.30}$ & 
1.05 & 5.49 & $0.60_{-0.20}^{+0.15}$ & 1.14 & 1.01 & 12876   \\ 
MACS 0744  & 1.26 & $15.10_{-3.01}^{+4.50}$ & $13.20_{-2.29}^{+3.18}$ & 11.99 & $13.15 \pm 0.81$ & $1.71_{-0.21}^{+0.29}$ & 
1.05 & 5.49 & $0.00^{+0.15}$ & 0.90 & 1.02 & 12875   \\ 
CLJ1226    & 1.00 & $10.50_{-1.94}^{+2.65}$ & $9.46_{-1.60}^{+2.03}$ & 11.84 & $19.29 \pm 1.25$ & $1.90_{-0.50}^{+0.60}$ & 
1.05 & 5.49 & $0.29_{-0.29}^{+0.36}$ & 0.92 & 1.03 & 12875   \\ 
\hline
All          &  --    &  --    &  --    &  --    &  $8.58  \pm 2.37$ & $1.3_{-0.1}^{+0.1}$ & 1.05 & 5.49 & $0.3_{-0.1}^{+0.1}$ & -- & -- & -- \\ 
%All          &  --    &  --    &  --    &  --    &  $7.94  \pm 2.10$ & $1.3_{-0.1}^{+0.1}$ & 1.05 & 5.49 & $0.3_{-0.1}^{+0.1}$ & -- & -- & -- \\ 
Cool Core    &  --    &  --    &  --    &  --    &  $3.55  \pm 1.53$ & $0.9_{-0.1}^{+0.1}$ & 1.05 & 5.49 & $0.6_{-0.1}^{+0.1}$ & -- & -- & -- \\
Disturbed    &  --    &  --    &  --    &  --    &  $13.81 \pm 1.55$ & $1.6_{-0.1}^{+0.1}$ & 1.05 & 5.49 & $0.0^{+0.1}$       & -- & -- & -- \\ 
%Disturbed    &  --    &  --    &  --    &  --    &  $12.56 \pm 2.07$ & $1.5_{-0.2}^{+0.1}$ & 1.05 & 5.49 & $0.0^{+0.1}$       & -- & -- & -- \\ 
% Well behaved &  --    &  --    &  --    &  --    &  $5.34 \pm 0.08$ & $1.2_{-0.1}^{+0.1}$ & 1.05 & 5.49 & $0.5_{-0.1}^{+0.1}$  & -- & -- & -- \\ 
\hline
All (A10)    &  --    &  --    &  --    &  --    &  $8.403 h_{70}^{-3/2}$ & 1.18 & 1.05 & 5.49 & 0.31 & -- & -- & -- \\
Cool core (A10) &  --    &  --    &  --    &  --    &  $3.249 h_{70}^{-3/2}$ & 1.13 & 1.22 & 5.49 & 0.78 & -- & -- & -- \\
Disturbed (A10) &  --    &  --    &  --    &  --    &  $3.202 h_{70}^{-3/2}$ & 1.08 & 1.41 & 5.49 & 0.38 & -- & -- & --
\enddata
\tablecomments{Results from our pressure profile analysis. $Y_{sph}$ is calculated using the tabulated value of $R_{500}$.
  $^a$Values of $R_{500}$ and $P_{500}$ are taken from \citet{mantz2010} and \citet{sayers2013} respectively.
  We have assumed A10 values of $\alpha$ and $\beta$.
    The findings from A10 are reproduced in the last three rows. The $h_{70}$ dependence is included for explicit replication
    of A10 results; all $P_0$ values have this dependence (the assumed cosmologies are the same).}
\end{deluxetable*}

%%%%%%%%%%%%%%%%%%%%%%%%%%%%%%%%%%%%%%%%%%%%%%%%%%%%%%%%%%%%%%%%%%%%%%%%%%%%%%%%%%%%%%%%%%%%%%%%%%%%%%%%%%%
%%%                                             END OF THAT TABLE!                                      %%%
%%%%%%%%%%%%%%%%%%%%%%%%%%%%%%%%%%%%%%%%%%%%%%%%%%%%%%%%%%%%%%%%%%%%%%%%%%%%%%%%%%%%%%%%%%%%%%%%%%%%%%%%%%%

%%%%%%%%%%%%%%%%%%%%%%%%%%%%%%%%%%%%%%%%%%%%%%%%%%%%%%%%%%%%%%%%%%%%%%%%%%%%%%%%%%%%%%%%%%%%%%%%%%%%%%%%%%%
%%%                                                NEXT SECTION                                         %%%
%%%%%%%%%%%%%%%%%%%%%%%%%%%%%%%%%%%%%%%%%%%%%%%%%%%%%%%%%%%%%%%%%%%%%%%%%%%%%%%%%%%%%%%%%%%%%%%%%%%%%%%%%%%

\section{Integrated Compton $Y$ Scaling Relations}

%\textcolor{red}{I need to introduce integrated Y a bit more I think.}

We calculate integrated Compton $Y$ values at $R_{500}$ due to the expected minimal scatter at intermediate 
radii \citep[e.g.][]{kravtsov2012}. We use $R_{500}$ derived from X-ray observations \citep{mantz2010}, 
and calculate $Y_{sph}$, given by:
\begin{equation}
  Y_{sph}(R) = \frac{\sigma_T}{m_e c^2} \int_0^R P(r') 4 \pi r'^2 dr' 
  \label{eqn:ysph}
\end{equation}
and $Y_{cyl}$, which is given by:
\begin{equation}
  Y_{cyl}(R) = \frac{\sigma_T}{m_e c^2} \int_0^R  2 \pi r dr \int_r^{R_b} \frac{2 r' P(r') dr'}{\sqrt{r'^2 - r^2}},
  \label{eqn:ycyl}
\end{equation}
where we adopt $R_b = 5 R_{500}$ as in \citetalias{arnaud2010}.
The error bars on $Y_{sph}(R_{500})$ and $Y_{cyl}(R_{500})$ are found by calculating the respective quantities 
from the pressure profile fits over the 1000 noise realizations, and taking the values encompassing the middle 68\%. 
We take $M_{500}$ from \citet{mantz2010}, who arrive at $M_{500}$ in the following steps:
(1) take the measured $f_{gas}(r_{2500})$ from \citet{allen2008} and extrapolate it to $f_{gas}(r_{500})$ by using
simulations
(2) determine the deprojected gas mass profile from their X-ray data, and
(3) combine the deprojected gas mass profile with the value of $f_{gas}(r_{500})$ to solve for
$M_{500}$ (and $R_{500}$).
%(1) calculate $R_{500}$ using 
%a ratio of $R_{500} / R_{2500} \sim 2.3$ assuming a NFW profile with concentration parameter $c=4$, (2) calculate 
%$M_{gas,500}$, the total gas mass enclosed in $R_{500}$ from deprojected gas mass (non-parametric) profiles, (3) 
%determine the gas mass fraction, $f_{gas}(r_{500})$, by fitting a power law model to $f_{gas}(r)$ from simulations, and (4)
%calculate $M_{500}$ as $(M_{gas}(r_{500})) ((1+B) f_{gas}(r_{500}))^{-1}$, where $B = 0.03 \pm 0.06$ is a systematic fractional 
%bias.
\citet{mantz2010} note that the dominant source of systematic uncertainty associated with $M_{500}$ comes from the 
uncertainty in the assumed $f_{gas}(r_{2500}) = 0.1104$, which was used in calibrating $f_{gas}(r_{500}) \approx 0.115$.

\begin{figure}[!ht]
  \begin{center}
  \includegraphics[width=0.5\textwidth]{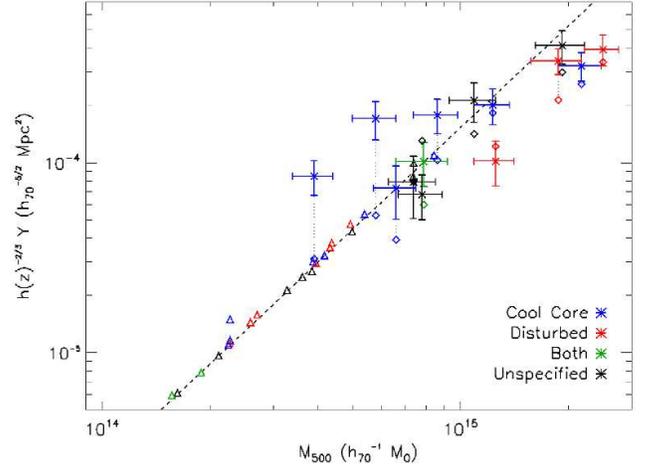}
  \end{center}
  \caption{$Y_{sph,SZ}(R_{500})$ as calculated in this work (Table~\ref{tbl:pressure_profile_results}),
    and $M_{500}$ as calculated from \citet{mantz2010} are shown as asterisks with error bars.
    The scaling relation (dashed line) and triangles
  are from \citet{arnaud2010} and \citet{pratt2010}. The diamonds are $Y_{sph,X}(R_{500})$ as calculated from the gNFW fits
  to the ACCEPT2 pressure profiles. MACS 1311 and MACS 0429 are the notable outliers above the scaling relation.}
% The labelled clusters are those whose $Y_{sph}(R_{500})$ values lie
%  more than $2\sigma$ from the scaling relation from \citet{arnaud2010}, the dot-dashed line.
% Currently Figure 6. (Nov. 2016)
  \label{fig:ysph_scaling}
\end{figure}

We compare our $Y_{sph}(R_{500}) - M_{500}$ relation to that of A10 in Figure~\ref{fig:ysph_scaling}.
The $Y_{sph} - M_{500}$ scaling relation calculated in \citetalias{arnaud2010} is given as:
\begin{equation}
%  h(z)^{-2/3} Y_{sph}(x R_{500}) = A_x \left{[}\frac{M_{500}}{3 \times 10^{14} h_{70}^{-1} M_{\odot}} \right{]}^{\alpha} ,
  h(z)^{-2/3} Y_{sph}(x R_{500}) = A_x \left[ \frac{M_{500}}{3 \times 10^{14} h_{70}^{-1} M_{\odot}} \right] ^{\alpha} ,
  \label{eqn:ysph_scaling}
\end{equation}
where $\alpha = 1.78$, $A_X = 2.925 \times 10^{-5} I(x) h_{70}^{-1}\text{Mpc}^2$, and $I(1) = 0.6145$.
We find six of fourteen clusters that are more than $2 \sigma$ in $Y_{sph}$ from the scaling relation.
When we consider the mass uncertainty, that number drops to three. While our sample size is small, the
tendency of cool core clusters to lie above the scaling relation and of disturbed clusters to lie below the
scaling relation is interesting. Regardless of cluster type, our sample does show a more shallow $Y_{sph,500}-M_{500}$ slope
($1.06\pm0.13$) than the predicted self similar slope ($5/3$) or $1.78$ found in \citetalias{arnaud2010}. 
This is consistent with the slope found for the BOXSZ sample by \citet{czakon2015} for $Y_{cyl,2500} - M_{2500}$ of $1.06\pm0.12$.

%%%%%%%%%%%%%%%%%%%%%%%%%%%%%%%%%%%%%%%%%%%%%%%%%%%%%%%%%%%%%%%%%%%%%%%%%%%%%%%
\section{Combining SZ and X-ray Data}
\label{sec:xray_comp}
%%%%%%%%%%%%%%%%%%%%%%%%%%%%%%%%%%%%%%%%%%%%%%%%%%%%%%%%%%%%%%%%%%%%%%%%%%%%%%%

The observed SZ and X-ray signal from galaxy clusters differ in their 
dependence upon the physical properties in the intracluster medium (ICM). This difference
has, in the past, been exploited to make calculations of the Hubble parameter, $H_0$, 
assuming spherical geometry of galaxy clusters. Alternatively, one could derive the ICM electron
temperature without X-ray spectral information, estimate effects such as helium sedimentation, or,
relax the spherical assumption and estimate cluster elongation along the line of sight. Unfortunately,
these cannot all be independently constrained.
%inferred quantities to calculate cluster elongation
%along the line of sight, helium sedimentation, or independently derive the ICM electron temperature.
%These are all degenerate (i.e. these cannot all be independently constrained). 
Helium sedimentation 
will produce a higher $P_X$ relative to $P_{SZ}$. However, within the predicted range of helium
sedimentation \citep[e.g.][]{peng2009}, we lack sufficient sensitivity to constrain it. Thus, we investigate
cluster geometry and electron temperature individually, and conclude that differences in the SZ and X-ray 
spherically derived pressure profiles are unlikely to be explained exclusively by either cluster
elongation or ICM temperature distribution.

We compare our SZ data (primarily the pressure profiles) to the ACCEPT2 catalog\footnote{ACCEPT2 includes 
any publicly available \emph{Chandra} observations, thus increasing the sample size and integration times 
relative to ACCEPT, as reported by \citet{baldi2014}. A public release
of ACCEPT2 is anticipated in the near future.}
We correct for the difference in cosmologies assumed in our SZ analysis and that used in ACCEPT2.

%When we assume the X-ray derived temperatures, we find most clusters to be elongated along the line of sight,
%which is perhaps a result of the CLASH sample selection. If, on the other hand, we assume the clusters are
%spherical, then our SZ data combined with ACCEPT2 data imply temperatures systematically higher than
%X-ray derived temperatures (this augmentation occurs at larger radii). Such a trend could be indicative of
%a systematic calibration offset in either (or both) dataset(s): the SZ calibration may be biased high, 
%or the X-ray calibration may be biased low.

%with a geometric mean of $2.7$.
%%% Add some general discussion of SZ vs. ACCEPT2?

\subsection{Ellipsoidal Geometry}
\label{sec:ellgeo}

%%% Many comments from CLS.
The geometry of a cluster along the line of sight can be calculated by comparing SZ and
X-ray pressure profiles.  If we assume azimuthal symmetry in the plane of the sky with scale radius $\theta_{proj}$
and a scale radius along the line-of-sight of $\theta_{los}$, then we denote the elongation/compression 
along the line-of-sight with an axis ratio $c = \theta_{los} / \theta_{proj}$, where $c > 1$ implies that the cluster is 
longer along the line-of-sight than in the plane of the sky.
%The X-ray signal is proportional to $n_e^2 \Lambda_{ee}(T,Z)$,
The X-ray surface brightness is proportional to $\int n_e^2 \Lambda(T,Z) \, dl \propto \int (P/T)^2 \Lambda(T,Z) \, dl$,
where $Z$ is the abundance of heavy elements and $\Lambda$ is the X-ray cooling function, 
%while SZ signal is proportional to $n_e T$.
while the SZ signal is proportional to $\int P dl$ (Equation~\ref{eqn:compton_y}).
%We make the simplifying assumption that we can concern ourselves solely with the electron distributions
%as determined by SZ and X-ray observations ($n_{e,SZ}$ and $n_{e,X}$ respectively)
%so that we find $n_{e,SZ} / n_{e,X} = c^{1/2}$. If the temperature distribution determined by ACCEPT2 is
%assumed to be true, then we have $P_{SZ} / P_{X} = c^{1/2}$
The temperature $T$ can be derived from X-ray. Initially, we will assume that the cluster is spherically
symmetric, and derive the pressure profile from the X-ray observations (giving $P_{X}$), and from the SZ
observations (giving $P_{SZ}$). If the pressure profiles disagree, one explanation would be the elongation
of the cluster along the line-of-sight. In this case, the elongation is given by
\begin{equation}
  c = (P_{SZ} / P_{X})^2.
  \label{eqn:axis_ratio}
\end{equation}

%We take $n_e$ as the true electron density, and let 
%it be a function of ellipsoidal radius, $E^2 = \frac{x^2}{a^2} + \frac{y^2}{b^2} + \frac{z^2}{c^2}$, 
%and $\rho = \sqrt{\frac{x^2}{a^2} + \frac{y^2}{b^2}}$ is the radius in the plane of the sky.
%We take our observable as $O(\rho) = \int_{-\infty}^{\infty} Q(\rho, z) dz$, with $Q(\rho, z)$ being our source function.
%For the spherical case (subscripted with $S$), we assume $a=b=c=1$. In finding the geometry along the line
%of sight, we allow $c$ to vary, but keep $a=b=1$. Given the our observable is fixed,
%we have $Q_s(\rho, z) = Q(\rho, z) / c$. For X-ray observations, our source function has the proportionality
%$Q_X \propto n_e^2 \Lambda_{ee}(T_e,Z)$, where $Z$ is the abundance of heavy elements, while for SZ, the source function 
%has the proportionality $Q_{SZ} \propto n_e T_e$. 

To estimate the ellipticity of clusters, we wish to compare the amplitudes, as fit to X-ray and SZ data,
  of a given pressure profile shape per cluster. Thus, we fit the ACCEPT2 pressure profiles with a gNFW pressure
  profile, with $\alpha$ and $\beta$ fixed at their A10 values: 1.05 and 5.49, respectively. 
  The resultant gNFW profile is then integrated along the line of sight (LOS) to create a Compton $y$
  map, and then filtered as discussed in \citet{romero2015a}. We refer to this filtered map,
  per cluster, as the ``ACCEPT2 model''. Allowing the amplitude to vary, we take $P_{SZ}$
  as the amplitude (renormalization) of this ACCEPT2 model when fit to the SZ data, whereby we have
  effectively set $P_X$ to 1 in Equation~\ref{eqn:axis_ratio}. Similarly, we define
  $P_B$ as the fitted amplitude (renormalization) of the ACCEPT2 model to just Bolocam data.
%These are tabulated in Table~\ref{tbl:accept_gnfw}. 

The axis ratio is calculated as $c = P_{SZ}^{2}$, and its associated uncertainty is calculated as 
$\sigma_{c}^2 = 4 P_{SZ}^{4} ((\sigma_{SZ}/P_{SZ})_{tot}^2 + (\sigma_{X}/P_{X})^2)$, 
where $(\sigma_{SZ}/P_{SZ})_{tot}^2 = (\sigma_{SZ}/P_{SZ})_{st.}^2 + 0.11^2$ includes the total (statistical and
calibration uncertainties of Bolocam and $(\sigma_{X}/P_{X}) = 0.10$ is the calibration uncertainty of ACCEPT2. 
Table~\ref{tbl:accept_gnfw} presents relevant fitted gNFW parameters used in calculating
the cluster geometry.

\begin{deluxetable*}{l | ccc | ccc | cc | c}
\tabletypesize{\footnotesize}
\tablecolumns{10}
\tablewidth{0pt} 
% Currently Table 6. (Nov. 2016)
\tablecaption{ACCEPT2 gNFW Fitted Parameters and Comparison to SZ data \label{tbl:accept_gnfw}}
\tablehead{
    Cluster & $P_0$ & $C_{500}$ & $\gamma$ & $P_{SZ}$ & $k$ & $P_{B}$ & $c$ & $\sigma_{c}$ & $\Delta P_{SZ,B} / \sigma_{P_{SZ}}$ 
}
%\begin{table}
%  \centering
%  \begin{tabular}{l l l l l | l l l | l l | l }
%    Cluster & $P_0$ & $C_{500}$ & $\gamma$ & $P_{SZ}$ & k & $P_{B}$ & $c$ & $\sigma_c$ & $\Delta P$ ($\sigma$) \\
%    \hline   
\startdata
  Abell 1835 & 10.7 &  1.4 & 0.44 & $0.83\pm0.03$ & 1.15 & $0.82\pm0.03$ & 0.69 & 0.16 &  0.48 \\
   Abell 611 &  3.3 &  0.9 & 0.62 & $1.31\pm0.09$ & 0.92 & $1.35\pm0.10$ & 1.71 & 0.45 &  0.44 \\
   MACS 1115 & 13.7 &  1.5 & 0.35 & $0.84\pm0.05$ & 1.14 & $0.80\pm0.05$ & 0.70 & 0.17 &  0.71 \\
   MACS 0429 &  3.8 &  1.0 & 0.71 & $1.30\pm0.11$ & 0.64 & $1.48\pm0.11$ & 1.70 & 0.47 &  1.56 \\
   MACS 1206 &  3.7 &  1.0 & 0.49 & $1.12\pm0.04$ & 1.01 & $1.11\pm0.04$ & 1.24 & 0.29 & -0.03 \\
   MACS 0329 &  4.7 &  1.2 & 0.59 & $1.61\pm0.09$ & 0.90 & $1.64\pm0.09$ & 2.58 & 0.64 &  0.41 \\
    RXJ 1347 & 22.8 &  2.4 & 0.40 & $0.95\pm0.02$ & 1.18 & $0.94\pm0.02$ & 0.89 & 0.20 &  0.37 \\
   MACS 1311 & 19.2 &  1.6 & 0.26 & $1.28\pm0.12$ & 0.85 & $1.40\pm0.12$ & 1.64 & 0.47 &  0.96 \\
    MACS1423 & 11.2 &  1.8 & 0.51 & $1.26\pm0.11$ & 0.81 & $1.39\pm0.12$ & 1.58 & 0.44 &  1.12 \\
   MACS 1149 &  3.3 &  0.9 & 0.23 & $1.24\pm0.06$ & 0.70 & $1.28\pm0.06$ & 1.54 & 0.37 &  0.77 \\
   MACS 0717 & 10.2 &  1.5 & 0.00 & $1.36\pm0.04$ & 0.71 & $1.39\pm0.04$ & 1.85 & 0.43 &  0.54 \\
   MACS 0647 &  3.6 &  0.9 & 0.54 & $1.29\pm0.05$ & 1.09 & $1.27\pm0.05$ & 1.67 & 0.40 & -0.38 \\
   MACS 0744 &  0.6 &  0.6 & 0.93 & $1.04\pm0.06$ & 0.94 & $1.05\pm0.06$ & 1.08 & 0.27 &  0.24 \\
    CLJ 1226 & 20.6 &  1.3 & 0.04 & $0.64\pm0.04$ & 1.15 & $0.60\pm0.04$ & 0.41 & 0.11 &  0.97 
\enddata
\tablecomments{$P_0$, $C_{500}$, and $\gamma$ as determined by fitting the ACCEPT2 pressure profiles.
%gNFW fits to the ACCEPT2 pressure profiles. 
  $P_{SZ}$ denotes the fitted amplitude (renormalization) of the ACCEPT2 model to the SZ data. $P_{B}$
  denotes the fitted amplitude (renormalization) of the ACCEPT2 model to just Bolocam data. We fix the gNFW
  parameters $\alpha = 1.05$ and $\beta = 5.49$.  The elongation $c$ is the ratio between the scale radius
  along the line-of-sight and the projected scale radius (taken to be azimuthally symmetric in the plane of
  the sky). Positive values in the column $\Delta P_{SZ,B} / \sigma_{P_{SZ}}$ indicate that the core is more
  spherical than the extended cluster.}
% The column $\Delta k$ ($\sigma$) lists the 
%  significances of a more spherical core, as compared to the outer regions. $\Delta k$ was calculated
%  as the difference between the $k$ in this table (column 5), and the values listed in
%  Table~\ref{tbl:pressure_profile_results}. A negative $\Delta k$ value
%  signifies that the core is more ellipsoidal than the outer regions.}
\end{deluxetable*}

This investigation has made the assumption that the geometry of a given cluster is globally consistent.
That is, one ellipsoidal geometry applies to all regions of the cluster. However, a cluster should appear 
more spherical towards the center, where baryons have
condensed \citep[e.g.][and references therein]{kravtsov2012}. Also, the DM and baryonic distributions
need not align (one need only look at the Bullet cluster \citep{Markevitch2004} for a dramatic example).
This is not a particular concern to this analysis as we are comparing quantities based on the baryonic
distribution, but would be more of a concern when including lensing. 

Across our sample, we find an average pressure ratio $\langle P_{SZ} \rangle = 1.14 \pm 0.09$, where
  we have included the calibration uncertainties in this calculation. We note that the cluster-to-cluster
  scatter in the pressure ratios is  0.25, which is larger than our uncertainty.
%%%a portion ($0.07$) of the uncertainty is from the scatter in the values and the dominant source of uncertainty is the statistical
%%%uncertainties of individual clusters.
  That average pressure ratio corresponds to $\langle c \rangle = 1.31 \pm 0.22$, where again, our cluster-to-cluster
  scatter is quite large (0.58) compared to our uncertainty. 
Using cosmological smooth particle hydrodynamics (SPH) simulations, 
\citet{battaglia2012} find average 2D (random projection) minor-to-major axis ratios $\simeq 0.95$ based on
gas pressure distributions at $\sim R_{500}$ over all cluster masses at $z=0$. This ratio has some dependence on cluster
mass and redshift, where in both cases the deviations from unity grow with increasing mass and with increasing redshift.
%As we have a heterogenous sample, there is no easy comparison with the expected scatter. 
%As a starting point, we take the uncertainty for the $z = 0.5$ bin to be $\sim 0.05$. This would then support potential
%values $0.8 < c < 1/0.8$, which produces tension for our results.

Working with a smaller sample than that in \citet{battaglia2012} size (16 clusters) and higher resolution, 
\citet{lau2011} use a cosmological simulation with adaptive mesh refinement (AMR) code to
investigate the shape of gas and dark matter, assuming different baryonic physics in two separate runs:
a radiative (CSF) and non radiative (NR) run. While comparable 2D projections of the gas density or pressure are
not tabulated in \citet{lau2011}, they find smaller 3D minor-to-major axis ratios of the gas density
than in \citet{battaglia2012}. We may conclude that simulations support average elongation values $0.9 < c < 1/0.9$,
which is in reasonable agreement with our derived average elongation $\langle c \rangle = 1.31 \pm 0.22$.

%in relaxed clusters at $R_{500}$ and $z=0.6$, they find minor-to-major 
%axis ratios of the gas distribution of $0.80 \pm 0.04$ and $0.72 \pm 0.05$ in their CSF and NR runs, respectively.
%For all clusters at $z = 0.6$, these values are $0.69 \pm 0.05$ and $0.66 \pm 0.04$ and correspond to elongations of 
%1.45 and 1.5, respectively. 

Observationally, using SZ and X-ray data on a sample of 25 clusters, \citet{defilippis2005} 
find a median projected elongation of $1.24 \pm 0.09$,
and median elongation along the line of sight ($c$) of $1.08 \pm 0.17$, where two clusters have $c > 2.0$,
and three clusters have $1.5 < c < 2.0$.
Accounting for our uncertainties, only MACS 0329 and CLJ 1226 are outside (by $1\sigma$) of the range of elongations
found in the literature. While this is true, our investigations here are only concerned with elongations along the
line of sight, for which we are dominated by clusters with $c > 1$. This could be due
to a systematic bias of $P_{SZ}$ high, or $P_{X}$ low, or even a selection bias within the CLASH sample.
%Noting that these last values correspond to major-to-minor axis ratios (elongations) 
%of 1.45 and 1.5, and our uncertainties, only MACS 0329 and CLJ 1226 are outside of range found in the literature.
The CLASH sample contains X-ray (20) and lensing (5) selected clusters and was not explicitly designed to
be orientation unbiased. It is, therefore, not too surprising that we find indications that many of the clusters 
in our sample are elongated along the line of sight ($c > 1$).
Abell 1835 is not in the CLASH sample, but is a notably well studied cool core cluster, i.e. it is the subject 
of many studies on the basis of its cool core.

We take the difference, $\Delta P_{SZ,B} = (P_{SZ} - P_{B})$ to be indicative that the gas in the core has a different 
elongation than ICM at moderate to large radii. In particular, for $P_{B} < 1$, then $P_{SZ} > P_{B}$ is indicative of a
more spherical core and for $P_{B} > 1$, then a more spherical core will have $P_{SZ} < P_{B}$. As $P_{SZ}$ and $P_{B}$
are not independent, we appendent, we approximate the uncertainty in $\Delta P_{SZ,B}$ as $\sigma_{P_{SZ}}$ and report the 
pseudo-significances $(\Delta P_{SZ,B} / \sigma_{P_{SZ}}$) of core sphericity 
(relative to the region outside the center) in Table~\ref{tbl:pressure_profile_results}. 
While none of our determinations individually are above $3\sigma$, 
it is nonetheless interesting to note the tendency for core sphericity.

%One way to infer a difference in geometries between the inner and outer regions is to use the (multiplicative) calibration
%offset between Bolocam and MUSTANG in our fits. In almost all cases, we find that $k$ tends to be inversely related to 
%$P_{B}$, which suggests that the central pressure distribution is more spherical than the outer pressure distribution, 
%as shown in Table~\ref{tbl:accept_gnfw}, under $\Delta k$($\sigma$). However, we must attempt to account for the true
%calibration offset. Taking our values from Table~\ref{tbl:pressure_profile_results} to be our best estimate of the true
%calibration offset, we calculate $\Delta k$($\sigma$) using the uncertainty prior of $12$\% and modulate the
%sign such that positive $\Delta k$($\sigma$) indicates that the core is more spherical than the cluster at large scales.

%A simple, but not very robust, estimate of this significance of the signature
%is found by comparing to the calibration offset values found in Table~\ref{tbl:pressure_profile_results}, and
%finding those clusters that show a preference in $\Delta k$ towards a more spherical center. Recalling that $k$ 
%has a prior on it of $12$\%, we can calculate significances shown in Table~\ref{tbl:accept_gnfw}.

%%%%%%%%%%%%%%%%%%%%%%%%%%%%%%%%%%%%%%%%%%%%%%%%%%%%%%%%%%%%%%

\subsection{Temperature profiles}
\label{sec:temp_profiles}

%%%%%%%%%%%%%%%%%%%%%%%%%%%%%%%%%%%%%%%%%%%%%%%%%%%%%%%%%%%%%%

If we assume a given geometry (known ellipticity), then instead of solving for the ellipticity, we can
derive a temperature profile, making use of the direct pressure constraints from SZ observations and the
electron density constraints from X-ray observations. That is, we calculate
\begin{equation}
  T_{SZ} = \frac{P_{SZ}}{n_{e,X}},
  \label{eqn:telec}
\end{equation}
where $P_{SZ}$ is the pressure derived from pressure profile fits to the SZ data (Section~\ref{sec:pp_constraints}) and
$n_{e,X}$ is the deprojected electron density derived from X-ray data by the ACCEPT2 collaboration.
For each bin, we assign radial values as the arithmetic mean of its radial bounds.
%(i.e. $r = (r_i + r_{i+1})/2$), where $r_i$ are the bounds between bins.
Binned values of $P_{SZ}$ are then calculated from the fitted gNFW profile for each radial value 
for the corresponding bins used for $n_{e,X}$.

\begin{figure*}
  \centering
  \begin{tabular}{ccc}
    % Previously: 8_Aug_2016
   \epsfig{file=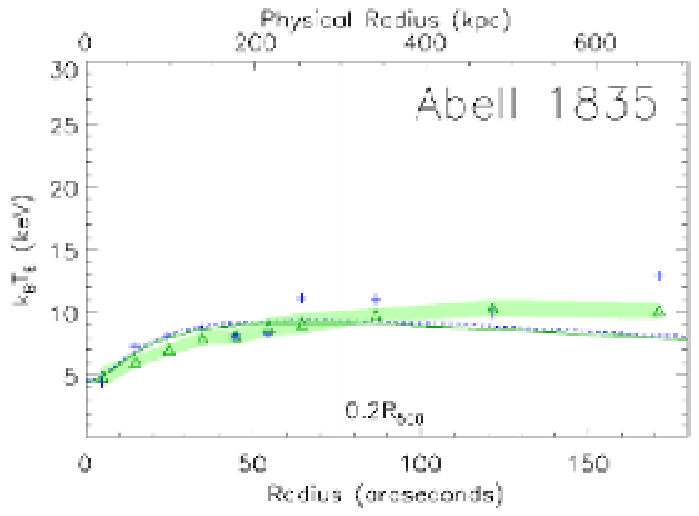,width=0.33\linewidth,clip=}   &
   \epsfig{file=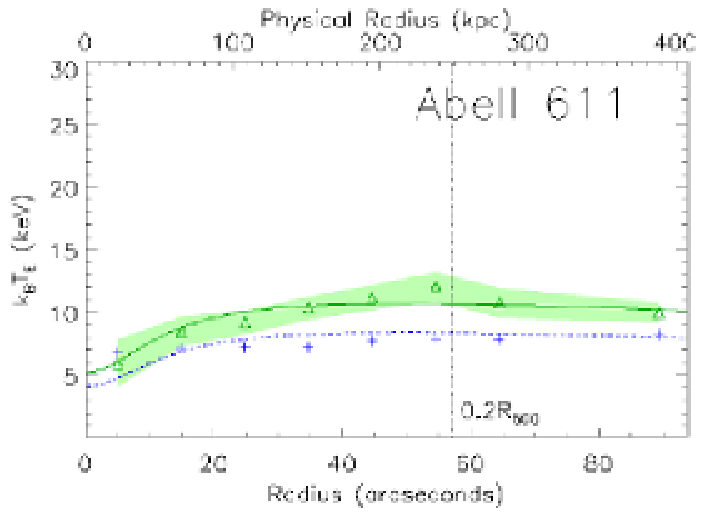,width=0.33\linewidth,clip=}    &
   \epsfig{file=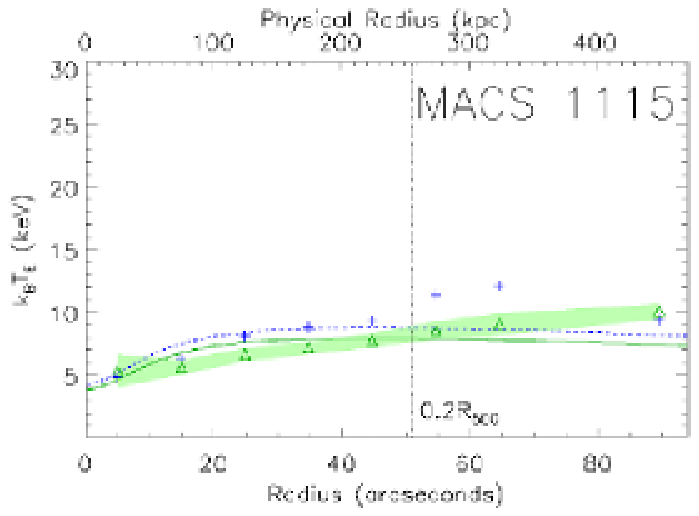,width=0.33\linewidth,clip=}   \\
   \epsfig{file=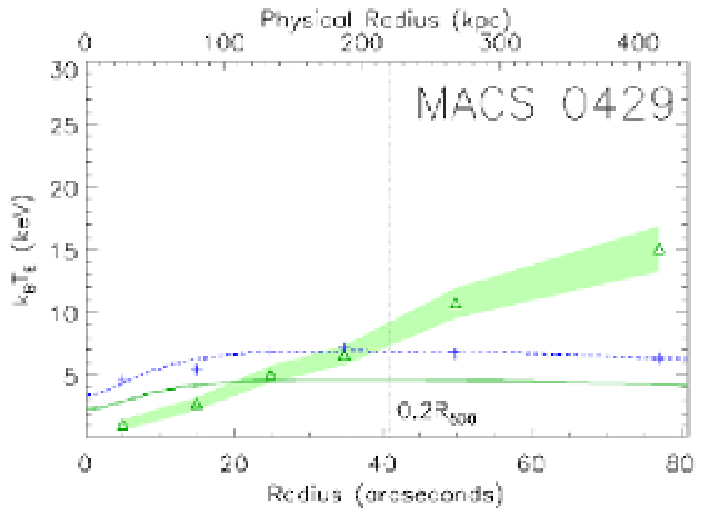,width=0.33\linewidth,clip=}   &
   \epsfig{file=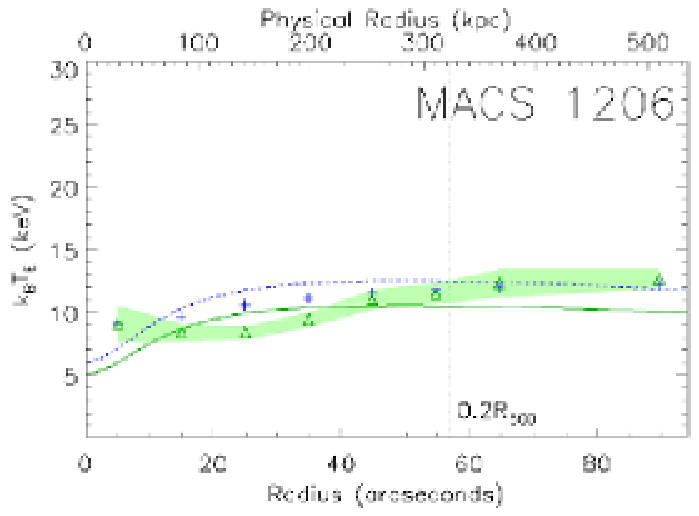,width=0.33\linewidth,clip=}   &
   \epsfig{file=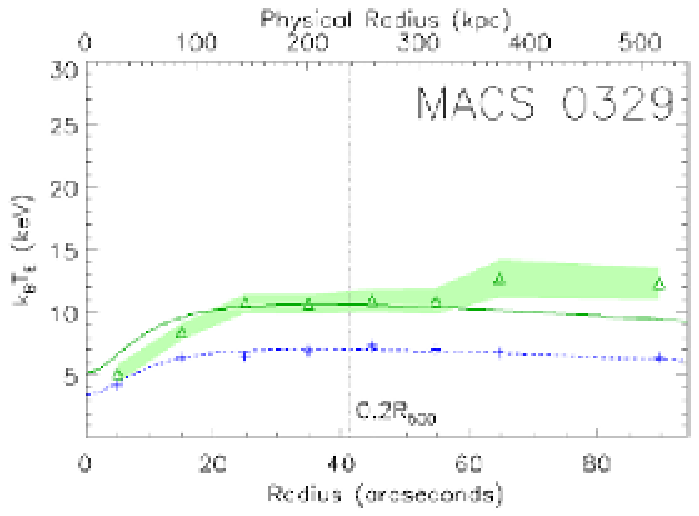,width=0.33\linewidth,clip=}   \\
   \epsfig{file=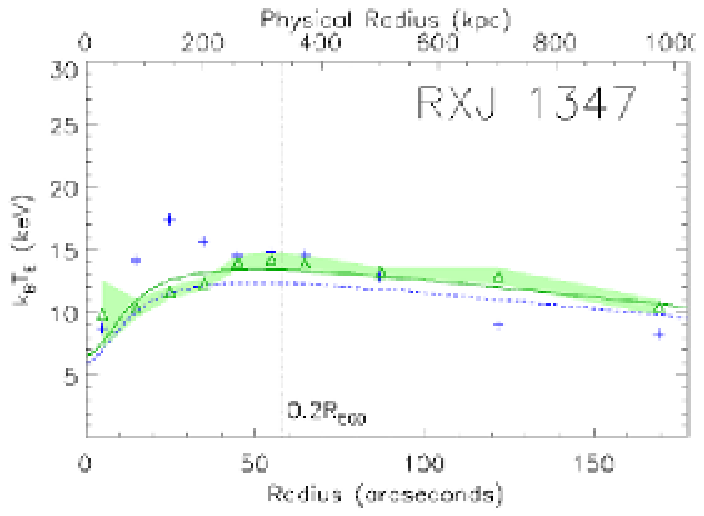,width=0.33\linewidth,clip=} &
   \epsfig{file=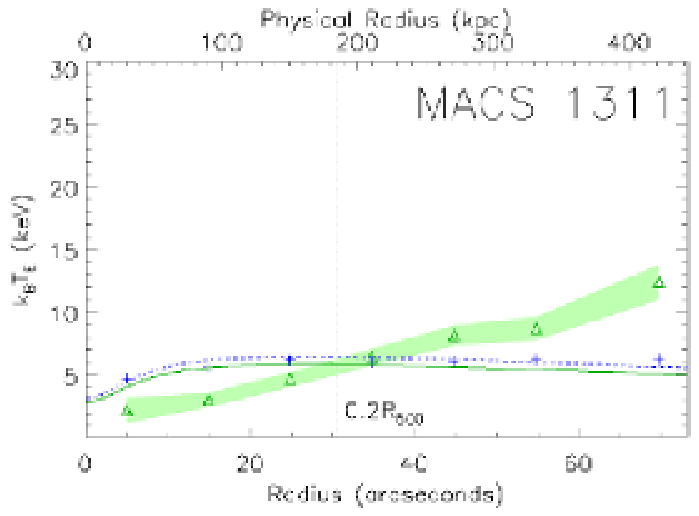,width=0.33\linewidth,clip=}   &
   \epsfig{file=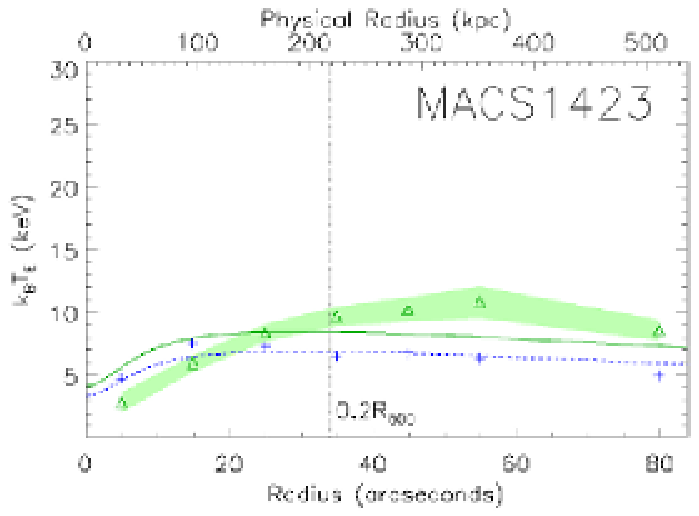,width=0.33\linewidth,clip=}   \\
   \epsfig{file=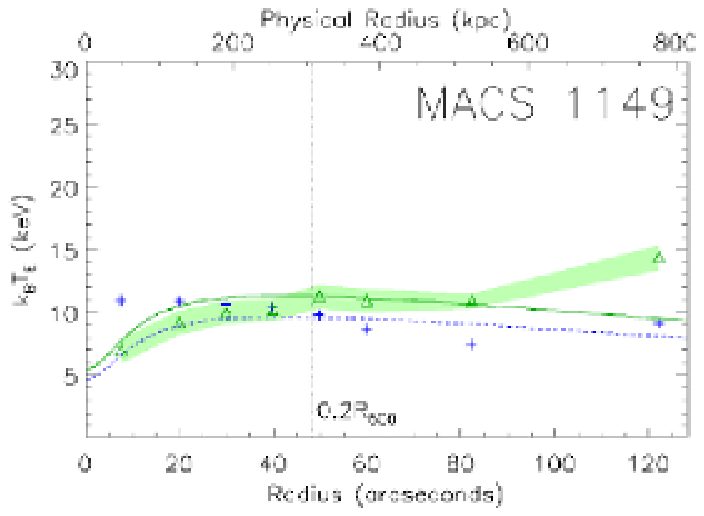,width=0.33\linewidth,clip=}   &
   \epsfig{file=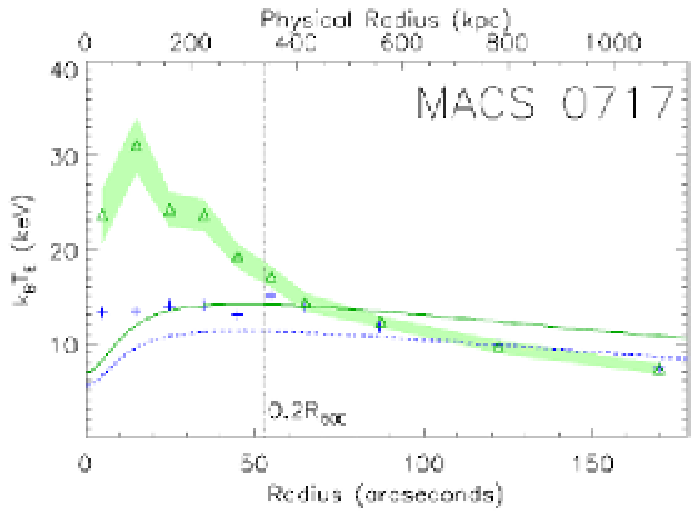,width=0.33\linewidth,clip=}   &
   \epsfig{file=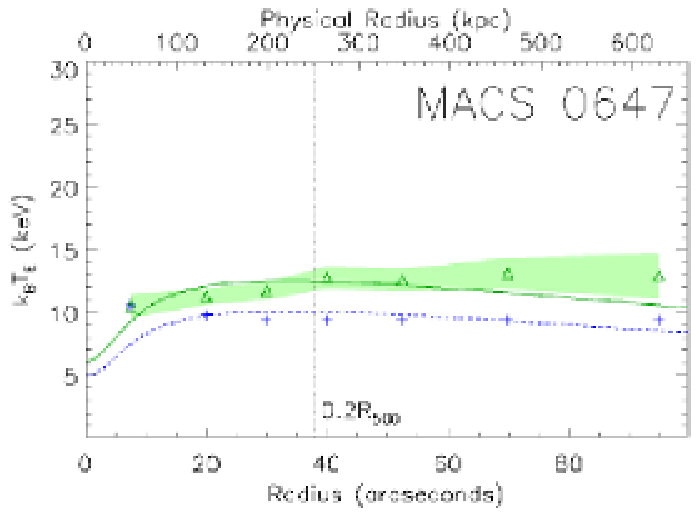,width=0.33\linewidth,clip=}   \\
   \epsfig{file=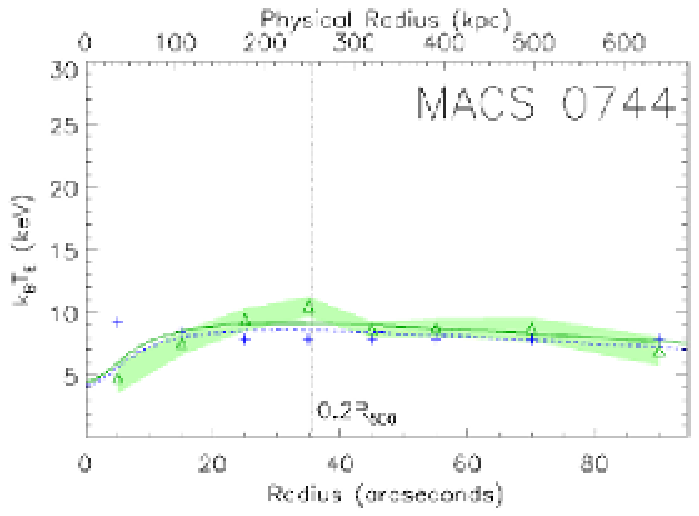,width=0.33\linewidth,clip=}   &
   \epsfig{file=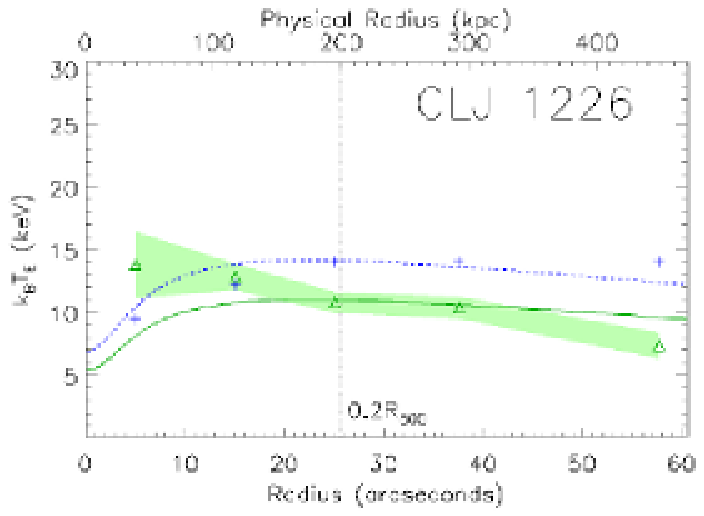,width=0.33\linewidth,clip=}  &
  \end{tabular}
  \caption{Temperature Profiles. The green triangles are derived as $T_{SZ} = P_{SZ} / n_{e,X}$, and the shaded green
    indicates $1\sigma$ uncertainties, including calibration uncertainties. The blue points are X-ray 
    spectroscopically derived temperatures from ACCEPT2 and associated error bars.
    %The solid lines are our fitted Vikhlinin temperature models and the dashed lines are our
    %fitted Bulbul temperature models.
    The solid green curve is $T_{mg}$ normalized to $T_{SZ}$, while the dashed blue curve is
    $T_{mg}$ normalized to $T_{X}$.}
% Currently Figure 7. (Nov. 2016)
  \label{fig:tprofs_all}
\end{figure*}

Our SZ and X-ray derived temperature profiles (Figure~\ref{fig:tprofs_all}) reveal, on average, 
larger temperatures than the spectroscopically derived temperatures from ACCEPT2.
%Generally, both the V06 and B10 models are easily fit to our data (Table~\ref{tbl:temperature_profile_results}),
%as both often have $\tilde{\chi}^2 < 1$. Our fits do account for covariance. We note that the V06 model typically
%performs worse than B10 in our fits (by the goodness of fit parameter, $\chi^2$.) 
%This may be due to fixing $b$ and $c$ (especially $c$) as these are fixed power laws that incorporate the
%behavior of $n_e$ with radius. However, we are still fitting the same number of parameters in the two models, thus
%the B10 model offers more flexibility for the same number of fitted parameters. 
As an additional means of comparison, we fit an average profile derived in \citet{vikhlinin2006b} for the gas
mass weighted temperature:
\begin{equation}
  \frac{T(r)}{T_{mg}} = 1.35 \frac{(x_r/0.045)^{1.9} + 0.45}{(x_r/0.045)^{1.9} + 1} \frac{1}{(1+(x_r/0.6)^2)^{0.45}},
  \label{eqn:tmg}
\end{equation}
where $x_r = r / R_{500}$. Thus, since we take $R_{500}$ as known, the shape of the profile is fixed. The values
fit to the ACCEPT2 temperatures are reported in Table~\ref{tbl:cluster_properties}. We fit $T_{mg}$ to our
$T_{SZ}$ profiles and $T_{X}$ profiles from ACCEPT2, which we take as respective gas mass temperature proxies.
We compute the ratio  $T_{mg,SZ} / T_{mg,X}$ of the two fitted gas mass weighted temperature proxies.
We find that $\langle T_{mg,SZ} / T_{mg,X} \rangle = 1.06$ with a RMS scatter of $0.23$.

From Figure~\ref{fig:tprofs_all}, we see that the shape of $T_{mg}$ is generally quite consistent with the
spectroscopic X-ray temperatures, while it is, in some cases, not reflective of the shape of $T_{SZ}$.
Despite the difference in shapes between $T_{X}$ and $T_{SZ}$, it is of moderate surprise that the shape of
  $T_{mg}$ fits similar temperatures between $T_{mg,X}$ and the $T_{mg,SZ}$.

In contrast to our results, which indicate on-average higher values of Tsz than Tx, we note that
  \citet{rumsey2016} find the opposite trend when comparing SZ data from the Arcminute Microkelvin Imager (AMI)
  with Chandra X-ray data for a subsample of the CLASH clusters (10 of 25), 7 of which overlap with our sample.
  However, \citet{rumsey2016} use a much different technique to
  constrain $T_{SZ}$, based solely on the SZ data with strong priors on cluster parameters such as $f_{gas}$. In
  addition, the potential systematics in the 15 GHz interferometric SZ data used by Rumsey et al. (2016) are
  largely distinct from those related to our higher frequency bolometric SZ images. As a result, it is not
  possible to make a direct comparison of the results to better ascertain the cause of the discrepancy.

%\afterpage{
%\clearpage
\thispagestyle{empty}

\begin{deluxetable}{l | ccc | c}
\tabletypesize{\footnotesize}
\tablecolumns{4}
\tablewidth{0pt} 
% Currently Table 7. (Nov. 2016)
\tablecaption{Normalized Gas Mass Weighted Temperatures \label{tbl:Tmg_normalization_fits}}
\tablehead{
  Cluster &   $T_{mg,SZ}$ & $\chi^2$ &  DOF & $T_{mg,X}$
}
\startdata
Abell 1835 & $ 7.29\pm0.17$ &  29.70 &     9 &  7.49 \\
 Abell 611 & $ 8.56\pm0.31$ &   2.65 &     7 &  6.71 \\
 MACS 1115 & $ 6.35\pm0.19$ &  31.55 &     7 &  7.04 \\
 MACS 0429 & $ 3.71\pm0.25$ &  96.30 &     5 &  5.56 \\
 MACS 1206 & $ 8.44\pm0.22$ &  31.10 &     7 & 10.00 \\
 MACS 0329 & $ 8.52\pm0.27$ &  15.67 &     7 &  5.64 \\
  RXJ 1347 & $10.73\pm0.16$ &  14.32 &     9 &  9.86 \\
 MACS 1311 & $ 4.68\pm0.24$ &  76.77 &     6 &  5.18 \\
 MACS 1423 & $ 6.77\pm0.27$ &  35.39 &     6 &  5.50 \\
 MACS 1149 & $ 9.04\pm0.26$ &  30.10 &     7 &  7.70 \\
 MACS 0717 & $11.35\pm0.27$ & 212.62 &     9 &  9.06 \\
 MACS 0647 & $10.00\pm0.28$ &   7.91 &     6 &  8.06 \\
 MACS 0744 & $ 7.30\pm0.26$ &   5.70 &     7 &  6.85 \\
  CLJ 1226 & $ 8.78\pm0.39$ &  13.27 &     4 & 11.30 \\
%%%%%%%%%%%%%%%%%%%%%%%%%%%%%%%%%%%%%%%%%%%%%%%%%%%%%%%
%Abell 1835 & $ 9.17\pm0.29$ &   4.07 &  9 & 7.49 \\
% Abell 611 & $10.81\pm0.52$ &  10.55 &  7 & 6.71 \\
% MACS 1115 & $ 8.92\pm0.37$ &   0.26 &  7 & 7.04 \\
% MACS 0429 & $ 4.99\pm0.42$ &  63.18 &  5 & 5.56 \\
% MACS 1206 & $12.73\pm0.52$ &   1.75 &  7 & 10.0 \\
% MACS 0329 & $11.66\pm0.53$ &   9.22 &  7 & 5.64 \\
% RXJ  1347 & $12.14\pm0.35$ &  73.54 &  9 & 9.86 \\
% MACS 1311 & $ 6.57\pm0.41$ &  22.40 &  6 & 5.18 \\
% MACS 1423 & $ 8.87\pm0.47$ &  20.46 &  6 & 5.50 \\
% MACS 1149 & $11.99\pm0.49$ &   5.53 &  7 & 7.70 \\
% MACS 0717 & $ 9.80\pm0.37$ & 247.16 &  9 & 9.06 \\
% MACS 0647 & $14.10\pm0.66$ &   6.79 &  6 & 8.06 \\
% MACS 0744 & $ 9.32\pm0.48$ &  19.75 &  7 & 6.85 \\
%  CLJ 1226 & $11.07\pm0.74$ &  30.91 &  4 & 11.3 
\enddata
\tablecomments{The gas mass-weighted temperature proxies, $T_{mg,SZ}$ and $T_{mg,X}$, are calculated by fitting
  a fixed profile shape, $T_{mg}$ (Equation~\ref{eqn:tmg}), to $T_{SZ}$ and $T_{X}$. The $\chi^2$ and degrees of freedom
  for the $T_{mg,SZ}$ fits are tabulated. We find that $T_{mg,SZ}$ is generally larger than $T_{mg,X}$.}
  %Covariance is not taken into account.
  %The ratio between the fitted gas mass temperatures, $T_{mg,SZ} / T_{mg,X}$,
  % is also presented for a global comparison of temperature profiles.}
\end{deluxetable}

Given the typically long exposure times (our sample has total \emph{Chandra} exposure times, $19.5 < t_{exp} < 134.1$ ks)
required to derive spectroscopic X-ray temperatures, it is likely that 
deriving temperatures from SZ pressure profiles and X-ray electron densities will be more commonplace as SZ instruments
have progressed rapidly in recent years. We consider the how the uncertainties of two temperature derivations 
($\sigma_{T_{SZ}}$ and $\sigma_{T_X}$) compare within our sample. We find that $\sigma_{T_{SZ}}$ is generally about
twice as large as $\sigma_{T_X}$. Furthermore, $\sigma_{T_{SZ}}$ is dominated by both the statistical and systematic
uncertainties associated with $P_{SZ}$, where the fractional uncertainties in our SZ pressure profiles are roughly a factor 
of 3 larger than the fractional uncertainties in the X-ray electron densities.

Thus far, we have not taken into consideration systematic errors within $\sigma_{T_X}$. 
The systematic errors on $T_x$ are not well quantified, despite evidence that these systematic
errors can be notable \citep[e.g.][]{donahue2014}. We are thus interested in finding for what fractional systematic error
do the uncertainties in $\sigma_{T_{SZ}}$ and $\sigma_{T_X}$ become comparable and find that a systematic uncertainty 
of 20\% on spectroscopic X-ray temperatures results in $\sigma_{T_{SZ}} \sim \sigma_{T_X}$ over our sample.

We should additionally revisit our uncertainties on $T_{SZ}$ to consider the impact of the uncertainty of the 
cluster geometry on $\sigma_{T_{SZ}}$. With consideration for elongation along the line-of-sight,
$T_{SZ} = (P_{SZ} / n_{e,X}) c^{1/2}$, which will results in an additional fractional error term: $(\sigma_{c} / 2 c)$
to be added in quadrature. From \citet{battaglia2012,lau2011}, we can likely expect this term to be of order $0.1$,
which is the same as the statistical and systematic uncertainties on $P_{SZ}$.

\subsection{Discussion: Comparison Between SZ- and X-ray- Derived Quantities}
\label{sec:discuss}

We find overall agreement in ensemble constraints of the pressure profile between our SZ pressure profiles and
those fitted to ACCEPT2 \citetalias{baldi2014} pressure profiles (Figure~\ref{fig:pp_sets}). 
When calculating elongation along the line-of-sight, we find an average axis ratio 
$\langle c \rangle = 1.38 \pm 0.58$. In our temperature analysis, we find the average gas mass weighted temperature 
ratio $\langle T_{mg,SZ} / T_{mg,X} \rangle = 1.06 \pm 0.23$.
%The temperature ratio does incorporate SZ data
%at large scales (and within Bolocam's sensitivity), whereas the axis ratio does. 

While these average values show consistency between the SZ and X-ray quantities, the SZ pressure is, on average, 
generally larger than the X-ray pressure, especially at larger radii. In our elongation analysis, this pressure 
difference is manifest as the majority of our clusters showing elongation along the line of sight, which we find
is largely consistent with numerical simulations (Section~\ref{sec:ellgeo}). Alternatively, in our temperature
analysis, we find that $T_{SZ}$ is generally larger than $T_X$, especially at larger radii. Differences in temperatures 
could indicate a bias of spectroscopic X-ray temperatures to lower temperatures, as emission will be dominated by 
the cooler (denser) regions. Moreover, \emph{Chandra} is not sensitive to higher energy photons and therefore 
constraints on gas hotter than $k_B T \gtrsim 10$ keV are generally poor.

In two clusters, MACS 0717 and CLJ 1226, we attribute the differences in SZ and X-ray pressure profiles to be 
primarily driven by differences in temperature. The triple merging cluster MACS 0717 (Section~\ref{sec:results_m0717}) 
does not present a clear shock in SZ or X-ray within the central region, but it may be that the merger activity 
is primarily along the line of sight. The notable enhancement of 
SZ-to-X-ray spectroscopic temperature in the center is undoubtedly due to merger activity, and bears credence as
other studies have found hot (roughly 20 keV in \citet{sayers2013,adam2016b}, and 34 keV in \citet{mroczkowski2012}) gas
in the region about subcluster C, which would contribute to temperature enhancements at small radii. In CLJ 1226,
the average temperature values we derive are not significantly different than those in ACCEPT2, but the slope is
reversed. In particular, the ACCEPT2 temperature in CLJ 1226 rises from 10 keV in the core to 15 keV at $r \sim 200$ kpc.
In contrast, $T_{SZ}$ shows a more characteristic, declining temperature profile.
It is unclear what would cause this difference.
We believe that this difference in slope accounts for a non-trivial change in the fitted pressure profile to
ACCEPT2 (Section~\ref{sec:ellgeo}), which drives the corresponding SZ-fitted normalizations ($P_{SZ}$ and $P_B$) low.

A third cluster with notable differences in the pressure profiles is MACS 0429. The SZ and X-ray pressure profiles have
considerably different shapes . While this may be due, in part, to an increase in temperature with radius, we do not 
contend that this increase is as dramatic as that shown in Figure~\ref{fig:tprofs_all}.
It is possible that the intrinsic weakness of the decrement of this cluster, combined with the unusual strength of the
central source ($S_{90} = 8$ mJy), has exceeded the capabilities of our point source treatment (Section~\ref{sec:pp_error}) 
and could thus be biasing the results on this particular cluster. 
However, this should primarily affect the inner pressure profile, and Bolocam is constraining the pressure at moderate 
to large radii to be well above that found in ACCEPT2. 

%Beyond what we have considered thus far, 
Clumping may also be responsible for raising $T_{SZ}$ relative to  $T_{X}$ at larger radii. 
Clumping is expected to increase with radius, and thus may account for some of the discrepancy between our inferred 
temperature and the X-ray spectroscopically derived temperatures. \citet{battaglia2015} find that clumping
is more pronounced for more massive clusters. For the most massive bin of clusters considered, which is most applicable to our
sample, the density clumping ($C_{2,\rho} = \langle \rho^2 \rangle / \langle \rho \rangle ^2$) at $R_{500}$ is roughly 1.2. 
Some SZ/X-ray constraints \citep[e.g.][]{morandi2013b,morandi2014} find clumping factors $C_{2,\rho} \sim 2$ at
$R_{200} \sim 1.6 R_{500}$, are are within agreement with simulations. A clumping factor of 1.2 can account for 
biasing the $T_X$ low relative to $T_{SZ}$ by roughly $\sim5$\%.

%%%%%%%%%%%%%%%%%%%%%%%%%%%%%%%%%%%%%%%%%%%%%%%%%%%%%%%%%%%%%%%%%%%%%%%%%%%%%%%%%%%%%%%%%%%%%%%%%%%%%%%%%%%%%%%%
%%%%%%%%%%%%%%%%%%%%%%%%                    CONCLUSIONS!!!                           %%%%%%%%%%%%%%%%%%%%%%%%%%%
%%%%%%%%%%%%%%%%%%%%%%%%%%%%%%%%%%%%%%%%%%%%%%%%%%%%%%%%%%%%%%%%%%%%%%%%%%%%%%%%%%%%%%%%%%%%%%%%%%%%%%%%%%%%%%%%

\section{Conclusions}
\label{sec:conclusions}

We developed an algorithm to jointly fit gNFW pressure profiles to clusters observed via the SZ
effect with MUSTANG and Bolocam. We applied this algorithm to 14 clusters and found the profiles are 
consistent with a universal pressure profile found in \citet{arnaud2010}. Specifically, the 
pressure profile is of the form:
\begin{equation*}
  \Tilde{P} = \frac{P_0}{(C_{500} X)^{\gamma} [1 + (C_{500} X)^{\alpha}]^{(\beta - \gamma)/\alpha}},
%  \label{eqn:norm_gnfw}
\end{equation*}
where we fixed $\alpha$ and $\beta$ to values found in \citet{arnaud2010}. A comparison to previous
determinations of pressure profiles is shown in Figure~\ref{fig:pp_sets}. Within the radii where we
have the greatest constraints ($0.03 R_{500} \lesssim r \lesssim R_{500}$), the pressure profile from this
work is comparable to the other pressure profiles. This is further evidenced in the parameters themselves,
as seen in Table~\ref{tbl:pressure_profile_results}, especially in comparison to \citetalias{arnaud2010} 
parameter values.

With the high resolution of MUSTANG, we were able to identify and remove point sources. MUSTANG is also
sensitive to substructure, which we modeled and incorporated in our fitting algorithm. In the MUSTANG maps, 
we found that substructure in the central regions of clusters is not a rare occurrence, as four of our 14 
clusters have clearly identified substructure, and two more have potential substructure. However, the substructure
only impacts the fitted pressure profile above a 10\% level for RXJ1347 and MACS 0744, where the substructure
(shocks) occurs very near to the core ($\theta \lesssim 20$\asec).

We find general agreement between the SZ and X-ray pressure profiles for the ensemble of our sample. 
%we found discrepancies between the SZ and X-ray derived
%pressure profiles for individual clusters and investigated the potential to explain these discrepancies
%as being due to cluster geometry and ICM temperature (Section~\ref{sec:xray_comp}). 
Additionally, we investigated cluster geometry by taking the ratio between spherically derived pressure profiles 
as fit to SZ and X-ray data and we found that the clusters have an average axis ratio
$\langle c \rangle = 1.38 \pm 0.58$ (individual axis ratios are tabulated in Table~\ref{tbl:accept_gnfw}).
This suggests that most of these clusters in our sample are elongated along the line of sight. This may not be
surprising for a heterogeneously selected sample such as CLASH, for which several clusters were chosen for their
strong lensing magnifications. We extended our analysis to estimate the relative cluster geometry in the core 
(from MUSTANG), compared to the larger scale ICM (from Bolocam) and we found some 
hint that the cores tend to be more spherical than the ICM at larger radii. 
%\textcolor{red}{larger scale ICM}. However, our assumption is that the
%cluster geometry is one (or two) ellipsoids and this only explains a scalar offset in the pressure profiles.

When we assumed spherical symmetry and independently calculated temperature, $T_{SZ}$, from SZ pressure and electron
density, we found an average gas mass weighted temperature ratio $\langle T_{mg,SZ} / T_{mg,X} \rangle = 1.06 \pm 0.23$.
Furthermore, our profiles of $T_{SZ}$ reveal a trend towards higher temperatures than $T_X$ at larger radii.
We argue that higher $T_{SZ}$ temperatures should be expected in clusters where merging activity will heat the gas
beyond the sensitivity range of X-ray instruments (for \emph{Chandra}, this is roughly $k_B T \gtrsim 10$ keV).

Cluster geometry appears to play a significant role in yielding different SZ- and X-ray-derived pressure profiles
within our sample; however, it is implausible that it is the sole factor to finding larger SZ pressures than
X-ray pressures. Other relevant factors include deviations from ellipsoidal geometry; different sensitivities to
hot gas in SZ and X-ray observations; and, at large radii, clumping of the ICM.
%Other relevant factors may include systematic biases, either instrumental (calibration, or 
%limits on temperature sensitivity), or astrophysical, such as 

%We conclude that cluster geometry and ICM temperature appear crucial in accounting for the differences
%between SZ and X-ray derived pressure profiles. 
%However, we do not believe that either of these (with strict ellipsoids) are sufficient to explain the differences observed. 
%We argue that deviations from the ellipsoidal geometry, such as the geometry structures seen in MACS 1115
%(Figure~\ref{fig:mustang_maps_sample}), will also be important in explaining the discrepancies observed.

%As SZ and X-ray surveys discover many thousands of new clusters over the next few years, it will be imperative
%to jointly analyse clusters, especially at higher redshift, to determine ICM temperatures. This study shows
%the feasibility, alongside the challenges, of determining temperature profiles by joining SZ and X-ray datasets.

Finally, as we look forward to the future of galaxy cluster surveys 
(e.g. SPT3G, ACTpol, WFIRST, SPHEREx, Euclid, LSST, and eRosita),
%\citep[e.g. eRosita, SPT3G, ACTpol][]{borm2014,benson2014,thornton2016}
we expect ICM temperature derivations from SZ intensity and X-ray surface brightness (density) to be more common. 
In our study, the temperatures that we derived in this manner, $T_{SZ}$, are dominated by 
uncertainties in the SZ measurements. The fractional uncertainties in our SZ pressure profiles are roughly a factor 
of 3 larger than the fractional uncertainties in the X-ray electron densities. Despite this, we find $T_{SZ}$ 
uncertainties $\sim 2$ times larger than the statistical spectroscopic X-ray temperature uncertainties. In light of
new SZ instruments \citep[e.g. MUSTANG-2, NIKA2, and ALMA][]{dicker2014a,calvo2016,kitayama2016} coming online 
with vastly improved mapping speeds, our results are encouraging for the prospects of physically characterizing the
ICM of newly discovered systems with rapid follow-up programs.

\section*{Acknowledgements}

While at NRAO and UVA, support for CR was provided through the Grote Reber Fellowship at NRAO. 
Support for CR, PK, and AY was provided by the Student Observing Support (SOS) program. 
Support for TM was partially provided by the National Research 
Council Research Associateship Award at the U.S.\ Naval Research Laboratory. Basic research in radio 
astronomy at NRL is supported by 6.1 Base funding. JS was partially supported by a
Norris Foundation CCAT Postdoctoral Fellowship and by NSF/AST-1313447.

The National Radio Astronomy Observatory is a facility of the National Science Foundation which is operated
under cooperative agreement with Associated Universities, Inc. The GBT observations used in this paper were
taken under NRAO proposal IDs GBT/08A-056, GBT/09A-052, GBT/09C-020, GBT/09C-035, GBT/09C-059, GBT/10A-056, 
GBT/10C-017, GBT/10C-026, GBT/10C-031, GBT/10C-042, GBT/11A-001, and GBT/11B-009 and VLA/12A-340.
We  thank the GBT operators Dave Curry, Greg Monk, Dave Rose, Barry Sharp, and Donna Stricklin for their
assistance. 

The Bolocam observations presented here were obtained form the Caltech Submillimeter Observatory, which,
when the data used in this analysis were taken, was operated by the California Institute of Technology under
cooperative agreement with the National Science Foundation. Bolocam was constructed and commissioned using funds
from NSF/AST-9618798, NSF/AST-0098737, NSF/AST-9980846, NSF/AST-0229008, and NSF/AST-0206158. Bolocam observations
were partially supported by the Gordon and Betty Moore Foundation, the Jet Propulsion Laboratory Research and
Technology Development Program, as well as the National Science Council of Taiwan grant NSC100-2112-M-001-008-MY3.

Access to ACCEPT2 data was possible due to the gracious assistance of Rachael Salmon. We thank the anonymous
referee for helpful comments.

%%%%%%%%%%%%%%%%%%%%%%%%%%%%%%%%%%%%%%%%%%%%%%%%%%%%%%%%%%%%%%%%%%%%%%%%%%%%%%%%%%%%%%%%%%%%%%%%%%%%%%%%%%%%%%%%
%%%%%%%%%%%%%%%%%%%%%%%%                       APPENDIX                              %%%%%%%%%%%%%%%%%%%%%%%%%%%
%%%%%%%%%%%%%%%%%%%%%%%%%%%%%%%%%%%%%%%%%%%%%%%%%%%%%%%%%%%%%%%%%%%%%%%%%%%%%%%%%%%%%%%%%%%%%%%%%%%%%%%%%%%%%%%%

\appendix

%%%%%%%%%%%%%%%%%%%%%%%%%%%%%%%%%%%%%%%%%%%%%%%%%%%%%%%%%%%%%%%%%%%%%%%%%%%%%%%
\section{Residual Maps}
\label{sec:residual_maps}
%%%%%%%%%%%%%%%%%%%%%%%%%%%%%%%%%%%%%%%%%%%%%%%%%%%%%%%%%%%%%%%%%%%%%%%%%%%%%%%

\begin{figure*}
  \begin{center}
  \begin{tabular}{cccc}
    \epsfig{file=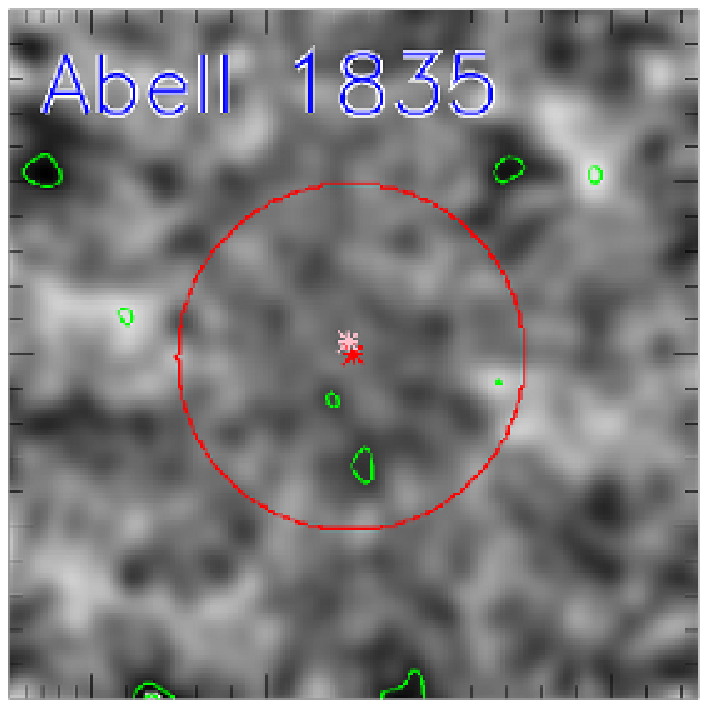,width=0.25\linewidth,clip=} &
    \epsfig{file=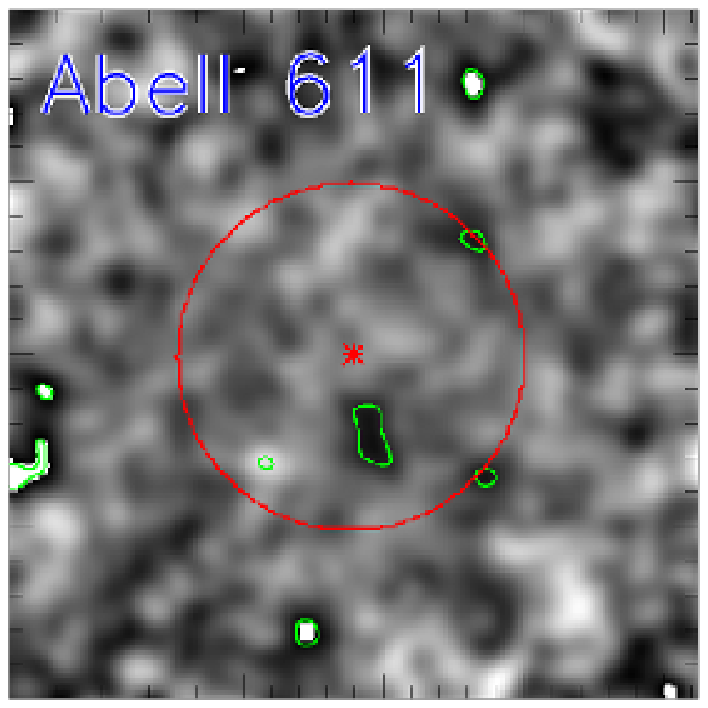,width=0.25\linewidth,clip=} &
    \epsfig{file=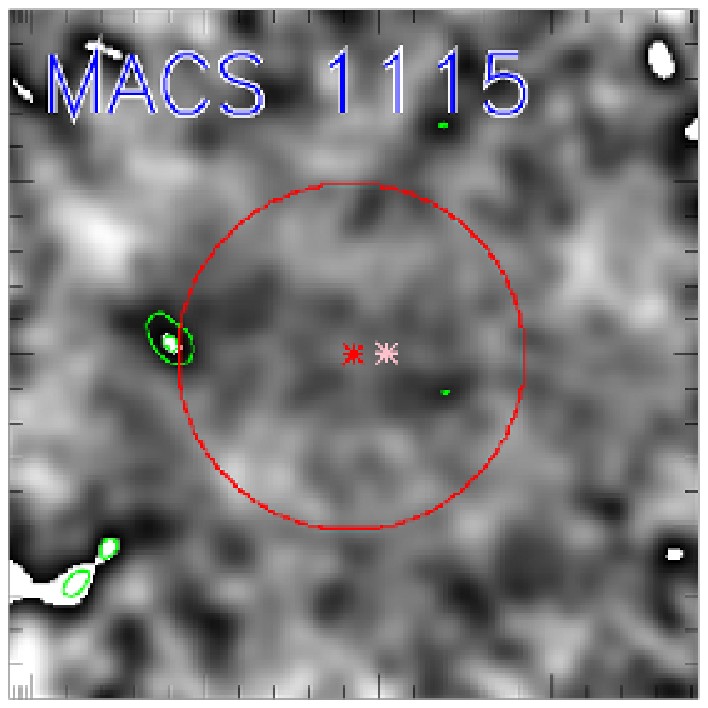,width=0.25\linewidth,clip=} &
    \epsfig{file=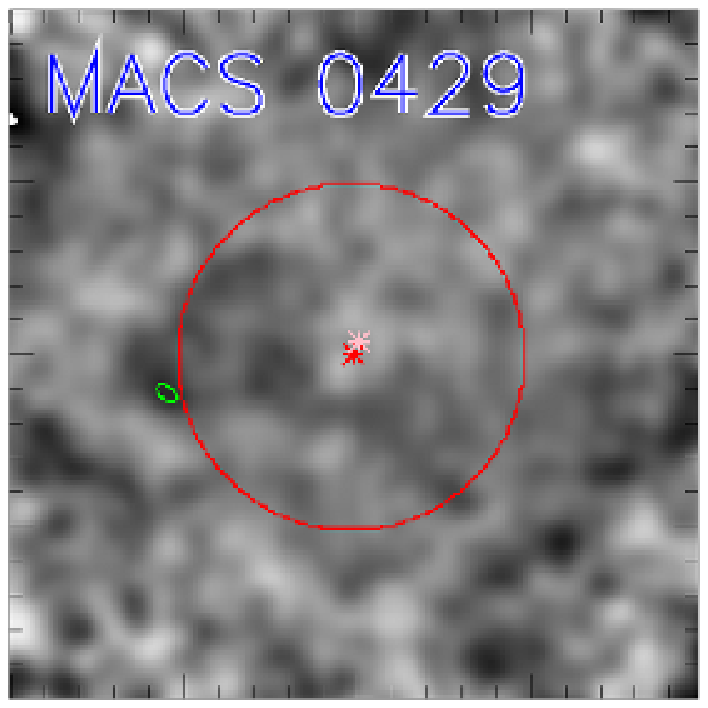,width=0.25\linewidth,clip=} \\
    \epsfig{file=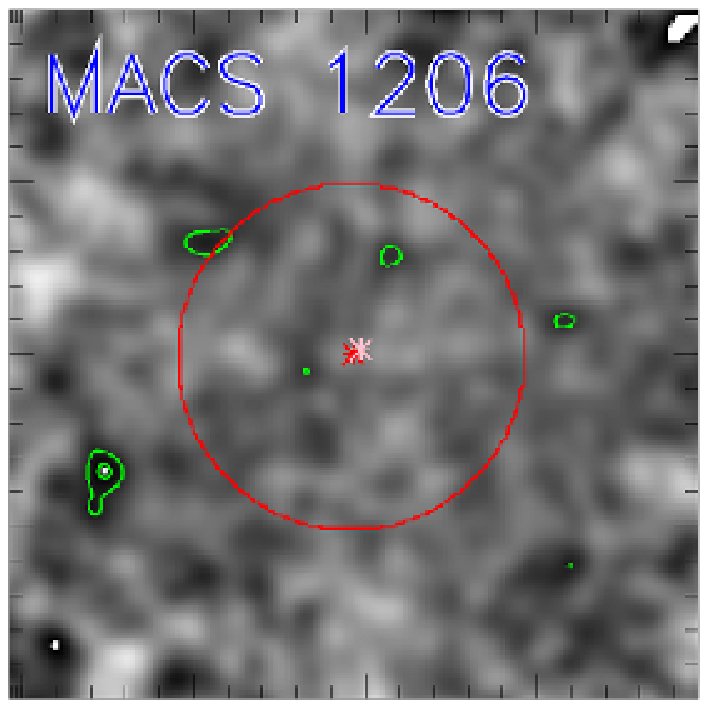,width=0.25\linewidth,clip=} &
    \epsfig{file=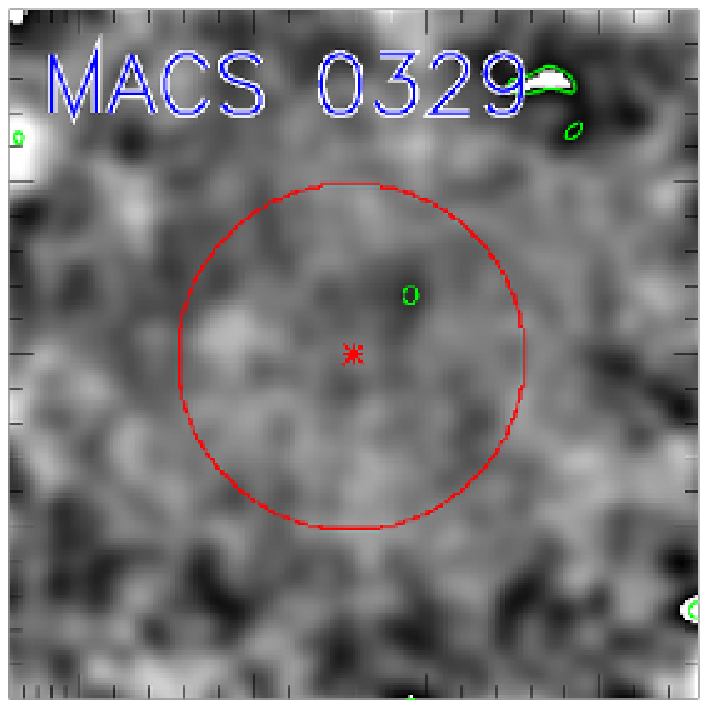,width=0.25\linewidth,clip=} &
    \epsfig{file=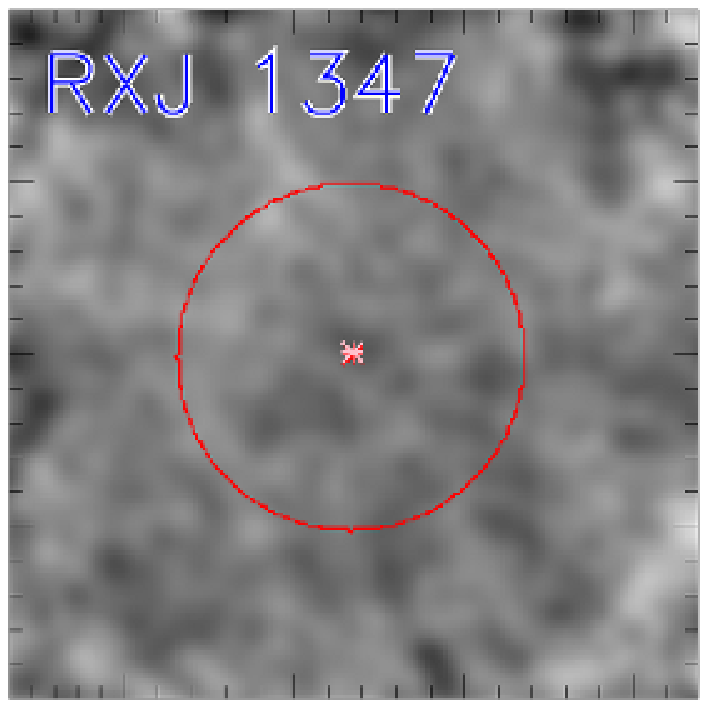,width=0.25\linewidth,clip=} &
    \epsfig{file=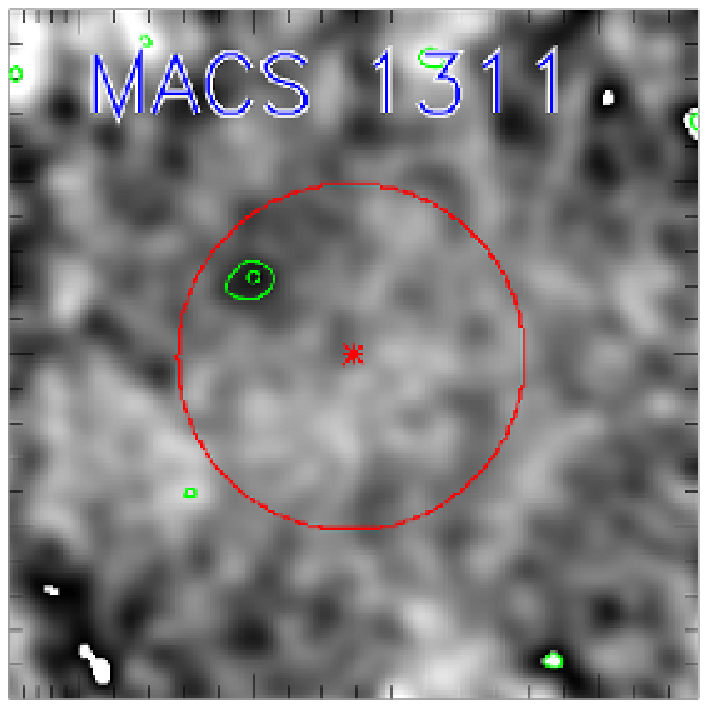,width=0.25\linewidth,clip=} \\ 
    \epsfig{file=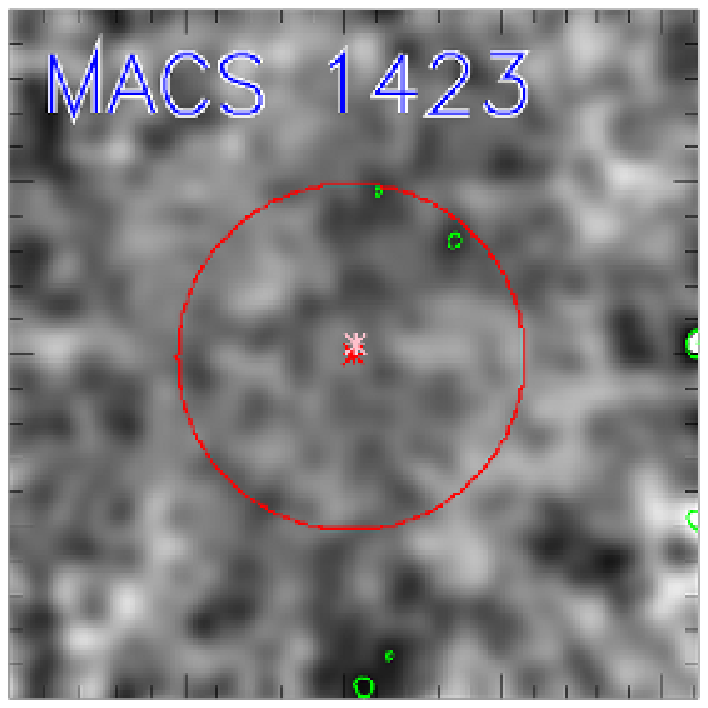,width=0.25\linewidth,clip=} &
    \epsfig{file=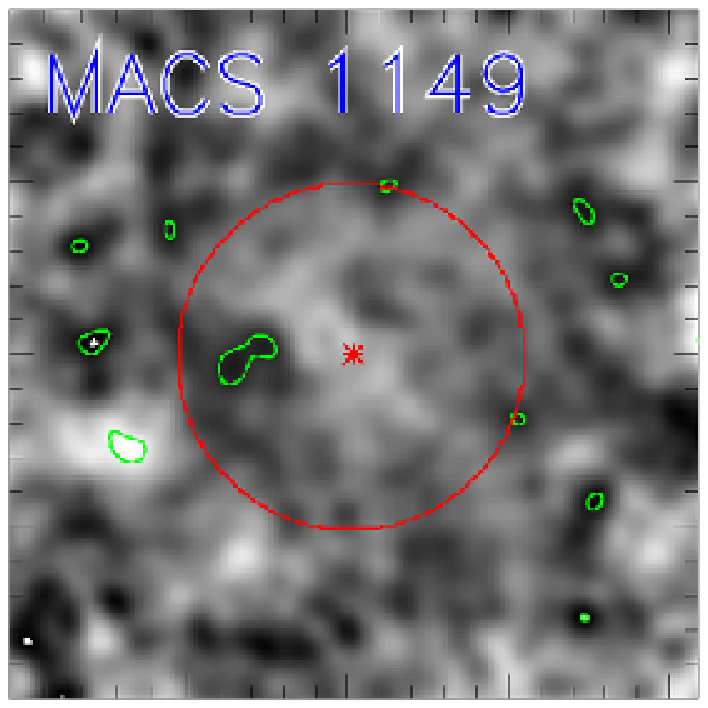,width=0.25\linewidth,clip=} &
    \epsfig{file=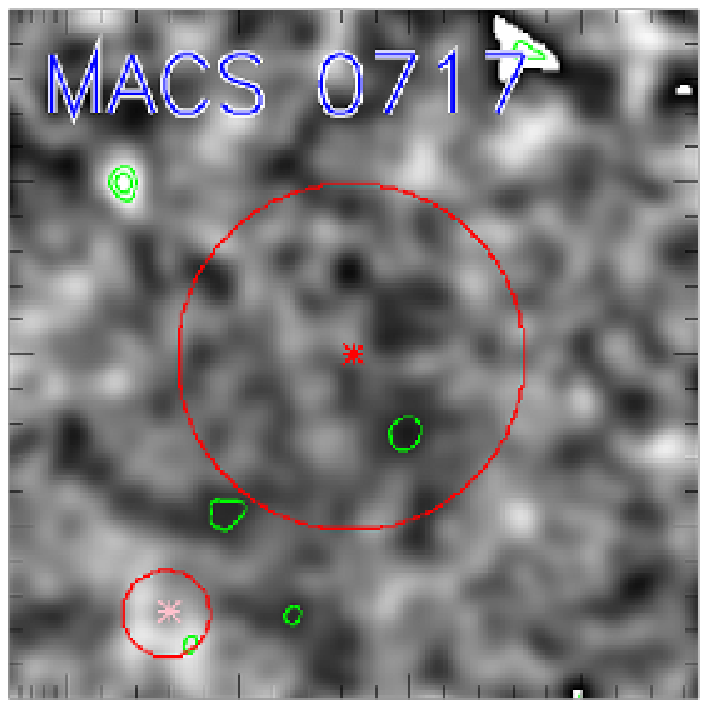,width=0.25\linewidth,clip=} &
    \epsfig{file=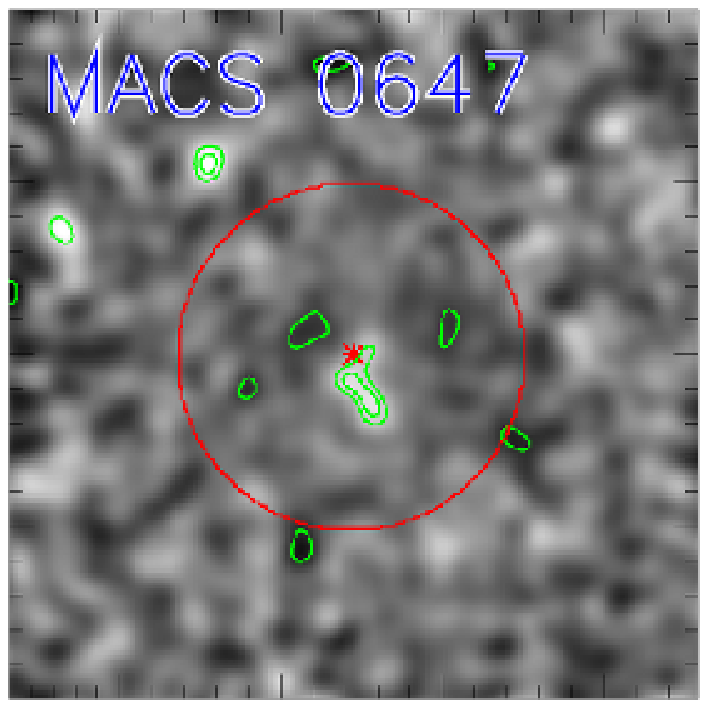,width=0.25\linewidth,clip=} \\
    \epsfig{file=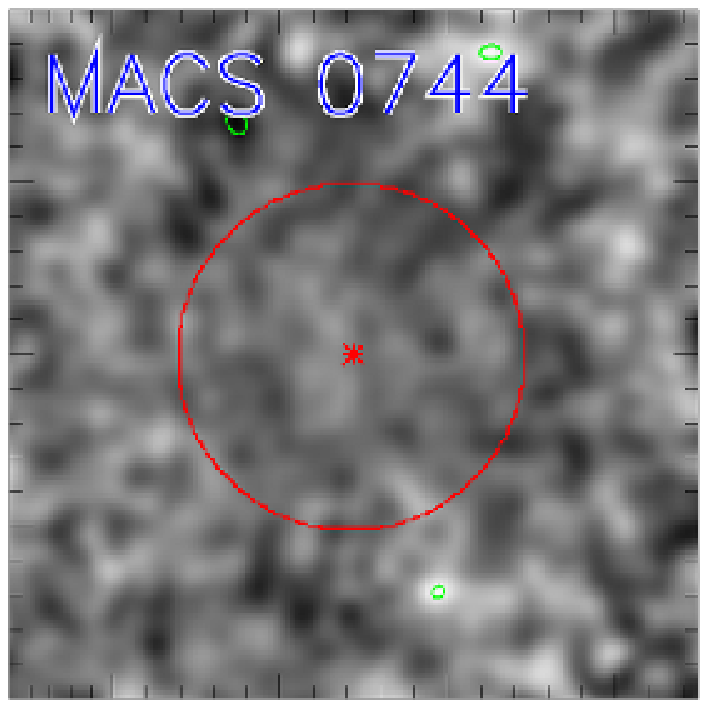,width=0.25\linewidth,clip=} &
    \epsfig{file=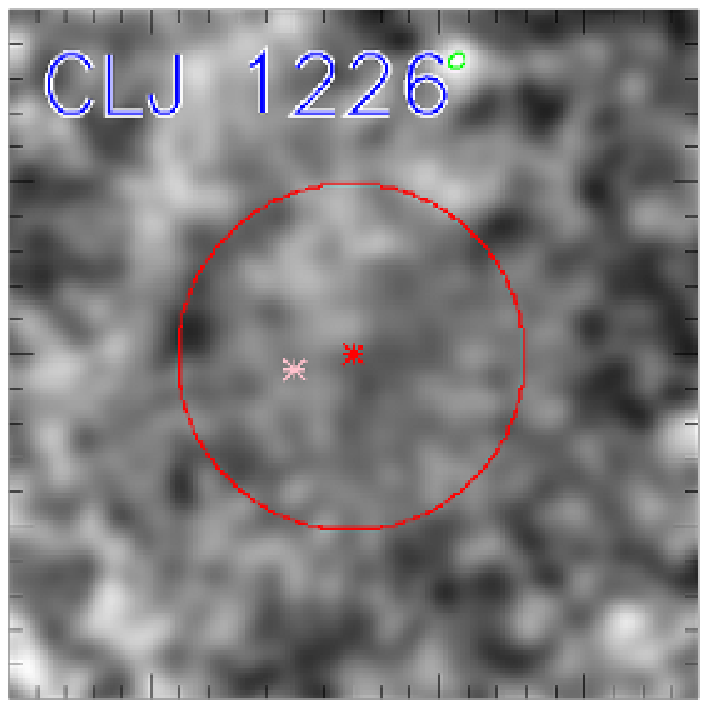,width=0.25\linewidth,clip=} &
 \end{tabular}
  \end{center}
  \caption{Residual MUSTANG flux maps of all clusters. The color scaling spans the range $\pm5\times$Noise$_{M}$, where 
    Noise$_{M}$ (for MUSTANG) is given in Table~\ref{tbl:cluster_obs}. As Noise$_{M}$ was calculated in the inner arcminute, 
    the increase in noise with radius is evident with this scaling. The contours are calculated from a signal-to-noise map 
    (i.e. noise-corrected) and start at $\pm3\sigma$, with $1\sigma$ intervals. The red asterisk is the 
    ACCEPT centroid; the pink asterisk is the point source centroid (if a point source was subtracted). 
    All relevant components, including any residual component, were fit and subtracted.}
% Currently Figure 8. (Nov. 2016)
  \label{fig:resid_maps}
\end{figure*}

%%%%%%%%%%%%%%%%%%%%%%%%%%%%%%%%%%%%%%%%%%%%%%%%%%%%%%%%%%%%%%%%%%%%%%%%%%%%%%%
\section{Notes on Individual Clusters}
\label{sec:ind_notes}
%%%%%%%%%%%%%%%%%%%%%%%%%%%%%%%%%%%%%%%%%%%%%%%%%%%%%%%%%%%%%%%%%%%%%%%%%%%%%%%

%%%%%%%%%%%%%%%%%%%%%%%%%%%%%%%%%%%%%%%%%%%%%%%%%%%%%%%%%%%%%%

\subsection{Abell 1835 (z=0.25)}
\label{sec:results_a1835}

%%%%%%%%%%%%%%%%%%%%%%%%%%%%%%%%%%%%%%%%%%%%%%%%%%%%%%%%%%%%%%

Abell 1835 is a well studied massive cool core cluster. The cool core was noted to have substructure in the central
10\asecs by \citet{schmidt2001}, and identified as being due the central AGN by \citet{mcnamara2006}. Abell 1835 has also
been extensively studied via the SZ effect \citep{reese2002,benson2004,bonamente2006,sayers2011,mauskopf2012}. The models adopted
were either beta models or generalized beta models, and tend to suggest a shallow slope for the pressure interior
to 10\asec. Previous analysis of Abell 1835 with MUSTANG data \citep{korngut2011} detected the SZ effect decrement, but not
at high significance, which is consistent with a featureless, smooth, broad signal. Our updated MUSTANG reduction
of Abell 1835, shown in Figure \ref{fig:mustang_maps_sample}, has the same features as in \citet{korngut2011}.

%%%%%%%%%%%%%%%%%%%%%%%%%%%%%%%%%%%%%%%%%%%%%%%%%%%%%%%%%%%%%%

\subsection{Abell 611 (z=0.29)}
\label{sec:results_a611}

%%%%%%%%%%%%%%%%%%%%%%%%%%%%%%%%%%%%%%%%%%%%%%%%%%%%%%%%%%%%%%

The MUSTANG map (Figure~\ref{fig:mustang_maps_sample}) shows an enhancement
south of the X-ray centroid, and the Bolocam map shows elongation towards the south-southwest. Weak lensing maps
are suggestive of a southwest-northeast elongation \citep{newman2009, zitrin2015}.
Using the density of galaxies, \citet{lemze2013} find a core and a halo which align with the elongation seen 
in the SZ. We note that AMI \citep{hurley-walker2012} and \citet{defilippis2005} also see an elongation in the
same direction in the plane of the sky. However, more recent AMI observations \citep{rumsey2016} show almost
no elongation. Along the line of sight, \citet{defilippis2005} calculate an elongation
$c = 1.05 \pm 0.37$. Despite these notions of elongation, within the sample investigated in
\citet{hurley-walker2012}, Abell 611 is the most relaxed cluster in their sample and that
the X-ray data presented from \citet{laroque2006} is very circular and uniform.
Despite being relaxed, Abell 611 is not listed as a cool core cluster (nor disturbed) \citep{sayers2013}.
%Recent work by Siegal et al., in prep.,
%\citet{siegel2016} 
%finds that both SZ (Bolocam) and X-ray (\emph{Chandra}) data are well described
%by a spherical ICM supported entirely by thermal pressure.

In an analysis of the dark matter distribution, \citet{newman2009} find that the core (logarithmic) slope of the
cluster is shallower than an NFW model, with $\beta_{DM} = 0.3$, where the dark matter distribution has been characterized
by yet another generalization of the NFW profile:
\begin{equation}
  \rho(r) = \frac{\rho_0}{(r/r_s)^{\beta_{tot}}(1 + r/r_s)^{3-\beta_{tot}}}
\end{equation}
They find the distribution of dark matter within Abell 611 to be inconsistent with a single NFW model. 

%%%%%%%%%%%%%%%%%%%%%%%%%%%%%%%%%%%%%%%%%%%%%%%%%%%%%%%%%%%%%%

\subsection{MACS 1115 (z=0.36)}
\label{sec:results_m1115}

%%%%%%%%%%%%%%%%%%%%%%%%%%%%%%%%%%%%%%%%%%%%%%%%%%%%%%%%%%%%%%

MACS 1115 is listed as a cool core cluster \citep{sayers2013}. It is among seven CLASH clusters that show
unambiguous ultraviolet (UV) excesses attributed to unabsorbed star formation rates of 5-80 $M_{\odot} $yr$^{-1}$
\citep{donahue2015}. MUSTANG detects a point source in MACS 1115, which is coincident with its BCG. 
%The NVSS, at 1.4 GHz, finds the flux of the point source to be $16.2$ mJy.
MACS 1115 is fit by a fairly steep inner pressure profile slope to the SZ data (Table~\ref{tbl:pressure_profile_results}).
Adopting the Bolocam centroid, the inner pressure profile slope is notably reduced, yet the goodness of fit is
not significantly changed. In particular, the Bolocam image shows a north-south elongation (particularly to the
north of the centroids). In contrast, weak and strong lensing \citep{zitrin2015} show a more southeast-northwest
elongation.

%%%%%%%%%%%%%%%%%%%%%%%%%%%%%%%%%%%%%%%%%%%%%%%%%%%%%%%%%%%%%%

\subsection{MACS 0429 (z=0.40)}
\label{sec:results_m0429}

%%%%%%%%%%%%%%%%%%%%%%%%%%%%%%%%%%%%%%%%%%%%%%%%%%%%%%%%%%%%%%

MACS 0429 has been well studied in the X-ray \citep{schmidt2007,comerford2007,maughan2008,allen2008,mann2012}
MACS 0429 is identified as a cool core cluster \citep[cf.][]{mann2012,sayers2013}. The bright point source in 
the MUSTANG image is the cluster BCG, which is noted as having an excesses UV emission \citep{donahue2015}.
Of the point sources observed by MUSTANG, this has the shallowest spectral index between 90 GHz and 140 GHz
of $\alpha_{\nu} = 0.55$.
%At 90 GHz, we find the flux density as $7.67 \pm 0.84$ mJy. The point source subtracted from the Bolocam data
%is a $6.0 \pm 1.8$ mJy source at 140 GHz. At 1.4 GHz, NVSS finds the point source to have a flux density of
%$138.8 \pm 4.2$ mJy \citep{condon1998}. 

Despite MACS 0429's stature as a cool core cluster, its pressure profile
(Table~\ref{tbl:pressure_profile_results}) is surprisingly shallow in the core, and shows elevated pressure relative to
X-ray derived pressure at moderate radii. The offset between the Bolocam centroid \citep{sayers2013} and ACCEPT
\citep{cavagnolo2009} centroid is 100 kpc, which is notably larger than the X-ray-optical separations of the cluster
peaks and centroids reported in \citet{mann2012} of 12.8 and 19.5 kpc respectively. \citet{siegel2016}
report an excess in SZ pressure (Bolocam) relative to X-ray (\emph{Chandra}) pressure at moderate to large radii.

%\afterpage{
%\clearpage
%\thispagestyle{empty}
%\begin{figure}
%  \centering
%  \includegraphics[width=0.85\textwidth]{analysis_MACS0429_flux_figure_with_centroid_ptsub_mnsub_9_Jul_2015}
%  \includegraphics[width=0.85\textwidth]{analysis_MBO_Contours_m0429_lens_22_Jan_2015.eps}
%  \caption{MACS 0429}
%  \label{fig:macs_0429params}
%\end{figure}
%\clearpage
%}

%%%%%%%%%%%%%%%%%%%%%%%%%%%%%%%%%%%%%%%%%%%%%%%%%%%%%%%%%%%%%%

\subsection{MACS 1206 (z=0.44)}
\label{sec:results_m1206}

%%%%%%%%%%%%%%%%%%%%%%%%%%%%%%%%%%%%%%%%%%%%%%%%%%%%%%%%%%%%%%

MACS 1206 has been observed extensively \citep[e.g.][]{ebeling2001,ebeling2009,gilmour2009,umetsu2012,
zitrin2012a,biviano2013,sayers2013}. It is not categorized as a cool core or a disturbed cluster
\citep{sayers2013}. Using weak lensing data from Subaru, \citet{umetsu2012} find that the major-minor 
axis ratio of projected mass is $\gtrsim 1.7$ at $1\sigma$. They infer that this high ellipticity and 
alignment with the BCG, optical, X-ray, and LSS shapes are suggestive that the major axis is aligned 
close to the plane of the sky. In \citet{young2015}, substructure is identified that corresponds to an 
optically-identified subcluster, which may either be a merging subcluster, or a foreground cluster. 
In this analysis, the SZ signal observed by MUSTANG is well modelled by a residual component (coincident 
with the subcluster) and a spherical bulk ICM component. We note that the Bolocam contours of MACS 1206
do not exhibit much ellipticity. We do find that MACS 1206 has a major-minor axis 
ratio of $1.24 \pm 0.29$ (Table~\ref{tbl:accept_gnfw}), where the major axis is along the line of sight.

%The point source was found to have a flux density of $0.77 \pm 0.06$ mJy with the best fit model in
%\citet{young2015}. In this analysis, we find it to have a flux density of $0.75 \pm 0.08$ mJy. A proposal 
%has been accepted for \emph{XMM-Newton} observations of this substructure (PI: Sarazin).

%\afterpage{
%\clearpage
%\thispagestyle{empty}
%\begin{figure}
%  \centering
%  \includegraphics[width=0.85\textwidth]{analysis_cres/JF_Conf_Intervals_2params_both_default_speedy_9_Feb_2015_m1206.eps}
%  \includegraphics[width=0.85\textwidth]{analysis_cres/PPP_arnaud_v3_log-log_30_Mar_2015_m1206.eps}
%  \caption{MACS 1206}
%  \label{fig:macs_1206params}
%\end{figure}
%\clearpage
%}

%%%%%%%%%%%%%%%%%%%%%%%%%%%%%%%%%%%%%%%%%%%%%%%%%%%%%%%%%%%%%%

\subsection{MACS 0329 (z=0.45)}
\label{sec:results_m0329}

%%%%%%%%%%%%%%%%%%%%%%%%%%%%%%%%%%%%%%%%%%%%%%%%%%%%%%%%%%%%%%

MACS 0329 has the distinction of being listed as both a cool core and disturbed cluster. Although it has
been classified as relaxed \citep{schmidt2007}, substructure has been noted \citep{maughan2008}, and it earns
its cool core and disturbed classifications based on central weighting of X-ray luminosity and comparing
centroid offsets between optical and X-ray data \citep{sayers2013}. The elongation of the weak lensing and
strong lensing are towards the northwest and southeast of the centroid.

MACS 0329 has two systems with multiple images: one at $z = 6.18$ and the other at $z = 2.17$. The Einstein
radii for these two systems are $r_E = 34$\asecs and $r_E = 28$\asec, respectively \citep{zitrin2012b}, which is
noted as being typical for relaxed, well-concentrated lensing clusters.

%\afterpage{
%\clearpage
%\thispagestyle{empty}
%\begin{figure}
%  \centering
%  \includegraphics[width=0.85\textwidth]{analysis_cres/JF_Conf_Intervals_2params_both_default_speedy_9_Feb_2015_m0329.eps}
%  \includegraphics[width=0.85\textwidth]{analysis_cres/PPP_arnaud_v3_log-log_23_Feb_2015_m0329.eps}
%  \caption{MACS 0329}
%  \label{fig:macs_0329params}
%\end{figure}
%\clearpage
%}

%%%%%%%%%%%%%%%%%%%%%%%%%%%%%%%%%%%%%%%%%%%%%%%%%%%%%%%%%%%%%%

\subsection{RXJ1347 (z=0.45)}
\label{sec:results_rxj1347}

%%%%%%%%%%%%%%%%%%%%%%%%%%%%%%%%%%%%%%%%%%%%%%%%%%%%%%%%%%%%%%

RXJ1347 is one of the most luminous X-ray clusters, and has been well studied in radio, SZ, lensing, optical
spectroscopy, and X-rays \citep[e.g.][]{schindler1995,allen2002, pointecouteau1999,komatsu2001,kitayama2004,
gitti2007b,ota2008,bradac2008,miranda2008}. X-ray contours have long suggested RXJ1347 is a relaxed system
\citep[e.g.][]{schindler1997}, and it is classified as a cool core cluster \citep[e.g.][]{mann2012,sayers2013}. 

Indeed, the first sub-arcminute SZ observations \citep{komatsu2001,kitayama2004} saw an enhancement to
the southeast of the cluster X-ray peak, which was suggested as being due to shock heating. This enhancement
was confirmed by MUSTANG \citep{mason2010}. Further measurements were made with CARMA \citep{plagge2013},
which find the 9\% of the thermal energy in the cluster is in sub-arcminute substructure. Most recently,
\citet{kitayama2016} has observed this cluster with ALMA to a resolution of 5\asec.
At low radio frequencies \citep[][237 MHz and 614 MHz]{ferrari2011},
\citep[][1.4 GHz]{gitti2007a} find evidence for a radio mini-halo in the core of RXJ1347. The cosmic ray electrons
are thought to be reaccelerated because of the shock and sloshing in the cluster \citep{ferrari2011}.

We observe a point source (coincident with the BCG) with flux density of $7.40 \pm 0.58$ mJy. Previous analysis of 
the MUSTANG data found the point source flux density as 5 mJy \citep{mason2010}. The difference in the flux 
densities is likely accounted by (1) the different modeling of point sources; primarily that we filter the double 
Gaussian, (2) we simultaneously fit the components, and (3) we assume a steeper profile in the core than the beta 
model assumed in \citet{mason2010}. Lower frequency radio observations found the flux density of the source to be 
$10.81 \pm 0.19$ mJy at 28.5 GHz \citep{reese2002}, and $47.6 \pm 1.9$ mJy at 1.4 GHz \citep{condon1998}. The BCG 
is observed to have a UV excess\citep{donahue2015}. 

Despite the classification of being a cool core cluster, it is also observed that there are hot regions, initially
constrained as $k_BT > 10$ keV \citep[e.g.][]{allen2002,bradac2008}, and more recently constrained to even hotter 
temperatures \citep[$k_BT > 20$ keV][]{johnson2012}, indicative of an unrelaxed cluster. \citet{johnson2012} also 
interpret the two cold fronts as being due to sloshing, where a subcluster has returned for a second passage.

Several previous studies have found similar evidence for compression along the line of sight in this cluster
\citep[e.g.][and references therein]{plagge2013}. However, the compression we find in this study is less 
severe as in \citet{plagge2013}.
%\afterpage{
%\clearpage
%\thispagestyle{empty}
%\begin{figure}
%  \centering
%  \includegraphics[width=0.85\textwidth]{analysis_cres/JF_Conf_Intervals_2params_both_default_speedy_9_Feb_2015_rxj1347.eps}
%  \includegraphics[width=0.85\textwidth]{analysis_cres/PPP_arnaud_v3_log-log_3_Feb_2015_rxj1347.eps}
%  \caption{RXJ1347}
%  \label{fig:rxj1347params}
%\end{figure}
%\clearpage
%}

%%%%%%%%%%%%%%%%%%%%%%%%%%%%%%%%%%%%%%%%%%%%%%%%%%%%%%%%%%%%%%

\subsection{MACS 1311 (z=0.49)}
\label{sec:results_m1311}

%%%%%%%%%%%%%%%%%%%%%%%%%%%%%%%%%%%%%%%%%%%%%%%%%%%%%%%%%%%%%%

MACS 1311 is listed as a cool core cluster \citep[e.g.][]{sayers2013}, and appears to have quite circular
contours in the X-ray and lensing images, yet has evidence for some disturbance, given its classification
in \citet{mann2012}. However, the SZ contours from Bolocam show some enhancement  to the west, and has
a notable centroid shift ($27.7$\asec, 167 kpc) westward from the X-ray centroid. When fitting pressure profiles
to this cluster, it appears that the enhanced SZ pressure at moderate radii ($r \sim 100$\asec) is due
to this enhancement, especially when noting that we use the X-ray centroid. Adopting the Bolocam centroid
does not change the pressure profile much, and we still observe a pressure enhancement at moderate radii.
In contrast, in their analysis, \citet{siegel2016} find that X-ray (\emph{Chandra}) and SZ (Bolocam) data 
are in good agreement with a spherical ICM model which is supported primarily with thermal pressure.

%\afterpage{
%\clearpage
%\thispagestyle{empty}
%\begin{figure}
%  \centering
%  \includegraphics[width=0.85\textwidth]{analysis_cres/JF_Conf_Intervals_2params_both_default_speedy_3_May_2015_m1311.eps}
%  \includegraphics[width=0.85\textwidth]{analysis_cres/PPP_arnaud_v3_log-log_26_Apr_2015_m1311.eps}
%  \caption{MACS 1311}
%  \label{fig:macs_1311params}
%\end{figure}
%\clearpage
%}

%%%%%%%%%%%%%%%%%%%%%%%%%%%%%%%%%%%%%%%%%%%%%%%%%%%%%%%%%%%%%%

\subsection{MACS 1423 (z=0.54)}
\label{sec:results_m1423}

%%%%%%%%%%%%%%%%%%%%%%%%%%%%%%%%%%%%%%%%%%%%%%%%%%%%%%%%%%%%%%

MACS 1423 is a cool core cluster \citep{mann2012,sayers2013}. While the Bolocam contours are quite concentric,
and suggestive of a relaxed cluster, the centroid is still offset from the X-ray peak by an appreciable angle 
($19.8$\asec, 126 kpc). While AMI \citep{rumsey2016} shows a perturbation/extension to the
southwest of the cluster, their analysis is supportive of MACS 1423 being a relaxed cluster.
Similar to MACS 1311, the pressure is slightly less than the ACCEPT2 X-ray derived pressure in the
core, and slightly greater at moderate radii. While this is expected for a centroid offset, we find that adopting
the Bolocam centroid again yields no substantial difference in the SZ pressure profile. Both our analysis and
that of \citet{siegel2016} find good agreement between SZ and X-ray pressure profiles. We observe a point source 
(the cluster BCG) with flux density of $1.36 \pm 0.13$ mJy, which is also observed to have a UV excess 
\citep{donahue2015}. 

%\afterpage{
%\clearpage
%\thispagestyle{empty}
%\begin{figure}
%  \centering
%  \includegraphics[width=0.85\textwidth]{analysis_cres/JF_Conf_Intervals_2params_both_default_speedy_9_Feb_2015_m1423.eps}
%  \includegraphics[width=0.85\textwidth]{analysis_cres/PPP_arnaud_v3_log-log_24_Feb_2015_m1423.eps}
%  \caption{MACS 1423}
%  \label{fig:macs_1423params}
%\end{figure}
%\clearpage
%}

%%%%%%%%%%%%%%%%%%%%%%%%%%%%%%%%%%%%%%%%%%%%%%%%%%%%%%%%%%%%%%

\subsection{MACS 1149 (z=0.54)}
\label{sec:results_m1149}

%%%%%%%%%%%%%%%%%%%%%%%%%%%%%%%%%%%%%%%%%%%%%%%%%%%%%%%%%%%%%%

MACS 1149 is classified as a disturbed cluster \citep[e.g.][]{mann2012,sayers2013}, and lensing studies have found
that a single DM halo does not describe the cluster well, but rather at least four large-scale DM hales are used to
describe the cluster \citep{smith2009}. A large radial velocity dispersion \citep[1800 km s$^{-1}$][]{ebeling2007} is 
observed, indicative of merger activity along the line of sight. X-ray, SZ, and lensing (particularly 
strong lensing) all show elongation in the northwest-southeast direction. More recently, \citet{golovich2016} 
investigate the dynamics of the cluster, identifying three subclusters, with merger activity (velocities) primarily 
in the plane of the sky. SZ data from AMI \citep{rumsey2016} does not strongly indicate merger activity,
  arguably because the mass of the primary halo is much greater than the subhalos. Morphologically, the SZ map from
  AMI does show minor asphericity.

%We see a $3\sigma$ feature to the east of
%the centroids, but it is not clear that this is associated with any particular feature.

While the Bolocam map of this cluster shows a modest elongation in the northwest-southeast direction, it is
well modelled as a spherical cluster. Our SZ derived pressure profile roughly matches the shape of the X-ray 
derived pressure profile, with the SZ pressure consistently greater than the X-ray pressure. We calculate
that the axis along the line of sight is $1.54 \pm 0.37$ (Section~\ref{sec:ellgeo}) times greater than the 
axes in the plane of the sky. In the MUSTANG map, we see a $3\sigma$ feature to the east of the centroids, 
but it is not clear that this is associated with any particular feature.
%Although we do not find previous analysis of the elongation in the plane of the sky, we might
%expect this given (1) the inferred merger activity along the line of sight, and (2) the lensing strength of the cluster.

%\afterpage{
%\clearpage
%\thispagestyle{empty}
%\begin{figure}
%  \centering
%  \includegraphics[width=0.85\textwidth]{analysis_cres/JF_Conf_Intervals_2params_both_default_speedy_9_Feb_2015_m1149.eps}
%  \includegraphics[width=0.85\textwidth]{analysis_cres/PPP_arnaud_v3_log-log_22_Jan_2015_m1149.eps}
%  \caption{MACS 1149}
%  \label{fig:macs_1149params}
%\end{figure}
%\clearpage
%}

%%%%%%%%%%%%%%%%%%%%%%%%%%%%%%%%%%%%%%%%%%%%%%%%%%%%%%%%%%%%%%

\subsection{MACS 0717 (z=0.55)}
\label{sec:results_m0717}

%%%%%%%%%%%%%%%%%%%%%%%%%%%%%%%%%%%%%%%%%%%%%%%%%%%%%%%%%%%%%%

%\begin{figure}
%  \centering
%  \includegraphics[width=0.85\textwidth]{analysis_M0717_mroczkowski_fig1.eps}
%  \caption{From \citet{mroczkowski2012}.}
%  \label{fig:m0717_mroczkowski}
%\end{figure}

Despite MACS 1149's impressive merging activity, MACS 0717 is thought to be the most disturbed massive cluster at $z> 0.5$
\citep{ebeling2007}, which appears to be accreting matter along a 6-Mpc-long filament \citep{ebeling2004}, and has the
largest known Einstein radius \citep[$\theta_e \sim 55$\asec;][]{zitrin2009}. Four distinct components are identified
from X-ray and optical analyses \citep{ma2009}, and the lensing analyses \citep{zitrin2009,limousin2012} find agreement
in the location of these four mass peaks with those from the X-ray and optical. While the complex X-ray
  morphology is not evident in AMI \citep{rumsey2016} or Bolocam SZ maps, there is still asphericity in the maps.

%%% Rework (subclusters, but without figure...?)
%The four components are labelled in Figure~\ref{fig:m0717_mroczkowski}. 
There are four identified subclusters \citep[labeled A through D][]{ma2009}. They find that subcluster C is the
most massive component, while subcluster A is the least massive, and subclusters B and D are likely remnant cores. The
velocities of the components from spectroscopy are found to be $(v_A, v_B, v_C, v_D) = (+278_{-339}^{+295},+3238_{-242}^{+252},
-733_{-478}^{+486},+831_{-800}^{+843})$ km s$^{-1}$ \citep{ma2009}. The first indication of detection of the kSZ signal
towards these subclusters was presented in \citet{mroczkowski2012}, with a subsequent paper from \citet{sayers2013}
having the first significant detection and derived cluster velocities. Most recently, \citet{adam2016b} has mapped the 
kSZ signal and derived model-dependent subcluster velocities. 

MACS 0717 has also been observed at 610 MHz with the Giant Metrewave Radio Telescope (GMRT) which reveals both a radio
halo and a radio relic \citep{vanweeren2009}. This is interpreted as likely being due to a diffuse shock acceleration
(DSA). More recently, however, deep, higher resolution JVLA data have found a connection between a central
  radio source and the diffuse emission, and favor re-acceleration as the source of the relativistic electrons
  \citep{vanweeren2017}.

We observe a foreground radio galaxy, well outside the cluster centered, which we model as a point source here, 
with flux density of $2.08 \pm 0.25$ mJy at 90 GHz. 
This was previously reported with an integrated flux density of $2.8 \pm 0.2$ mJy and an extended shape 
14.\asec4 $\times$ 16.\asec1 \citep{mroczkowski2012}. However, an improved beam modeling has allowed us to model the 
foreground galaxy given a known beam shape. It is also worth noting that the MUSTANG data itself has been processed 
slightly differently from that presented in \citet{mroczkowski2012}; in this work the map is produced with a common 
calculated as the mean across detectors, whereas in \citet{mroczkowski2012} the common mode was calculated as the 
median across detectors.

%\afterpage{
%\clearpage
%\thispagestyle{empty}
%\begin{figure}
%  \centering
%  \includegraphics[width=0.85\textwidth]{analysis_cres/JF_Conf_Intervals_2params_both_default_speedy_9_Feb_2015_m0717.eps}
%  \includegraphics[width=0.85\textwidth]{analysis_cres/PPP_arnaud_v3_log-log_12_Mar_2015_m0717.eps}
%  \caption{MACS 0717}
%  \label{fig:macs_0717params}
%\end{figure}
%\clearpage
%}

%%%%%%%%%%%%%%%%%%%%%%%%%%%%%%%%%%%%%%%%%%%%%%%%%%%%%%%%%%%%%%

\subsection{MACS 0647 (z=0.59)}
\label{sec:results_m0647}

%%%%%%%%%%%%%%%%%%%%%%%%%%%%%%%%%%%%%%%%%%%%%%%%%%%%%%%%%%%%%%

MACS 0647 is at $z = 0.591$ and is classified as neither a cool core nor a disturbed cluster \citep{sayers2013}. 
It was included in the CLASH sample due to its strong lensing properties \citep{postman2012}.
Gravitational lensing \citep{zitrin2011}, X-ray surface brightness \citep{mann2012}, 
and SZ effect (MUSTANG, see Figure \ref{fig:mustang_maps_sample}, and Bolocam) maps all
show elongation in an east-west direction. \citet{rumsey2016} find a circular SZ morphology with AMI, and take
  the discrepancies in the SZ and X-ray temperatures as an indication of a recent head-on merger.
In the joint analysis presented here, we see that the spherical model provides an adequate fit to both datasets,
as evidenced in Table~\ref{tbl:pressure_profile_results}. Still, from Figure~\ref{fig:resid_maps}, it appears
that some elongation of the bulk ICM or residual feature would better describe the cluster center.
%and we note that the spherical assumption allows for a easier interpretation of the mass profile of the cluster.

%%%%%%%%%%%%%%%%%%%%%%%%%%%%%%%%%%%%%%%%%%%%%%%%%%%%%%%%%%%%%%

\subsection{MACS 0744 (z=0.70)}
\label{sec:results_m0744}

%%%%%%%%%%%%%%%%%%%%%%%%%%%%%%%%%%%%%%%%%%%%%%%%%%%%%%%%%%%%%%

MACS 0744 is neither classified as a cool core cluster nor a disturbed cluster \citep{mann2012,sayers2013}, but qualifies
as a relaxed cluster \citep{mann2012}. There is a dense X-ray core, and a doubly peaked red sequence of galaxies as found
by \citet{kartaltepe2008}. The gas is also found to be rather hot: $k_B T = 17.9_{-3.4}^{+10.8}$ keV, as determined by combining
SZ and X-ray data \citep{laroque2003}. AMI observations \citep{rumsey2016} show some elongation in the plane of the
  sky in their SZ map of this cluster, but they otherwise find that the cluster is in a relaxed state.

The data presented here is the same as in \citet{korngut2011}, but has been processed differently: again, the primary difference
is in the treatment of the common mode. Additionally, \citet{korngut2011} optimize over the low-pass filtering of the common mode
and do not implement a correction factor for the SNR map. The surface brightness significance of the shock feature is the same, 
but is perhaps less
bowed than the kidney bean shape seen previously.  The excess found in \citet{korngut2011}
marked the first clear detection of a shock in the SZ that had not been previously been known from X-ray observations. 
\citet{korngut2011}
reanalyze the X-ray data with the knowledge of the shocked region from MUSTANG, and calculate the Mach number of the shock
based on (1) the shock density jump, (2) stagnation condition between the pressures at the edge of the cold front and just
ahead of the shock, and (3) temperature jump across the shock, and find Mach numbers between 1.2 and 2.1, with a velocity of 
$1827_{-195}^{+267}$ km s$^{-1}$. The shocked region (region II in \citet{korngut2011}) is well modelled with $19.7$ keV gas.

%\afterpage{
%\clearpage
%\thispagestyle{empty}
%\begin{figure}
%  \centering
%  \includegraphics[width=0.85\textwidth]{analysis_cres/JF_Conf_Intervals_2params_both_default_speedy_9_Feb_2015_m0744.eps}
%  \includegraphics[width=0.85\textwidth]{analysis_cres/PPP_arnaud_v3_log-log_26_Feb_2015_m0744.eps}
%  \caption{MACS 0744}
%  \label{fig:macs_0744params}
%\end{figure}
%\clearpage
%}

%%%%%%%%%%%%%%%%%%%%%%%%%%%%%%%%%%%%%%%%%%%%%%%%%%%%%%%%%%%%%%

%%%%%%%%%%%%%%%%%%%%%%%%%%%%%%%%%%%%%%%%%%%%%%%%%%%%%%%%%%%%%%

\subsection{CLJ 1226 (z=0.89)}
\label{sec:results_clj1226}

%%%%%%%%%%%%%%%%%%%%%%%%%%%%%%%%%%%%%%%%%%%%%%%%%%%%%%%%%%%%%%

CLJ 1226 is a well studied high redshift cluster \citep[e.g.][]{mroczkowski2009,bulbul2010,adam2015}. 
\citet{adam2015} find a point source at RA 12:27:00.01 and Dec +33:32:42 with a flux density of 
$6.8 \pm 0.7 \text{ (stat.)} \pm 1.0 \text{ (cal.)}$ mJy at 260 GHz and $1.9 \pm 0.2 \text{ (stat.)}$ at 150 GHz. 
This is not the same point source seen in \citet{korngut2011}, which is reported as a point source
with $4.6\sigma$ significance in surface brightness, and can be fit in our current analysis as a point source 
with a flux density of $0.33 \pm 0.13$ mJy. A short VLA filler observation (VLA-12A-340, D-array, at 7 GHz) 
was performed to follow up this potential source. To a limit of $\sim 50 {\rm \mu Jy}$ nothing is seen, 
other than the clearly spatially distinct radio source associated with the BCG at the cluster center 
(1 mJy at 7 GHz and 3.2 mJy in NVSS). \citet{rumsey2016} find a point source of weak significance with a flux
density of $\sim 0.18$ mJy in CLJ 1226 (at 15 GHz); however, coordinates are not provided and the location indicated
on the maps would be consistent with either the point source found by \citet{adam2015} or \citet{korngut2011}.
In contrast, the point source found in \citet{adam2015} is fit to our 
data with a flux density of $0.36 \pm 0.11$ mJy. Given the slight increase in significance of the point source
from \citet{adam2015}, we adopt that point source location for our pressure profile analysis of CLJ 1226.
 
%%% K09 flux is 0.34 +/- 0.13 mJy in our maps. Now I've written it in. Jan 2016.

In the previous analysis of the MUSTANG data, \citet{korngut2011} find a ridge of significant substructure after 
subtracting a bulk SZ profile (N07, fitted to SZA data). They find that this ridge, southwest of the cluster
center, alongside X-ray profiles, are consistent with a merger scenario. \citet{rumsey2016} also take the
  descrepancy that they find between SZ and X-ray temperature as indicative a merger scenario. When comparing to
  merger simulations, they find CLJ 1226 could be consistent with a head-on minor merger. \citet{adam2015} found
  evidence for a disturbed core, but relaxed on large scales. However, in this work, we do not find any
significant substructure after fitting a bulk component, or other indication of merger activity.
%\textcolor{red}{[I will do an analysis with the point source found in Korngut+2011. The question is if I already 
%have the point source modelled (and to find it).}

\section{Data Products}

We have made MUSTANG data products for the sample of clusters analyzed in this paper available at: 
\protect{\url{https://safe.nrao.edu/wiki/bin/view/GB/Pennarray/MUSTANG_CLASH}}. Links to accompanying
Bolocam and ACCEPT data are available from this website as well. In particular, we have publicized the final
data maps, noise maps, and signal-to-noise (SNR) maps used in this analysis, as well as transfer functions
for individual clusters. Further documentation is available on the website.

%%%%%%%%%%%%%%%%%%%%%%%%%%%%%%%%%%%%%%%%%%%%%%%%%%%%%%%%%%%%%%%%%%%%%%%%%%%%%%%%%%%%%%%%%%%%%%%%%%%%%%%%%%%%%%%%
%%%%%%%%%%%%%%%%%%%%%%%%                    BIBLIOGRAPHY!!!                          %%%%%%%%%%%%%%%%%%%%%%%%%%%
%%%%%%%%%%%%%%%%%%%%%%%%%%%%%%%%%%%%%%%%%%%%%%%%%%%%%%%%%%%%%%%%%%%%%%%%%%%%%%%%%%%%%%%%%%%%%%%%%%%%%%%%%%%%%%%%

\bibliographystyle{apj}
\bibliography{mycluster}
\label{references}

\end{document}